\documentclass[aps,amsmath,prb,twocolumn,superscriptaddress,floatfix]{revtex4-2}
\usepackage{lineno}
\usepackage{graphicx}
\usepackage{dcolumn}
\usepackage{bm}
\usepackage{lipsum}
\usepackage{color}
\usepackage[table]{xcolor}
\usepackage{soul}
\usepackage{amsfonts,eucal,mathrsfs,amssymb}
\def \Helen #1{\textcolor{magenta}{#1}} 
\newcommand{\beginsupplement}{        \setcounter{table}{0}
	\renewcommand{\thetable}{S\arabic{table}}        \setcounter{figure}{0}
	\renewcommand{\thefigure}{S\arabic{figure}}               
}
\begin{document}

\preprint{}
\title{
Structure, control, and dynamics of altermagnetic textures}
\author{O. Gomonay}
\affiliation{Institut f\"{u}r Physik, Johannes Gutenberg-Universit\"{a}t Mainz, Staudingerweg 7, D-55099 Mainz, Germany}
\author{V. P. Kravchuk}
\affiliation{Leibniz-Institut f\"{u}r Festk\"{o}rper- und Werkstoffforschung, Helmholtzstraße 20, D-01069 Dresden, Germany}
\affiliation{Bogolyubov Institute for Theoretical Physics of the National Academy of Sciences of Ukraine, 03143 Kyiv, Ukraine}

\author{R. Jaeschke-Ubiergo}
\affiliation{Institut f\"{u}r Physik, Johannes Gutenberg-Universit\"{a}t Mainz, Staudingerweg 7, D-55099 Mainz, Germany}
\author{K. V. Yershov}
\affiliation{Leibniz-Institut f\"{u}r Festk\"{o}rper- und Werkstoffforschung, Helmholtzstraße 20, D-01069 Dresden, Germany}
\affiliation{Bogolyubov Institute for Theoretical Physics of the National Academy of Sciences of Ukraine, 03143 Kyiv, Ukraine}

\author{ T. Jungwirth}
\affiliation{Inst. of Physics Academy of Sciences of the Czech Republic, Cukrovarnick\'{a} 10,  Praha 6, Czech Republic}
\author{ L. Šmejkal}
\affiliation{Institut f\"{u}r Physik, Johannes Gutenberg-Universit\"{a}t Mainz, Staudingerweg 7, D-55099 Mainz, Germany}
\affiliation{Inst. of Physics Academy of Sciences of the Czech Republic, Cukrovarnick\'{a} 10,  Praha 6, Czech Republic}
\author{J. van den Brink}
\affiliation{Leibniz-Institut f\"{u}r Festk\"{o}rper- und Werkstoffforschung, Helmholtzstraße 20, D-01069 Dresden, Germany}
\affiliation{Institute for Theoretical Physics and W\"urzburg-Dresden Cluster of Excellence ct.qmat, TU Dresden, 01069 Dresden, Germany}

\author{ J. Sinova}
\affiliation{Institut f\"{u}r Physik, Johannes Gutenberg-Universit\"{a}t Mainz, Staudingerweg 7, D-55099 Mainz, Germany}
\affiliation{Department of Physics, Texas A\&M University, College Station, Texas 77843-4242, USA}

\begin{abstract} 
We present a phenomenological theory of altermagnets, that captures their unique magnetization dynamics and allows modelling magnetic textures in this new magnetic phase. Focusing on the prototypical d-wave altermagnets, e.g. RuO$_2$, we can explain intuitively the characteristic lifted degeneracy of their magnon spectra, by 
the emergence of an  effective sublattice-dependent anisotropic spin stiffness arising naturally from the phenomenological theory. 
We show that as a consequence the altermagnetic domain walls, in contrast to antiferromagnets, have a finite gradient of the magnetization,  
with its  strength and gradient direction
connected to the altermagnetic anisotropy, even for 180$^\circ$ domain walls. 
This gradient generates a ponderomotive force in the domain wall in the presence of a strongly inhomogeneous 
external magnetic field, which may be achieved through magnetic force microscopy techniques.
The motion of these altermagentic domain walls is also characterized by an anisotropic Walker breakdown, with much higher speed limits of propagation than ferromagnets but lower than antiferromagnets. 
\end{abstract}
\maketitle
\section{Introduction}
The recent discovery of a third  fundamental class of collinear magnetic order,
termed altermagnets, has opened new possibilities of
 unconventional magnetism \cite{Smejkal2020} with many  material candidates \cite{Smejkal2021a,Smejkal2022a,Guo2023b}. 
Altermagnets exhibit spin-polarized d/g/i-wave order in the non-relativistic band structure, distinct from conventional ferromagnets and antiferromagnets.  By combining the fast magnetic dynamics and robustness to external fields of antiferromagnets with the strong spin-dependent splitting of electronic bands typical of ferromagnets, they offer new functionalities in spintronic applications \cite{Smejkal2022a}. At the same time,  they showcase unique novel phenomena such as the the crystal anomalous Hall effect \cite{Smejkal2020,Feng2022,Sato2023a} and the spin splitter effect \cite{Naka2019,Gonzalez-Hernandez2021}, distinct spectroscopic signatures
\cite{Krempasky2024,Fedchenko2024,Lee2024,Osumi2023}, and  magnonic spectra with anisotropic lifted degeneracy \cite{Smejkal2023,Naka2019,Gohlke2023}. 

The properties of altermagnets arise directly from their defining spin symmetries \cite{Smejkal2021a}. 
Altermagnets have 
opposite spin sublattices connected by a rotation (proper or improper, symmorphic or nonsymmorphic), but not connected by a translation or a center of inversion, delimiting them sharply  from conventional collinear antiferromagnets and ferromagnets \cite{Smejkal2021a,Smejkal2022a}. 
These spin symmetries enforce simultaneously magnetic compensation and time-reversal symmetry ($\mathcal{T}$) breaking of the non-relativistic band structure in reciprocal space with alternating spin polarization, which give rise to their unique characteristics. 
It is clear that this unconventional spin-split band structure should affect the dynamics and magnetic textures of the localized moments in altermagnets.  However, the standard phenomenological formalism, highly successful in the description of conventional ferromagnets and antiferromagnets, is not able to capture these altermagnetic salient characteristics. 

In this paper we present a phenomenological theory that incorporates the specific features of altermagnets into a standard approach to magnetic dynamics. 
Our theory confirms the predicted non-relativistic spin splitting  of the magnon spectra \cite{Smejkal2023,Naka2019,Gohlke2023}. It also predicts 
several unique properties of altermagnetic textures:  inhomogeneous distribution of the magnetization inside the domain wall, the possibility to manipulate domain walls with magnetic force microscopy tools,
and anisotropic Walker breakdown in the altermagnetic domain wall motion. We explain intuitively these features by the emergence of an effective sublattice-dependent anisotropic spin stiffness, whose symmetry is in one-to-one correspondence with the altermagnetic spin splitting of electronic bands. These results of the phenomenological modelling are supported by spin lattice model simulations.

\section{Results}
\textit{Phenomenological model of a $d$-wave altermagnet}. 
The micromagnetic approach for studying magnetic dynamics and textures requires the modeling of magnetic energy guided by the principles of Landau theory. For ferro- and antiferromagnets, prevalent models traditionally focus solely on the spatial positions of magnetic atoms and the orientations of localized spins. In altermagnets, however, the configuration of the non-magnetic atoms and their associated bonds can play a key role in distinguishing them from antiferromagnets. An adequate description of the altermagnetic properties therefore requires an order parameter in addition to the dipolar order parameter of the magnetic vectors.  

Below we introduce such order parameter and construct an altermagnetic contribution into magnetic energy from symmetry considerations, requiring invariance with respect to the spin symmetry group. For illustration we consider a typical example of a $d$-wave altermagnet, e.g. RuO$_2$, whose spin symmetry point group includes the operation [$C_2|| C_4$] \cite{Smejkal2022a}. Here $C_2$ corresponds to rotation in spin space and the operation $C_4$ is a rotation in real space. 

\begin{figure}[t]
\centering
\includegraphics[width=1.0\columnwidth]{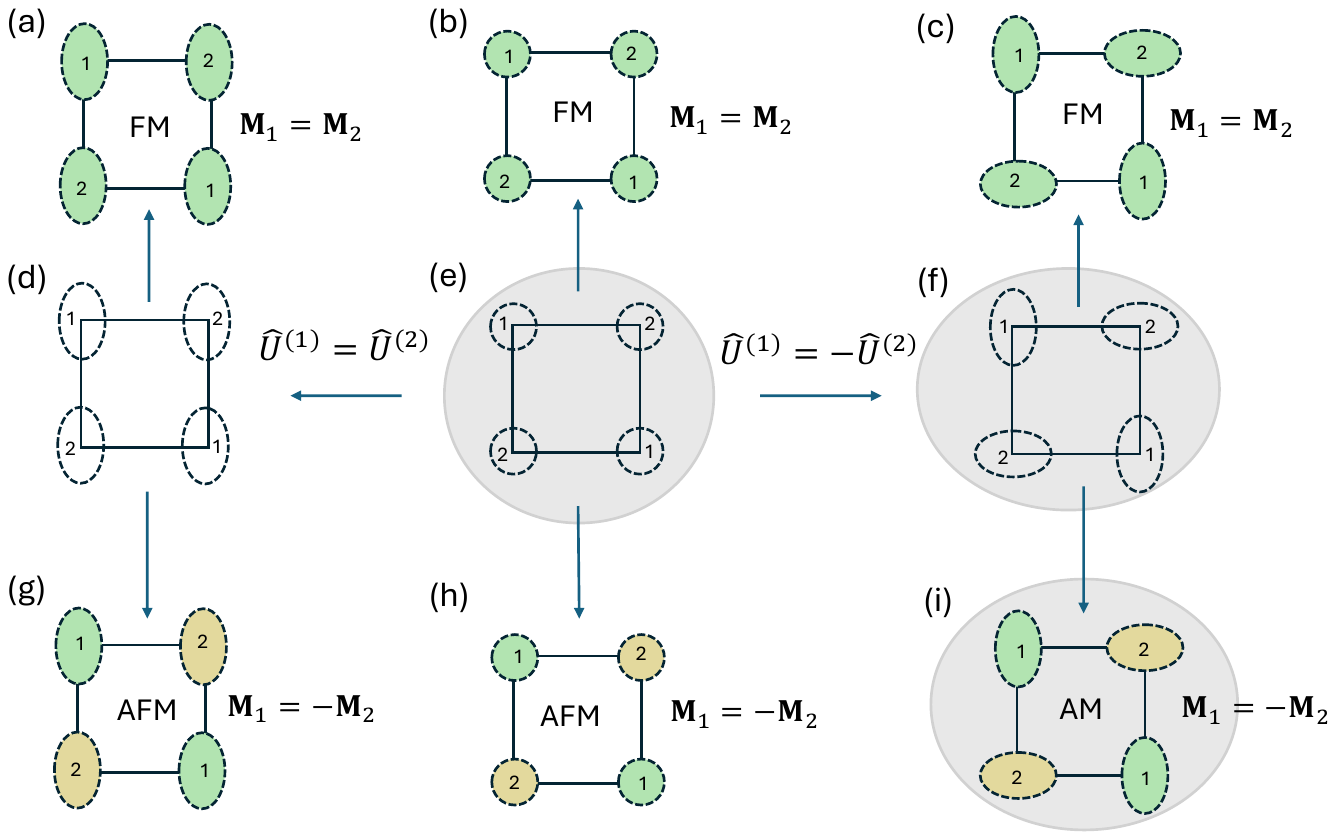}
\caption{Order parameters and possible structural and magnetic phases. All the ferromagnetic (a)-(c), structural (d),(f), and staggered magnetic  (g)-(i)  phases can be obtained  from the hypothetical high symmetry phase (e) with no magnetic order and isotropic local environment of magnetic atoms (shown with circles). (d),(f) Deformation of the local environment (ellipses) leads to two different structural phases with equal ($\hat{U}^{(1)}=\hat{U}^{(2)}$) or staggered ($\hat{U}^{(1)}=-\hat{U}^{(2)}$) local environment.  Magnetic ordering leads to either ferromagnetic ($\mathbf{M}_1=\mathbf{M}_2$, (a)-(c)) or staggered ($\mathbf{M}_1=-\mathbf{M}_2$ (g)-(i)) magnetic structures, depending on the sign of the exchange coupling between atoms 1 and 2. The combination of the staggered environment and the staggered magnetic order parameters leads to an altermangetic phase (i). The positions of the atoms are schematic, in real structures atoms 1 and 2 can be shifted in the vertical direction, e.g. in RuO$_2$.}
 \label{Landau}
\end{figure}

While the magnetic state of an antiferromagnetic system is described in terms of magnetic vectors (the magnetisation $\mathbf{m}=\mathbf{M}_1+\mathbf{M}_2$ and the N\'eel vector $\mathbf{n}=\mathbf{M}_1-\mathbf{M}_2$),
the order parameter describing altermagnetism must capture the reduced local symmetry of the magnetic density \cite{Smejkal2021a,Smejkal2022a}. Within the phenomenological Landau theory, 
 we start from the hypothetical high symmetry phase (non-ordered; see Fig.~\ref{Landau} (a)), 
and proceed to the degrees of freedom that describe {\em separately} the local environment deformations (Fig.~\ref{Landau} (b) and (c))
and the possible localized magnetic order (Fig.~\ref{Landau} (d)-(i)). 

To characterize the symmetry of the local environment, we introduce an additional set of variables (besides $\mathbf{M}_1$ and $\mathbf{M}_2$).  In our particular case these are symmetric second rank tensors $\hat{U}^{(1,2)}$  describing the possible deformation of each sublattice.  
We distinguish between the atoms as being type 1 and 2 
based on the sublattice that we assign them to, 
unconnected to their magnetic order.

In constructing the  Landau energy functional that connects to the altermagnetic order, we must combine
the new variables, $\hat{U}^{(1,2)}$, in a way that will be invariant in both the high and low symmetry phases that we seek to describe (Fig.~\ref{Landau} (a), (b), and (i)). 

In a simplified RuO$_2$, $\hat{U}^{(1,2)}$ can be seen as a deformation of the circular cage into an ellipse with the long axis along [110] (for sublattice 1) or [1$\bar{1}$0] (for sublattice 2). This structure is dictated by the Wyckoff point group $mmm$, where two of the mirror planes are (110) and (1$\bar{1}$0). Within the reference frame related to the crystallographic axes $x\|[100]$ and $y\|[010]$, the only non-trivial components of the $\hat{U}^{(1,2)}$ tensors are $U^{(1)}_{xy}$ and ${U}^{(2)}_{xy}$. The $C_4$ lattice rotation, which additionally permutes the Wyckoff positions 1 and 2, induces the transformation $U^{(1)}_{xy}\leftrightarrow -U^{(2)}_{xy}$. As a result, the combination $U^{(1)}_{xy}-U^{(2)}_{xy}$ is invariant with respect to the point group of the low symmetry phase Fig.~\ref{Landau}(b).

To obtain the altermagnetic terms of the non-relativistic energy functional   within this Landau phenomenological approach,  we look for invariant terms with respect to spin rotations. This enforces that there can only be scalar products of the vectors $\mathbf{m}$, $\mathbf{n}$, or their spatial derivatives. In addition, we must have terms that are invariant under the combined operations of $C_4$ lattice rotation with time reversal.
The simplest combination  invariant with respect to the $C_4$ lattice rotation and spin inversion (time reversal)
is then $(U^{(1)}_{xy}-U^{(2)}_{xy})(\partial_x\mathbf{m}\cdot\partial_y\mathbf{n}+\partial_y\mathbf{m}\cdot\partial_x\mathbf{n})$. By setting $U^{(1)}_{xy}-U^{(2)}_{xy}=1$ we get the additional energy term reflecting the altermagnetic symmetry in the magnetic dynamics as $A_\mathrm{ani}\left(\partial_x\mathbf{m}\cdot\partial_y\mathbf{n}+\partial_y\mathbf{m}\cdot\partial_x\mathbf{n}\right)$, where $A_\mathrm{ani}$ is a phenomenological coefficient that we will refer to as the anisotropic altermagnetic stiffness (AAS). Note that the same expression can be obtained directly from the Heisenberg Hamiltonian of an altermagnet by gradient decomposition, assuming smooth variation of the magnetic vectors (see sections I,II of Supplementary Materials for a derivation from a Heisenberg minimal model).

\begin{figure}[t]
\centering

\includegraphics[width=1.0\columnwidth]{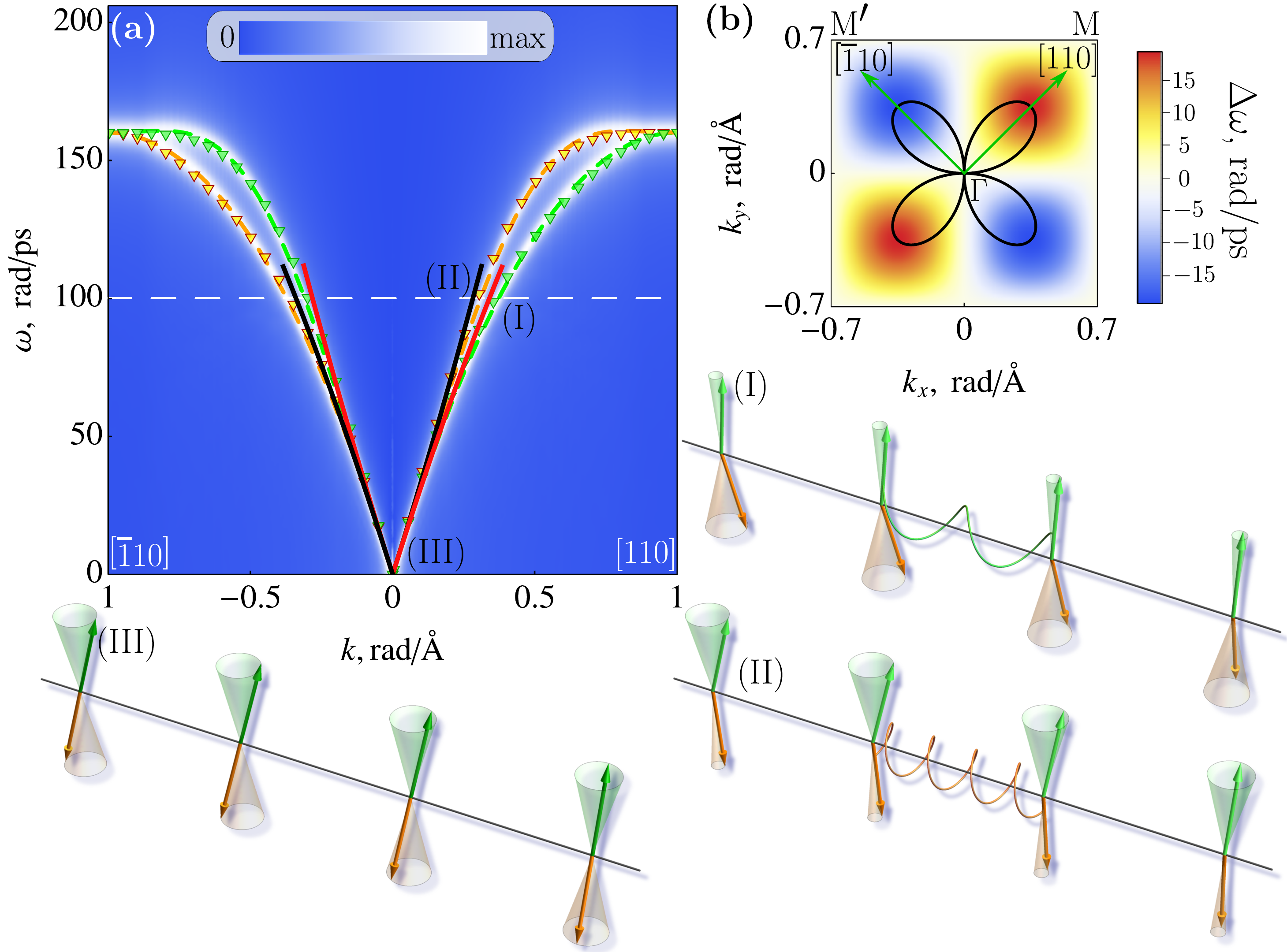}
\caption{ Splitting of magnon modes in an altermagnet. (a) Splitting of magnon modes in RuO$_2$ in two orthogonal directions calculated from DFT  \cite{Smejkal2023} (markers), atomistic {spin model} ({dashed lines}) and phenomenological {model (solid} lines){; the blue/white colors encode spectra obtained using spin lattice model simulations, see Supplementary Materials.}  
(I),(II) Spin-up and spin-down magnons with large $\mathbf{k}$ are determined by different stiffnesses (indicated by springs), while for small momenta magnons (marked by  III) 
 the dispersion is determined by the intersublattice exchange and the difference between the two magnon branches becomes negligible.
 (b) Angular (solid line; arbitrary units) and magnitude (color code) dependence of the  frequency splitting of two magnon modes.
 \label{fig_magnon_splitting}}
\end{figure}

To explain the physical origin of the $A_\mathrm{ani}$ coefficient, we turn for a moment to the representation in terms of magnetic sublattices. In this case the dominant inhomogeneous terms in the free energy density, $w_\mathrm{inh}$, are 
\begin{eqnarray}\label{eq_exchange_sublattice}
w_\mathrm{inh}&=&A_\mathrm{iso}\left[\left(\nabla\mathbf{M}_1\right)^2+\left(\nabla\mathbf{M}_2\right)^2\right]\nonumber\\
&+&2A_\mathrm{ani}\left[\partial_x\mathbf{M}_1\partial_y\mathbf{M}_1-\partial_x\mathbf{M}_2\partial_y\mathbf{M}_2\right],
\end{eqnarray}
where $A_\mathrm{iso}$ is isotropic stiffness typical for ferro- and antiferromagnets. As can be seen from Eq.~\eqref{eq_exchange_sublattice} both coefficients, $A_\mathrm{iso}$ and $A_\mathrm{ani}$, parametrize the \textit{intra}sublattice exchange. However, unlike antiferromagnets, the sublattices have different effective stiffness, and the difference depends on the direction of the inhomogeneity: 
it is zero along the [100] and [010] directions (corresponding to the nodal planes in reciprocal space)
maximal in the [110] and [1$\bar{1}$0] directions (corresponding to the maximum spin-splitting in reciprocal space),
and changes sign with rotation through 90$^\circ$ (see solid line in Fig.~\ref{fig_magnon_splitting}(b)). \Helen{}
We next illustrate the effects of AAS by considering magnetic dynamics of magnons and domain walls.

\textit{Magnons}. In a uniaxial antiferromagnet the magnons are spin-polarised along the easy magnetic axis and the two branches corresponding to two opposite spins are degenerate throughout the Brillouin zone \cite{Gomonay2018a,Rezende2019}. In contrast, in altermagnets the anisotropy of the exchange stiffness removes the degeneracy of the spin-up and spin-down modes in a large region of the Brillouin zone, except near its centre, edges and nodal planes \cite{Smejkal2023}.

In the context of the phenomenological theory, this effect can be understood by considering the spin-up and spin-down magnons with relatively large $\mathbf{k}$, which are mostly localised at different magnetic sublattices (see Fig.~\ref{fig_magnon_splitting}(I),(II), and Sec. IIB of Supplementary Material). As such, their velocities are determined by a different effective stiffness, ($A_\mathrm{iso}\pm A_\mathrm{ani}$) for $\mathbf{k}\|[110]$ and  $[1\bar{1}0]$ respectively), and their frequencies are therefore split. The splitting disappears near the $\Gamma$ point (BZ centre, Fig.~\ref{fig_magnon_splitting}(III)), where the magnons are evenly distributed between the two sublattices and the effective stiffness is mostly determined by isotropic inhomogeneous exchange $A_\mathrm{iso}$ and strong homogeneous \textit{inter}sublattice exchange. 
At the BZ edges, the two sublattices are equivalent, which forces the magnons at these momenta to be degenerate. 
At these BZ edge wave-vectors, the magnons can be considered as coherent oscillations with effective zero wave vector of the spins of one of the two sublattices, and as a result the AAS plays no role. 
 At the nodal planes, the AAS vanishes, aligning the magnetic dynamics of magnons with that of antiferromagnetic behavior. As expected from symmetry grounds, the degeneracy of the magnon modes matches the observed degeneracy in electronic band spin splitting~\cite{Smejkal2023}. 

Figure~\ref{fig_magnon_splitting}(a) shows the magnon spectra of RuO$_2$ calculated from the microscopic Hamiltonian \cite{Smejkal2023} (symbols), 
by means of the spin lattice model simulations (color code) and the fit (solid line) by the analytical dependence 
\begin{equation}\label{eq_spectra}
\omega_{\pm}=\sqrt{\omega_\mathrm{AFMR}^2+c^2k^2}\pm 2\gamma A_\mathrm{ani}M_s k_xk_y,
\end{equation}
 derived from the phenomenological model (see Methods). Here $\omega_\mathrm{AFMR}$ is the frequency of antiferromagnetic resonance \footnote{~Strictly speaking, in the case of altermagnets, we should call it altermagnetic resonance. However, since the difference between alter- and antiferromagnetic dynamics disappears at the $\Gamma$ point, we retain the traditional term antiferromagnetic resonance.} (AFMR), $c$ is the magnon velocity along the nodal planes, $\gamma$ is gyromagnetic ratio, $M_s/2$ is a saturation magnetization of a sublattice. 
 The dependence \eqref{eq_spectra} 
 is consistent with different microscopic models reported up to now \cite{Hodt2023,Brekke2023} in a wide range of $k$ values, including atomistic spin model, see Supplementary Material, Sec. I.B.
 The estimated values of the parameters for RuO$_2$,
 obtained by fitting {\it ab-initio} calculations \cite{Smejkal2023},
 are $c=35$ nm/ps, $\gamma A_\mathrm{ani}M_s=0.97$ rad$\cdot$nm$^2$/ps, which gives the relative splitting $\Delta\omega/\omega\geq 5\%$ starting from $k\geq 1$ rad/nm. The maximum value of the splitting obtained from the atomistic model is $\leq 15\%$, which allows the contribution of the AAS to be considered  as a pertubative correction. 

 \textit{Domain wall}. Next, we compare the antiferromagnetic and altermagnetic domain walls separating domains with opposite orientation of the N\'eel vector. We restrict our consideration to the Bloch-type wall, where the N\'eel vector rotates perpendicular to the inhomogeneity axis $\xi$. We also ignore the Dzyaloshinskii--Morya interactions~(DMI), which are allowed by symmetry and will be discussed below.  

\begin{figure}[t]
\centering
\includegraphics[width=1.0\columnwidth]{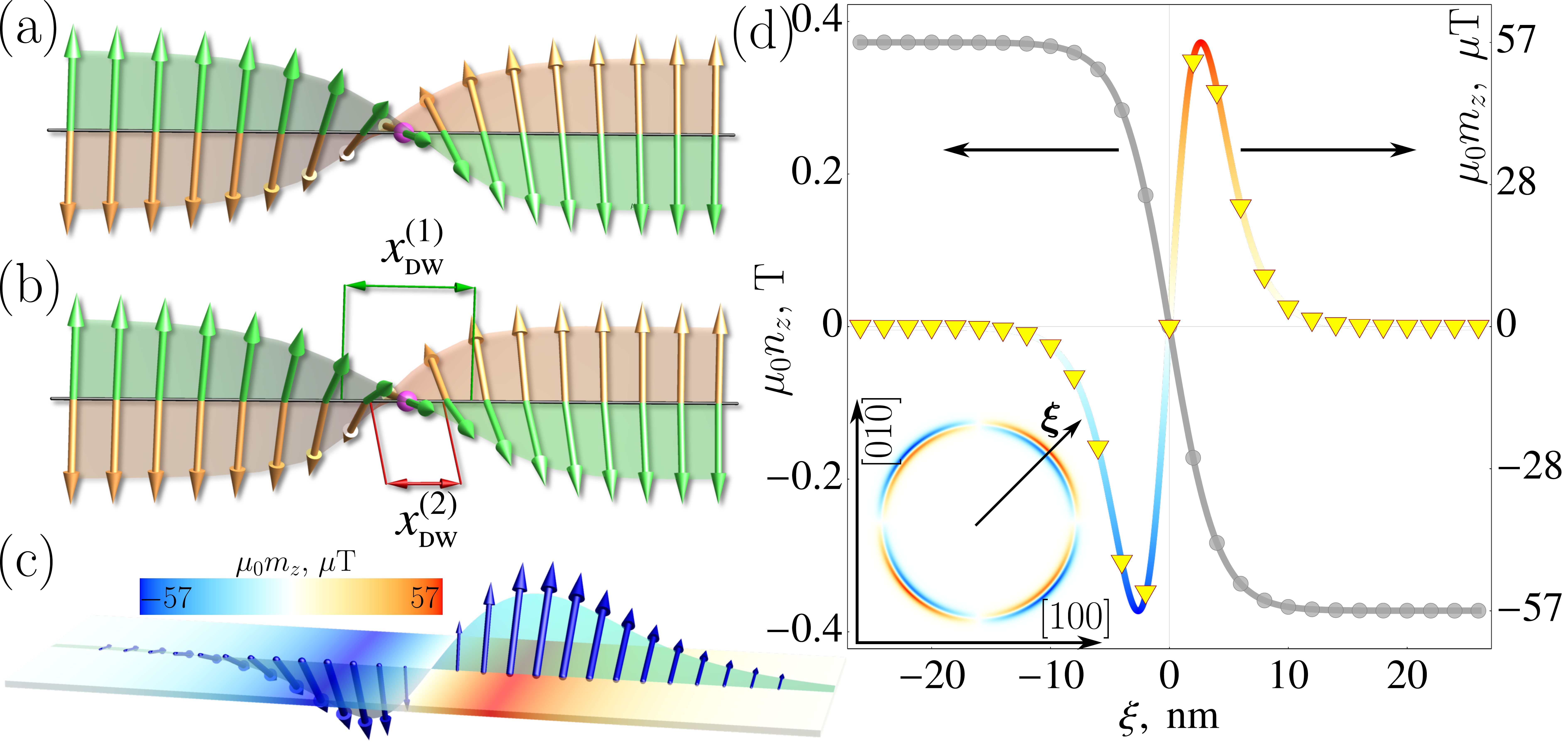}
\caption{Magnetization of the domain wall. Comparison of the static AF (a) and AM (b) domain walls (see text for explanations). (c) Distribution of the net magnetization inside the domain wall (exaggerated). (d) Calculated profiles of the N\'eel vector ($n_z$) and magnetization ($m_z$) components along the easy magnetic axis for the altermagnetic domain wall ($\xi\|[110]$) . Symbols: simulations, solid lines: analytical modelling.  Inset: distribution of $m_z$ depending on orientation of the domain wall with respect to crystallographic axes. Arrow ($\vec{\xi}$) shows one of the directions in which the altermagnetic effect is maximal. For estimations we take $\mu_0M_s=0.37$~T.
\label{fig_magnetization_DW}}
\end{figure}

In the direction of the nodal planes the distinction between altermagnetic and antiferromagnetic states disappears, rendering domain walls with $\xi$ parallel to the nodal plane equivalent to antiferromagnetic domain walls. 
In these walls, the intrasublattice exchange does not break the equivalence between sublattices. 
This allows the opposite orientation of the sublattice magnetisation, $\mathbf{M}_1(\uparrow)$ and $\mathbf{M}_2(\downarrow)$, prescribed by the antiparallel coupling, and thus exact compensation of the magnetic moments throughout the texture, see Fig.~\ref{fig_magnetization_DW}(a). However, in all other orientations of $\xi$ the individual domain walls, which are localised at each of the magnetic sublattices, have different widths, $x^{(1)}_\textsc{dw}\propto\sqrt{A_\mathrm{iso}+A_\mathrm{ani}}$ and $x^{(2)}_\textsc{dw}\propto\sqrt{A_\mathrm{iso}-A_\mathrm{ani}}$, due to the difference in intrasublattice exchange, see Fig.~\ref{fig_magnetization_DW}(b), resulting from the difference in the local environment, see Fig.~\ref {Landau}(i).  In this case, exact compensation within the domain wall is not possible, resulting in a finite inhomogeneous total magnetisation within the domain wall, see Fig.~\ref{fig_magnetization_DW}(c). 

\begin{figure}[t]
\centering
\includegraphics[width=1.0\columnwidth]{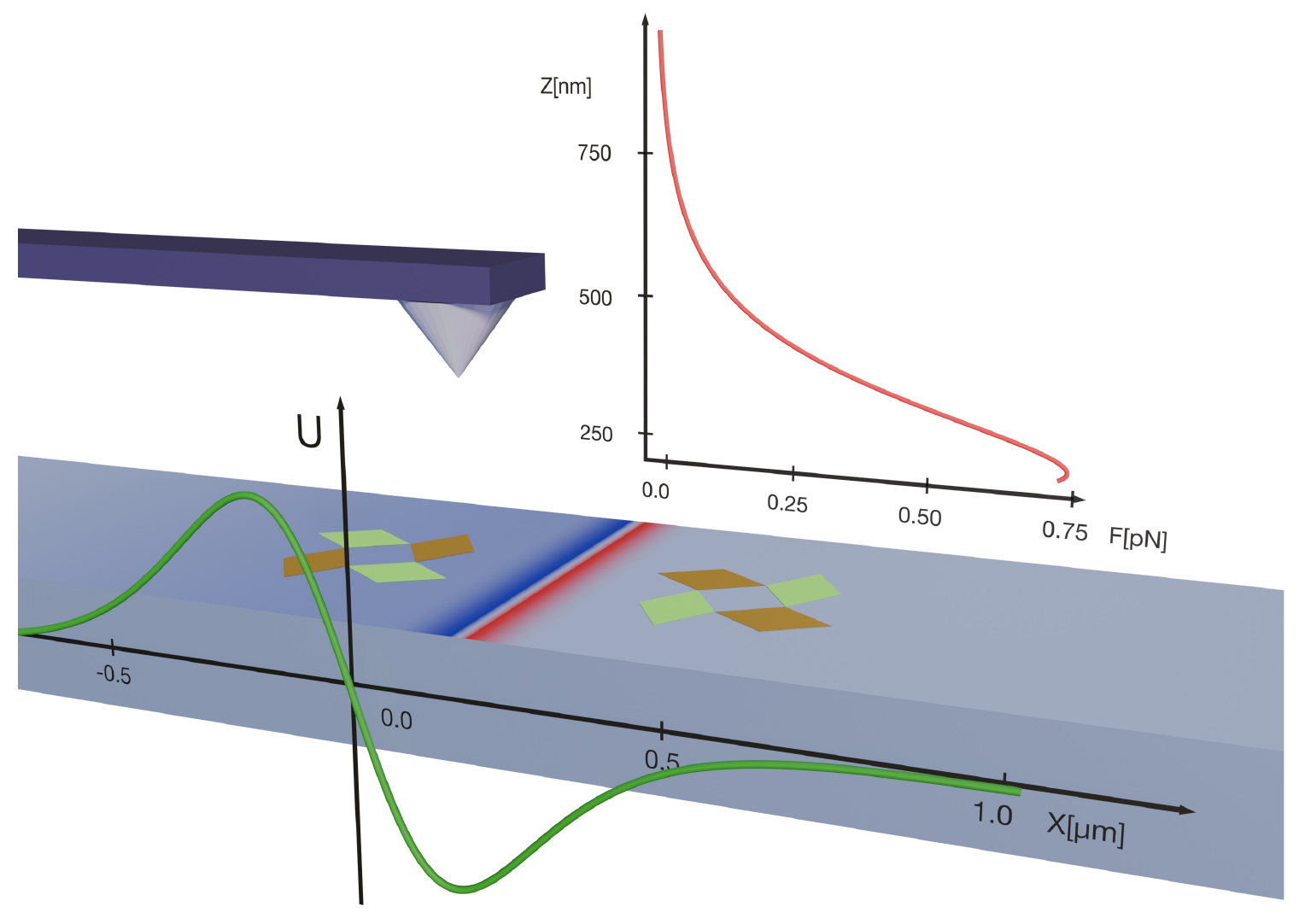}
\caption{Manipulation of the altermagnetic domain wall by a magnetic tip. The green and gold diamonds on either side of the wall show the distribution of the altermagnetic order parameters $U_{xy}^{(1)}$, $U_{xy}^{(2)}$ in each domain.  The color code shows the spatial distribution of $m_z$ (as in Fig. 2(c)). (Green line) The potential (Zeeman) energy  in the magnetic field of the tip as a function of the domain wall position.  The distance between the tip and the surface is 0.4~$\mu$m.  For the calculations we used the field distribution of a ferromagnetic Nd$_2$Fe$_{14}$B particle \cite{Urban2006}  with a radius of 100 nm and magnetization $\mu_0M_s=2.26$~T. The width of the altermagnetic domain wall is $x_\textsc{dw}=4$~nm.   
(Red line) The force generated at the tip by the pinned domain wall as a function of  the vertical distance $Z$.} \label{fig_tip}
\end{figure}

One of the magnetization components is parallel to the easy magnetic axis ($[001]$ for RuO$_2$) and has opposite signs on opposite sides of the domain wall (Fig.~\ref{fig_magnetization_DW}(c,d)). Although the total magnetization in this direction is zero, the domain wall has a non-zero multipole moment $\mathcal{M}_{z\xi}=\int m_z\xi d^3\mathbf{r}$ (according to classification \cite{Bhowal2022}) and couples with the magnetic field gradient $\partial_\xi B_z$.  This opens up the unique possibility of detecting and manipulating the position of the 180$^\circ$ domain wall with magnetic tips used in scanning probe microscopy \cite{Nazaretski2008,Bhallamudi2013}. In Fig.~\ref{fig_tip}(a) the green line shows the calculated profiles of the Zeeman energy of the domain wall in the magnetic field generated by a spherical magnetic particle.  The $X$ coordinate corresponds to the horizontal distance between the tip and the centre of the domain wall. Near the tip the energy shows a minimum corresponding to the equilibrium position $X_\mathrm{eq}$ of the non-pinned domain wall. The position of the minimum shifts from the tip as the vertical distance $Z$ between the tip and the sample surface increases (see section V of Supplementary Materials). 

The horizontal force that drives the domain wall motion to the equilibrium position is highly anisotropic due to the anisotropy of the exchange stiffness and shows the same angular dependence (see Fig.~\ref{fig_magnon_splitting}(b)). This is a unique feature of altermagnetic materials, as the magnetic field does not discriminate between the 180$^\circ$ antiferromagnetic domains and is unable to move the 180$^\circ$ antiferromagnetic domain walls. We have also calculated the vertical force acting on the tip in the presence of a pinned domain wall. The magnitude of this force depends on the vertical distance and decreases from $\approx 1$ to $0.01$~pN as $Z$ increases from 0.25 $\mu$m to 2 $\mu$m (see Supplementary Material for details). Notably, this range aligns with values that are experimentally accessible for detection \cite{Rugar1992}.  

We note that the magnetic field gradient can also induce the dynamics of an antiferromagnetic domain wall by modulating the effective anisotropy \cite{Yuan2018} or by coupling with small uncompensated magnetization \cite{Tveten2016}. However, in both cases the effective force is proportional to the anisotropy (spin-orbit interactions) and does not depend on the orientation of the domain wall, which is in contrast to the case of altermagnetism. In these cases, the origin of the small magnetization is usually attributed to a topological Weiss-Zumino (or Berry phase) parity-breaking term (see \cite{Haldane1983a, Fradkin1988, Papanicolaou1995,Ivanov1995a}. The absolute value scales as a small ratio of the interatomic distance to the domain wall width and is often neglected in continuous models.

\begin{figure}[t]
\centering
\includegraphics[width=1.0\columnwidth]{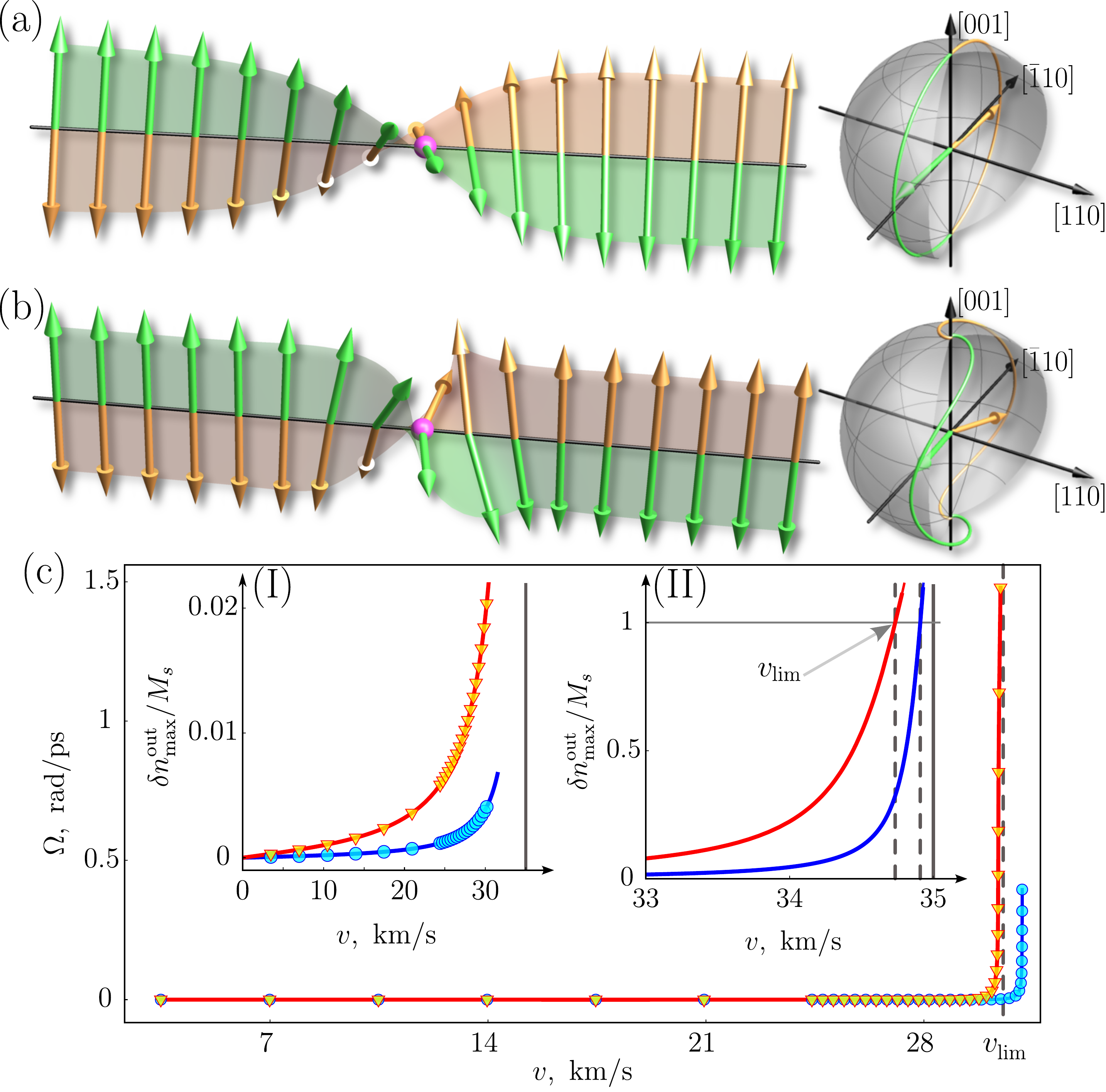}
\caption{Deformation of the moving domain wall and Walker breakdown: Snapshots of the static (a) and moving (b) domain wall. The motion induces canting of the sublattice magnetizations in the domain wall center (magenta spheres) due to the difference in intrasublattice stiffnesses. The canting creates intersublattice exchange torques that further rotates the magnetic moments out of the domain wall plane. The right panels show the spatial trajectories of the sublattice magnetizations in the static and moving domain wall. (c)~Frequency $\Omega$ of oscillations between Bloch and N\'eel type for the steady moving domain wall versus velocity, calculated for two values of AAS $\gamma A_\mathrm{ani}M_s$: 0.97 (red) and 0.2 (blue) rad$\cdot$nm$^2$/ps. The insets show the velocity dependence of the motion-induced component $\delta n_\mathrm{out}/M_s$. The moving domain wall loses stability when $\delta n_\mathrm{out}=M_s$ (Walker breakdown), as shown by the arrow line. Symbols are the results of spin lattice model simulations, solid lines are calculated from the analytical phenomenological model (see Section III of Supplementary Materials).  For calculations we use $x_\textsc{dw}=$4~ nm, the limiting magnon velocity in nodal directions is $c$=35 km/s.  \label{fig_Walker}}
\end{figure}

\textit{Moving domain wall and the Walker breakdown.}
Antiferromagnets exhibit significantly higher domain wall velocities than ferromagnets because they do not undergo the Walker breakdown  (see, e.g. \cite{Gomonay2016}). In ferromagnets the Walker breakdown occurs as a consequence of domain wall deformation during motion. A similar deformation of the domain wall and the resulting modification of the magnetic dynamics was reported long ago~\cite{Gomonai1990} for weak ferromagnets  and was explained by the presence of DMI. In particular, it was shown that the presence of a small magnetisation can reduce the effective velocity of the domain wall below the magnon velocity. We expect a similar deformation of the altermagnetic domain wall due to the different effective stiffnesses governing the individual domain wall widths and velocities. This effect is most pronounced in the centre of the domain wall (indicated by the magenta sphere in Fig.~\ref{fig_Walker}(a)), where at {$v=0$}, in a static state, the two sublattice magnetizations are strictly antiparallel. This exact compensation is broken at {$v\ne 0$} due to the motion of the sublattice domain walls with different thicknesses (Fig.~\ref{fig_Walker}(b)). The resulting canting induces internal torques that rotate $\mathbf{M}_1$ and $\mathbf{M}_2$ out of the domain wall plane. Note that this effect occurs in addition to the field/current induced canting in other directions that induces domain wall motion. The deformation of the steady moving domain wall is thus quantified by the out-of-plane component, $\delta n_\mathrm{out}$, of the N\'eel vector, shown in  Fig. 5 (b) of Supplementary Material.

Similar to the ferromagnetic domain wall, we can estimate the limiting velocity (Walker breakdown) from  the condition that the deformation reaches the possible maximum value $\delta n_\mathrm{out}=M_s$ (see Fig.~\ref{fig_Walker}(c)):
\begin{equation}\label{eq_limiting}
v^\mathrm{AM}_\mathrm{lim}\approx c\sqrt{1-\left(\frac{A^2_\mathrm{ani}M_s}{2A_\mathrm{iso}H_\mathrm{ex}x^2_\textsc{dw}}\right)^{1/3}}.
\end{equation}

Acceleration above $v^\mathrm{AM}_\mathrm{lim}$ induces internal oscillations between the Bloch and N\'eel type and slows down the translational motion of the domain wall. In atomistic spin-lattice simulations, the Walker breakdown (see Fig.~\ref{fig_Walker}(c)) appears as an exponential growth of the oscillation frequency, which allows the value $v_\mathrm{lim}$ to be determined more accurately. For the detailed analysis of the domain wall dynamics below and above the Walker breakdown, we refer the reader to Sec. III of the Supplementary Materials.  

The limiting velocity $v^\mathrm{AM}_\mathrm{lim}$ is smaller than the limiting velocity $c$ of the domain wall motion in antiferromagnets, but still much higher than the Walker velocity  \cite{Schryer1974a} $v^\mathrm{FM}_\mathrm{lim}$  in ferromagnets. This can be seen by comparing Eq.~\eqref{eq_limiting} with the expressions  $c=v^\mathrm{AFM}_\mathrm{lim}=\gamma \sqrt{A_\mathrm{iso}H_\mathrm{ex}/M_s}$ and  $v^\mathrm{FM}_\mathrm{lim}\approx\gamma \sqrt{A_\mathrm{iso}H_\mathrm{dip}/M_s}$, where $H_\mathrm{ex}$ is the intersublattice exchange field and $H_\mathrm{dip}$  is the dipolar anisotropy field in a ferromagnet.We assume that the magnetic anisotropy of a ferromagnets is of the same order of value as $H_\mathrm{dip}$. Considering  that the typical value of the exchange field is several orders of magnitude larger than $H_\mathrm{dip}$, we get $v^\mathrm{FM}_\mathrm{lim}\ll v^\mathrm{AM}_\mathrm{lim}\le c$.

\section{Discussion and conclusion}
 The emergence of altermagnets has given us materials 
 that amalgamate the strong spin splitting of electronic bands, characteristic of ferromagnets, with compensated magnetic order and exchange-enhanced magnetic dynamics commonly observed in antiferromagnets. In this study,  we have demonstrated that the magnetic properties of altermagnets  yield properties similar to those of ferromagnets and of antiferromagnets, as well as some unique to altermagnets alone.

In particular, altermagnetic textures inherit the fast internal dynamics characteristic of antiferromagnets. On the other hand, unlike in antiferromagnets, altermagnetic domain walls can be controlled by inhomogeneous magnetic fields and their velocity is constrained by the Walker breakdown, akin to ferromagnets. Lastly, the anisotropic splittings of the magnon bands represent distinctive altermagnetic characteristics. In contrast to electronic spin splitting, which for  low frequency responses is experimentally accessible primarily in conductors, these magnetic effects can also occur in insulators, making them an efficient tool for the identification of altermagnetic materials. 

We would like to emphasize that our model is quite general and can be applied to any of the candidates among the approximately 200 predicted d-type altermagnets.  However, in this paper we focused on RuO$_2$ motivated by the recent experimental confirmations of its altermagnetic nature, both in the bulk \cite{Fedchenko2024,Lin2024} and in the thin films \cite{Feng2022, Bose2022, Karube2022, Bai2022,Zhou2023,Jeong2024}. On the other hand, we also note that the bulk magnetic ordering is still being discussed; see e.g. \cite{Ryden1970a, Riga1977} and the more recent \cite{Hiraishi2024} experiments. The recent theory analysis as a function of defects in Ref.~\onlinecite{Smolyanyuk2024} clarifies this issue by relating the magnetic ordering in RuO$_2$ to the presence of the defect, which are more abundant in thin films, in agreement with the experimental results.

In our discussion of altermagnetic textures we have neglected DMI, expecting its effect to be relatively small compared to the AAS effects due to its relativistic nature.
Although both DMI and AAS can induce non-zero magnetization, the AAS induced magnetization depends on the gradients 
of the N\'eel vector and rotates with the N\'eel vector in spin space (See Eq.~\ref{eq_AAS_magnetization}). 
In contrast, the DMI-induced magnetization depends on the orientation of the N\'eel vector with respect to the crystallographic axes and can disappear for certain orientations (See Eq.~\ref{eq_DMI_magnetization}).

\begin{figure}[t]
 \centering
 \includegraphics[width=1.0\columnwidth]{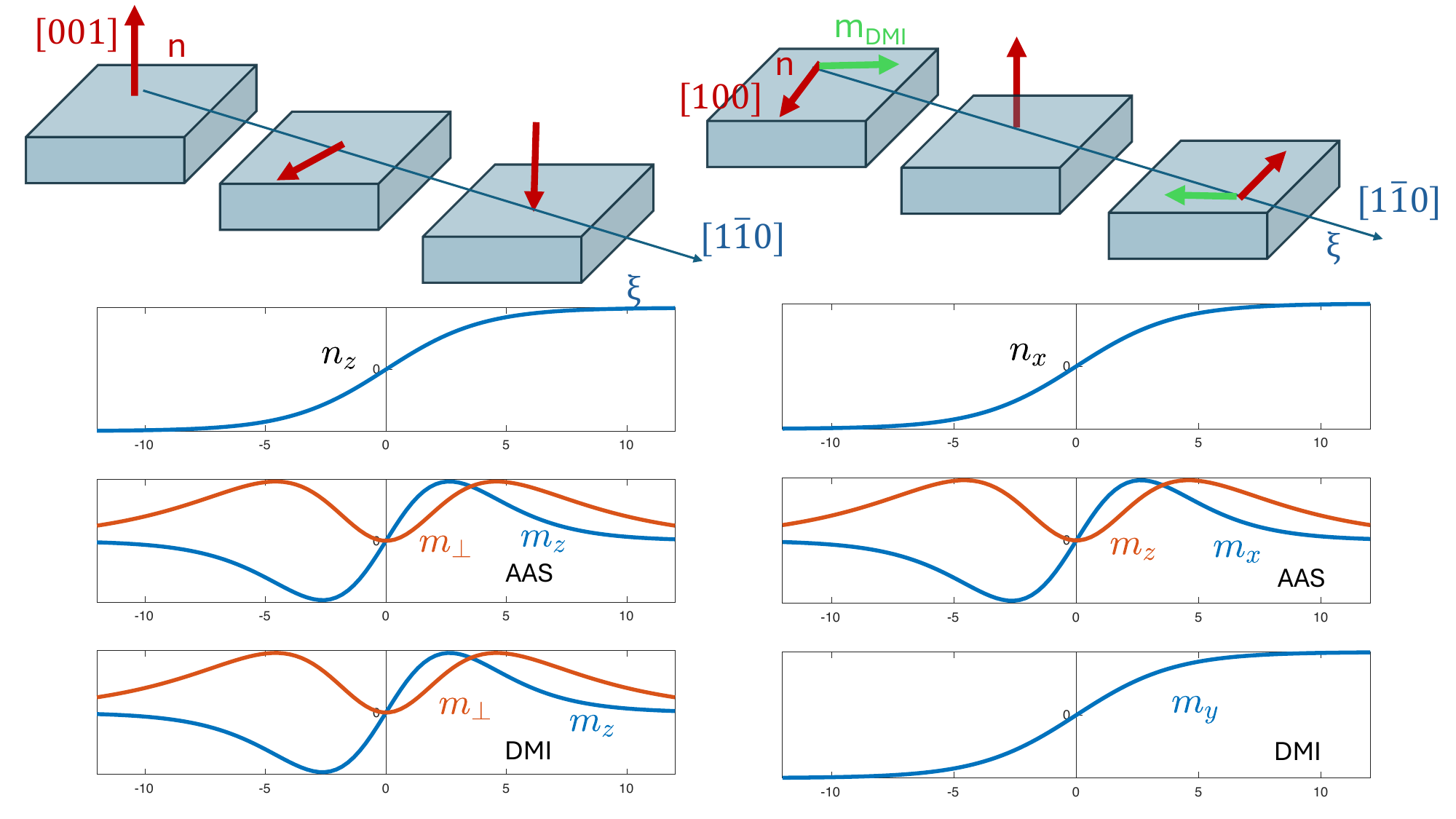}
 \caption{Difference between AAS and DMI effects in the textures. Left panel: Domain wall separating states with $\mathbf{n}\|[001]$ and $\mathbf{n}\|[00\bar{1}]$ (red arrows). Right panel: the same for $\mathbf{n}\|[100]$ and $\mathbf{n}\|[\bar{1}00]$. The graphs show the profile of the domain wall, nontrivial magnetization components induced by AAS and nontrivial magnetization components induced by DMI. In the case of the domain wall with $\mathbf{n}\|\langle 001\rangle]$, the AAS- and DMI- induced magnetizations can have the same or opposite sign depending on the direction of $\xi$.
  \label{fig_notes}}
 \end{figure}

When applicable, we can incorporate within our approach  the standard DMI terms in d-wave altermagnets, $D_\mathrm{long}(\mathbf{m}\cdot\mathbf{n})n_xn_y$ and $D_\mathrm{trans}(m_xn_y+m_yn_x)$, which are responsible for the potential longitudinal \cite{Dzialoshinskii1958} and transverse \cite{Dzyaloshinskii1957} weak ferromagnetism. To illustrate the difference between DMI-induced magnetization and AAS-induced magnetization, we consider a texture with the same spatial distribution but different orientations of the N\'eel vector and incorporate only the transverse DMI. 
Figure \ref{fig_notes} compares two cases: i) the N\'eel vector is along the easy axis [001]; ii) the N\'eel vector is along [100]. In both cases, the 
direction of the inhomogeneity is along [110] (maximum AAS effect).
 In case i) the {\em spatial} distribution of the DMI and AAS induced magnetizations is the same, the only difference being the value of the effect. In both cases the magnetizations are localized inside the domain wall. 
 In the second case, the DMI-induced magnetization within the domains is nonzero and perpendicular to the AAS-induced magnetization within the domain wall. The AAS-induced magnetization is localized only within the domain wall, as in case ii).
 From this realization, we can assume that 
 some prior estimations of DMI 
 (prior to the discovery of altermagnetism) were in fact an  overestimation of the DMI values, 
 with the contributions originating  from AAS not having been known and most likely dominating the magnitude of the effect. 
 
We have presented a phenomenological approach, supported by spin atomistic model calculations, that captures new rich dynamics of alteramagnetic textures and proposed means to control them. 
While other alternative approaches can be used based on the multipole series of the spin density in the real \cite{McClarty2023} and in the reciprocal space \cite{Bhowal2022}, these approaches require a conceptual modification of the equation of magnetic dynamics. Our phenomenological approach allows the description of magnetic dynamics and thermodynamics to be naturally extended to altermagnets, while retaining the N\'eel paradigm of magnetic sublattices.

\section{Methods}
To describe the altermagnetic textures and magnetic dynamics analytically we use a standard approach based on the Landau-Lifshitz equations for the N\'eel vector $\mathbf{n}$ and the magnetisation $\mathbf{m}$  (see, e.g. \cite{Hals2011}) :
\begin{eqnarray}\label{eq_dynamics_general}
\dot{\mathbf{m}}&=&\gamma\left(\mathbf{m}\times \frac{\delta W}{\delta\mathbf{m}}+\mathbf{n}\times \frac{\delta W}{\delta \mathbf{n}}\right),\\
\dot{\mathbf{n}}&=&\gamma\left(\mathbf{n}\times \frac{\delta W}{\partial \mathbf{m}}+\mathbf{m}\times \frac{\delta W}{\delta \mathbf{n}}\right),\nonumber
\end{eqnarray}
where $\gamma$ is the gyromagnetic ratio and $W=\int w d\mathbf{r}$ is the magnetic free energy of an altermagnet. In a $d$-wave altermagnet the free energy density $w$ is modelled as
\begin{eqnarray}\label{eq_free_energy}
w&=&\frac{1}{2M_s}H_\mathrm{ex}\mathbf{m}^2-\frac{1}{2M_s}H_\mathrm{an}n_z^2+w_\mathrm{inh},\\
w_\mathrm{inh}&=&\frac{1}{2}A_\mathrm{iso}(\nabla\mathbf{n})^2+A_\mathrm{ani}\left(\partial_x\mathbf{m}\cdot\partial_y\mathbf{n}+\partial_y\mathbf{m}\cdot\partial_x\mathbf{n}\right),\nonumber
\end{eqnarray}
where $M_s/2$ is a saturation magnetization of each of the magnetic sublattices, $H_\mathrm{ex}$ parametrizes the intersublattice exchange that keeps the sublattices antiparallel,
$A_\mathrm{iso}$ and $A_\mathrm{ani}$ are the isotropic and anisotropic exchange stiffnesses, respectively, $H_\mathrm{an}$ is the field of magnetic anisotropy. We assume uniaxial magnetic anisotropy with the easy axis along $z$. In Eq.~\eqref{eq_free_energy} we have neglected the small term $A_\mathrm{iso}\mathbf{m}^2$. The details of the calculations are explained in Supplementary Materials, Sections II-III.

For the analytical treatment we exclude the magnetization from Eqs.~\eqref{eq_dynamics_general}, \eqref{eq_free_energy}  and express it as a function of the N\'eel vector and its time and space derivatives (see Sections II and III of the Supplementary Materials). In the particular case of the equilibrium texture, the magnetization has two contributions, one induced by AAS:
\begin{equation}\label{eq_AAS_magnetization}
\mathbf{m}_\mathrm{AAS}=\frac{2A_\mathrm{ani}}{H_\mathrm{ex}M_s}\mathbf{n}\times \left(\partial_x\partial _y\mathbf{n}\times \mathbf{n}\right)
\end{equation}
and one induced by DMI:
\begin{equation}\label{eq_DMI_magnetization}
\mathbf{m}_\mathrm{DMI}=-\frac{D_\mathrm{trans}M_s}{H_\mathrm{ex}}
\left(
\begin{array}{c}
  n_y(1-2n_x^2/M^2_s)\\
     n_x(1-2n_y^2/M_s^2)\\
   -2n_xn_yn_z/M_s^2
\end{array}
\right)
\end{equation}

From Eqs.~\eqref{eq_AAS_magnetization} and \eqref{eq_DMI_magnetization} it follows that $\mathbf{m}_\mathrm{AAS}$ is related to the orientation of the N\'eel vector and its space gradients (non-relativistic), while $\mathbf{m}_\mathrm{DMI}$ depends on the projections of the N\'eel vector on the crystallographic axes (relativistic). Note that for our particular choice of the domain wall profile (hyperbolic tangent), Eq. (6) and (7) coincide for the first case shown in Fig.~\ref{fig_notes}.

The magnetic dynamics are described by the standard Landau--Lifshits equations for magnetic sublattices. The dynamical problem is considered as a set of $3N_1 N_2$ ordinary differential equations with respect to $3N_1 N_2$ time-dependent components of the sublattice magnetizations. The parameters $N_1$ and $N_2$ define the size of the system. For the given initial conditions, the set of time evolution is integrated numerically using the Runge--Kutta method in Python. Whereas the Landau-Lifshitz equation preserves the length of the magnetic moments, we additionally renormalize the magnetic moments at each time integration step to avoid the accumulation of numerical errors. More information can be found in Section IV of Supplementary Materials.

\section*{Author contributions}
O.G. and J.S. wrote the manuscript with contributions 
and comments from all the authors. O.G and V.K. developed and analysed the models, and performed analytical calculations. K.Ye. performed atomistic spin-lattice calculations. L.S and R.J-U. provided DFT data. All the authors participated in the discussions of the results and gave inputs throughout the project. 

\begin{acknowledgments}
O.G., J.S., and R.J.-U. acknowledge funding by the Deutsche Forschungsgemeinschaft (DFG, German Research Foundation)-TRR288-422213477 (project A09, A11, and B15). L.S. and T.J. acknowledge support by Grant Agency of the Czech Republic grant no. 19-28375X, the Ministry of Education of the Czech Republic grants CZ.02.01.01/00/22008/0004594 and the ERC Advanced Grant no. 101095925. J.v.d.B acknowledges financial support by the Deutsche Forschungsgemeinschaft (DFG, German Research Foundation), through SFB 1143 project A5 and the W{\"u}rzburg-Dresden Cluster of Excellence on Complexity and Topology in Quantum Matter-ct.qmat (EXC 2147, Project Id No. 390858490).
\end{acknowledgments}

%

%
\newpage

\renewcommand{\vec}[1]{\bm{#1}}
\newcommand{\dd}{\mathrm{d}}

\renewcommand{\thefigure}{S\arabic{figure}}
\renewcommand{\theequation}{S.\arabic{equation}}
\renewcommand{\thesection}{\Roman{section}}

\clearpage
\break
\onecolumngrid

\begin{center}
	\textbf{\large SUPPLEMENTARY MATERIALS}
\end{center}
\beginsupplement

    In these supplemental materials, we demonstrate  that the phenomenological continuous model discussed in the main text can be straightforwardly derived from the microscopic discrete Hamiltonian of double-layered altermagnets of RuO$_2$ type. We establish the relation between constants of the phenomenological and microscopic models. Introducing the Lagrangian for the phenomenological model, we compute the deformation of the moving domain wall and explain the Walker limit velocity $v_{\text{lim}}$ discussed in the main text. Using the method of collective variables, we demonstrate that for $v>v_{\text{lim}}$ the domain wall dynamics is necessarily of the oscillatory type.
\setcounter{section}{0}
\section{Discrete model of a double layer R$\text{u}$O$_2$ }

\subsection{Hamiltonian}

\begin{figure*}[h]
\includegraphics[width=0.7\textwidth]{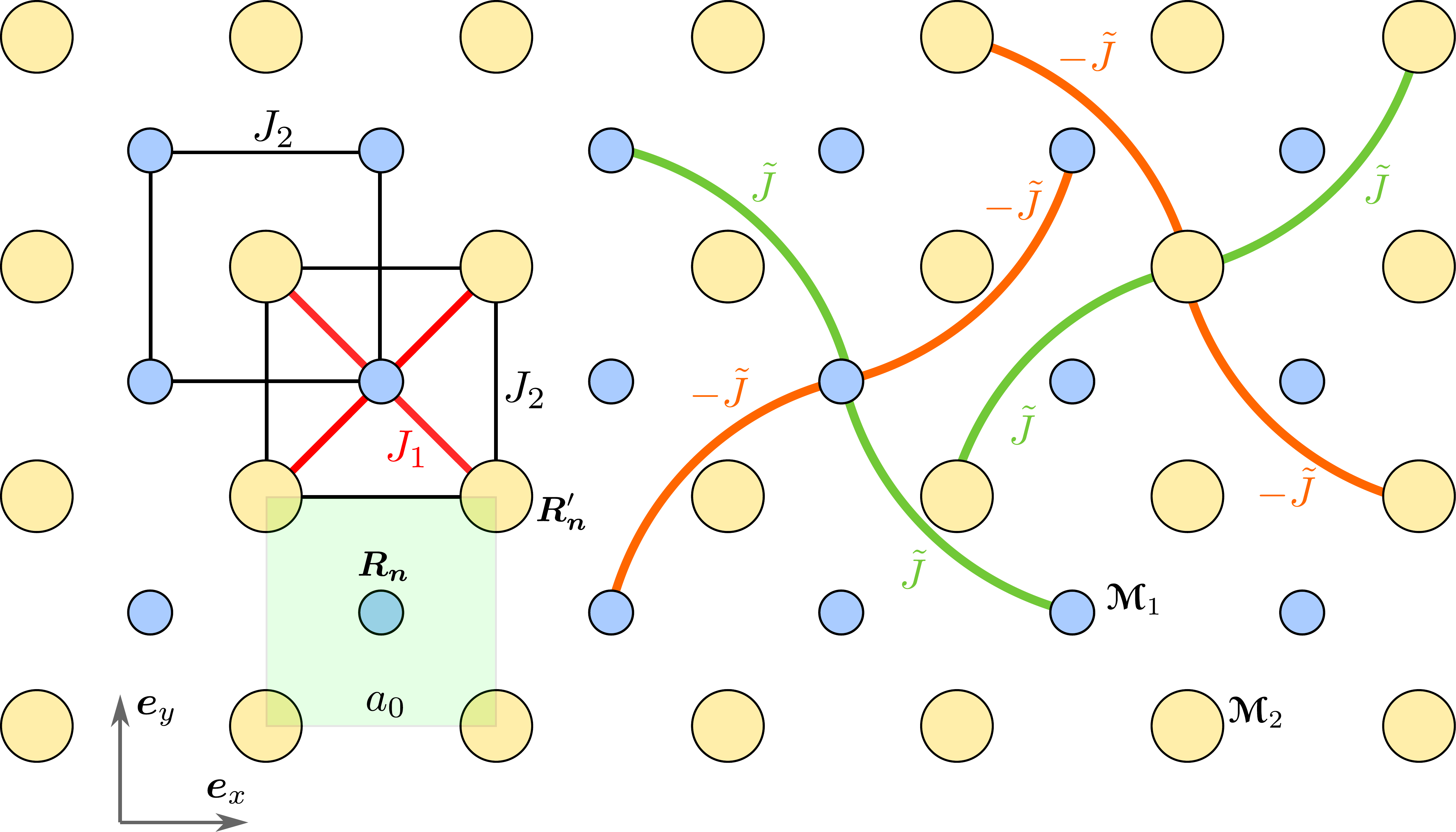}
\caption{Two layers of magnetic atoms carrying magnetic moments $\vec{\mu}_1=\mu_s\vec{\mathcal{M}}_1$ and $\vec{\mu}_2=\mu_s\vec{\mathcal{M}}_2$ with $\vec{\mathcal{M}}_{\nu}$ being the dimensionless unit vectors, are shown by small blue and big yellow markers, respectively. Here the inter-layer AFM ($J_1$) and intra-layer FM ($J_2$) nearest-neighbors exchange couplings are shown by the red and black bonds, respectively. The anisotropic exchange next-nearest-neighbors coupling ($\pm\tilde{J}$) acts within each of the layers possessing opposite symmetries in different layers, see green and orange bonds. Moments $\vec{\mathcal{M}}_1$ and $\vec{\mathcal{M}}_2$ are located in positions $\vec{R}_{\vec{n}}=a_0(n_1\vec{e}_x+n_2\vec{e}_y)$ and $\vec{R}_{\vec{n}}'=\vec{R}_{\vec{n}}+\frac{a_0}{2}(\vec{e}_x+\vec{e}_y)$, respectively. The unit cell of the two-layer system is shown by the green square with side $a_0$.}
\label{fig:discr-Model}
\end{figure*}

Here we consider a two-layer system formed by square lattices of magnetic moments $\vec{\mu}_{1}(\vec{R}_{\vec{n}})$ and $\vec{\mu}_{2}(\vec{R}'_{\vec{n}})$ and mutually shifted by half of the square diagonal, as shown in Fig.~\ref{fig:discr-Model}. 

In addition to the main antiferromagnetic (AFM) inter-layer and ferromagnetic (FM) intra-layer exchange couplings, we consider anisotropic exchange couplings within each layer. Constants of the anisotropic exchange switch signs in different layers leading to the altermagnetic effect. We also take into account the easy-axial anisotropy with the axis perpendicular to the layers. Hamiltonian of the system shown in Fig.~\ref{fig:discr-Model} is as follows
    \begin{equation}\label{eq:H-discr}
        \begin{split}
\mathcal{H}=&\sum\limits_{\vec{R}_{\vec{n}}}\biggl\{J_1\sum\limits_{\sigma\in\Xi_\times}\vec{\mathcal{M}}_1(\vec{R}_{\vec{n}})\cdot\vec{\mathcal{M}}_2(\vec{R}_{\vec{n}}+\delta\vec{R}_{\sigma})   -\frac{J_2}{2}\sum\limits_{\sigma\in\Xi_+}\sum\limits_{\nu=1,2}\vec{\mathcal{M}}_\nu(\vec{R}_{\vec{n}})\cdot\vec{\mathcal{M}}_\nu(\vec{R}_{\vec{n}}+\delta\vec{R}_{\sigma})-K\sum\limits_{\nu=1,2}\vec{\mathcal{M}}^2_{\nu z}(\vec{R}_{\vec{n}})\\
&+\frac{\tilde{J}}{2}\sum\limits_{\sigma\in\{\nwarrow,\searrow\}}\left[\vec{\mathcal{M}}_1(\vec{R}_{\vec{n}})\cdot\vec{\mathcal{M}}_1(\vec{R}_{\vec{n}}+2\delta\vec{R}_{\sigma})-\vec{\mathcal{M}}_2(\vec{R}_{\vec{n}})\cdot\vec{\mathcal{M}}_2(\vec{R}_{\vec{n}}+2\delta\vec{R}_{\sigma})\right]\\
&-\frac{\tilde{J}}{2}\sum\limits_{\sigma\in\{\nearrow,\swarrow\}}\left[\vec{\mathcal{M}}_1(\vec{R}_{\vec{n}})\cdot\vec{\mathcal{M}}_1(\vec{R}_{\vec{n}}+2\delta\vec{R}_{\sigma})-\vec{\mathcal{M}}_2(\vec{R}_{\vec{n}})\cdot\vec{\mathcal{M}}_2(\vec{R}_{\vec{n}}+2\delta\vec{R}_{\sigma})\right]\biggr\},
        \end{split}
    \end{equation}
where $\Xi_\times=\{\nearrow,\swarrow,\nwarrow,\searrow\}$, $\Xi_+=\{\rightarrow,\leftarrow,\uparrow,\downarrow\}$, and we assume that $J_1>0$, $J_2>0$, and $K>0$. We also introduced the following shift vectors $\delta\vec{R}_{\rightarrow}=-\delta\vec{R}_{\leftarrow}=a_0\vec{e}_x$,  $\delta\vec{R}_{\uparrow}=-\delta\vec{R}_{\downarrow}=a_0\vec{e}_y$; $\delta\vec{R}_{\nearrow}=-\delta\vec{R}_{\swarrow}=\frac{a_0}{2}(\vec{e}_x+\vec{e}_y)$, $\delta\vec{R}_{\nwarrow}=-\delta\vec{R}_{\searrow}=\frac{a_0}{2}(-\vec{e}_x+\vec{e}_y)$. We also utilized the assumption that the lattice is infinite, the latter enables us to do the change in the summation indices $\sum_{\vec{R}'_{\vec{n}}}F(\vec{R}'_{\vec{n}})=\sum_{\vec{R}_{\vec{n}}}F(\vec{R}_{\vec{n}})$ if $\vec{R}'_{\vec{n}}-\vec{R}_{\vec{n}}=\mathbf{const}$.

\subsection{Dispersion relation}

The dynamics of the system under consideration is governed by the set of coupled Landau-Lifshitz equations 
\begin{equation}\label{eq:LL}
\partial_t\vec{\mathcal{M}}_{\nu}=\frac{\gamma}{\mu_s}\left[\vec{\mathcal{M}}_{\nu}\times\frac{\partial\mathcal{H}}{\partial\vec{\mathcal{M}}_{\nu}}\right]
\end{equation}
formulated for each magnetic moment of the system. Here $\nu=1,2$ numerates the sublattices, $\gamma>0$ is the gyromagnetic ratio and the coupling between two equations in~\eqref{eq:LL} is provided by the Hamiltonian \eqref{eq:H-discr}. We start with the linearization of equations of motion \eqref{eq:LL} with respect to the in-plane components $\mathcal{M}_{1,2;x,y}$ on the top of the ground state $\vec{\mathcal{M}}_1^0=\vec{e}_z$, $\vec{\mathcal{M}}_2^0=-\vec{e}_z$:
\begin{equation}\label{eq:LL-lin}
	\begin{split}
		\partial_t\mathcal{M}_{1\alpha}=-\varepsilon_{\alpha\beta}\frac{\gamma}{\mu_s}\frac{\partial\mathcal{H}^{(2)}}{\partial \mathcal{M}_{1\beta}},\qquad\partial_t\mathcal{M}_{2\alpha}=\varepsilon_{\alpha\beta}\frac{\gamma}{\mu_s}\frac{\partial\mathcal{H}^{(2)}}{\partial \mathcal{M}_{2\beta}},
	\end{split}
\end{equation}
where $\alpha=x,y$, and the harmonic part of Hamiltonian~\eqref{eq:H-discr} is as follows
\begin{equation}\label{eq:H-harm}
\begin{split}
&\mathcal{H}^{(2)}=\sum\limits_{\vec{R}_{\vec{n}}}\sum\limits_{\alpha=x,y}\biggl\{J_1\sum\limits_{\sigma\in\Xi_\times}\mathcal{M}_{1\alpha}(\vec{R}_{\vec{n}})\mathcal{M}_{2\alpha}(\vec{R}_{\vec{n}}+\delta\vec{R}_{\sigma})-\frac{J_2}{2}\sum\limits_{\sigma\in\Xi_+}\sum\limits_{\nu=1,2}\mathcal{M}_{\nu\alpha}(\vec{R}_{\vec{n}})\mathcal{M}_{\nu\alpha}(\vec{R}_{\vec{n}}+\delta\vec{R}_{\sigma})\\
&+(2J_1+2J_2+K)\sum\limits_{\nu=1,2}\mathcal{M}_{\nu\alpha}^2(\vec{R}_{\vec{n}})+\frac{\tilde{J}}{2}\sum\limits_{\sigma\in\{\nwarrow,\searrow\}}\left[\mathcal{M}_{1\alpha}(\vec{R}_{\vec{n}})\mathcal{M}_{1\alpha}(\vec{R}_{\vec{n}}+2\delta\vec{R}_{\sigma})-\mathcal{M}_{2\alpha}(\vec{R}_{\vec{n}})\mathcal{M}_{2\alpha}(\vec{R}_{\vec{n}}+2\delta\vec{R}_{\sigma})\right]\\
&+\frac{\tilde{J}}{2}\sum\limits_{\sigma\in\{\nearrow,\swarrow\}}\left[\mathcal{\mathcal{M}}_{2\alpha}(\vec{R}_{\vec{n}})\mathcal{M}_{2\alpha}(\vec{R}_{\vec{n}}+2\delta\vec{R}_{\sigma})-\mathcal{M}_{1\alpha}(\vec{R}_{\vec{n}})\mathcal{M}_{1\alpha}(\vec{R}_{\vec{n}}+2\delta\vec{R}_{\sigma})\right]\biggr\}.
\end{split}
\end{equation}

\begin{figure*}
    \includegraphics[width=\textwidth]{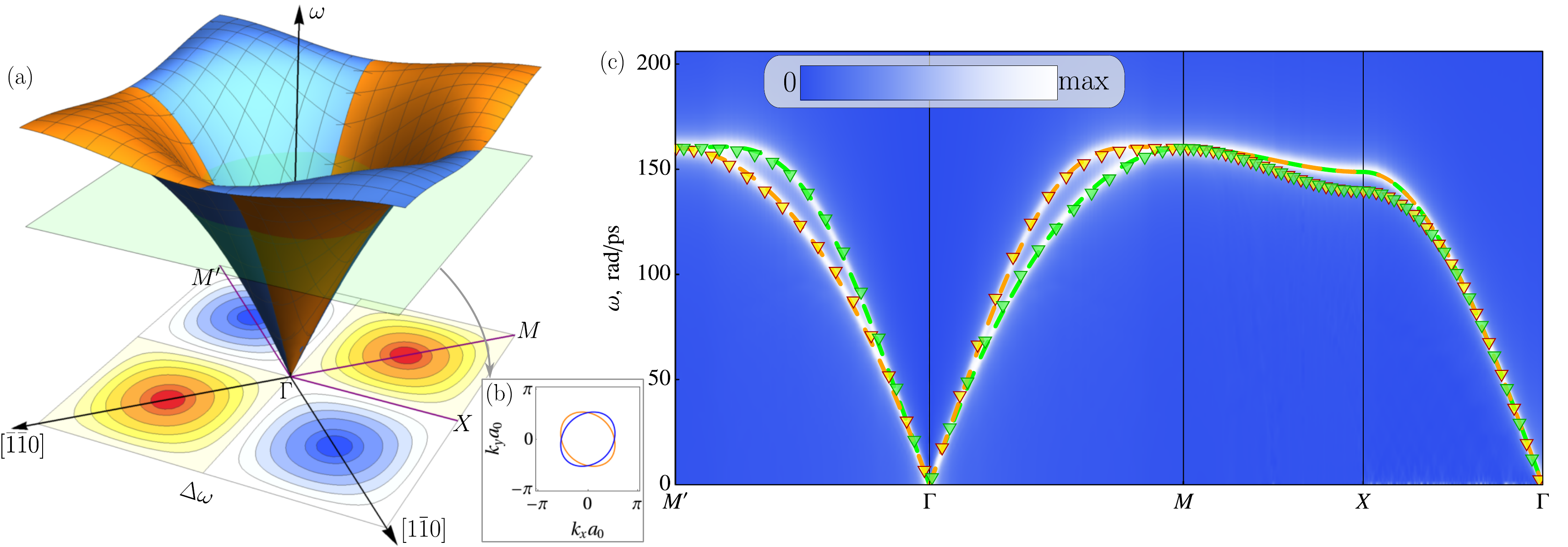}
    \caption{ The eigenfrequencies $\omega_{\pm}(\vec{k})$ determined by means of \eqref{eq:dispersion} and \eqref{eq:ABC} are shown within the first Brillouin zone (1.BZ) for parameters $\epsilon\approx 0.07$, $\kappa=0$, $\eta\approx 0.17$, see panel (a). The distribution of the splitting $\Delta\omega=\omega_+-\omega_-$ over the 1.BZ is shown on the bottom of plot (a). Panel (b) shows the isolines corresponding to constant frequency $\omega=0.8\omega_0$. Panel (c) shows dispersions $\omega_+(k)$ (dashed orange line) and $\omega_-(k)$ (dashed green line). The magnon spectra obtained by means of the spin-lattice simulations of the set of Landau-Lifshitz equations \eqref{eq:LL} with Hamiltonian \eqref{eq:H-discr} are shown by the color intensity. Parameters of the simulations are listed in Table.~\ref{tab:disp-iso}. Markers show the spectra obtained from DFT calculations.}\label{fig:disp}
\end{figure*}

Next, we use the Fourier transforms on the periodic lattice 
\begin{equation}\label{eq:FT-discr}
	\begin{split}
	f(\vec{R}_{\vec{n}})=\frac{1}{\sqrt{N}}\sum\limits_{\vec{k}\in1.\text{BZ}}\hat{f}(\vec{k})e^{i\vec{k}\cdot\vec{R}_{\vec{n}}},\qquad
	\hat{f}(\vec{k})=\frac{1}{\sqrt{N}}\sum\limits_{\vec{R}_{\vec{n}}}f(\vec{R}_{\vec{n}})e^{-i\vec{k}\cdot\vec{R}_{\vec{n}}}	
	\end{split}
\end{equation}
supplemented with the completeness relation $\sum_{\vec{R}_{\vec{n}}}e^{i(\vec{k}-\vec{k}')\cdot \vec{R}_{\vec{n}}}=N\delta_{\vec{k},\vec{k}'}$. Here $N$ is the number of magnetic moments in one sublattice. Applying \eqref{eq:FT-discr} to \eqref{eq:LL-lin}, we obtain the equations of motion in reciprocal space
\begin{equation}\label{eq:LL-lin-rec}
	\begin{split}
	&\partial_t\hat{\mathcal{M}}_{1\alpha}(\vec{k})=-\varepsilon_{\alpha\beta}\frac{\gamma}{\mu_s}\frac{\partial\mathcal{H}^{(2)}}{\partial\hat{\mathcal{M}}_{1\beta}(-\vec{k})},\qquad
	\partial_t\hat{\mathcal{M}}_{2\alpha}(\vec{k})=\varepsilon_{\alpha\beta}\frac{\gamma}{\mu_s}\frac{\partial\mathcal{H}^{(2)}}{\partial\hat{\mathcal{M}}_{2\beta}(-\vec{k})}.
\end{split}
\end{equation}
In reciprocal space, the Hamiltonian \eqref{eq:H-harm} is as follows
    \begin{subequations}\label{eq:H-harm-rec}
    \begin{align}
&\mathcal{H}^{(2)}=4J_1\sum\limits_{\vec{k}\in1.\text{BZ}}\biggl\{\mathcal{A}_{\vec{k}}\,\mathcal{M}_{1\alpha}(\vec{k})\mathcal{M}_{2\alpha}(-\vec{k})+\frac{\mathcal{B}_{\vec{k}}+\mathcal{C}_{\vec{k}}}{2}\mathcal{M}_{1\alpha}(\vec{k})\mathcal{M}_{1\alpha}(-\vec{k})+\frac{\mathcal{B}_{\vec{k}}-\mathcal{C}_{\vec{k}}}{2}\mathcal{M}_{2\alpha}(\vec{k})\mathcal{M}_{2\alpha}(-\vec{k})\biggr\},\\
\label{eq:ABC}&\mathcal{A}_{\vec{k}}=\cos\frac{a_0k_x}{2}\cos\frac{a_0k_y}{2},\qquad\mathcal{B}_{\vec{k}}=1+\frac{\kappa}{2}+\eta\left(\sin^2\frac{a_0k_x}{2}+\sin^2\frac{a_0k_y}{2}\right),\qquad\mathcal{C}_{\vec{k}}=\epsilon\sin(a_0k_x)\sin(a_0k_y),
\end{align}
\end{subequations}
where the summation over the repeating index $\alpha\in\{x,y\}$ is assumed and we introduce the following dimensionless parameters: $\kappa=K/J_1$, $\eta=J_2/J_1$, and $\epsilon=\tilde{J}/J_1$. With \eqref{eq:H-harm-rec}, the set of four equations \eqref{eq:LL-lin-rec} can be presented as
\begin{equation}\label{eq:LL-lin-rec1}
\partial_t\vec{\xi}=\omega_0\mathbb{M}\vec{\xi},\qquad\mathbb{M}=\begin{bmatrix}
0 & -(\mathcal{B}+\mathcal{C}) & 0 & -\mathcal{A} \\
\mathcal{B}+\mathcal{C} & 0 & \mathcal{A} & 0 \\
0 & \mathcal{A} & 0 & \mathcal{B}-\mathcal{C} \\
-\mathcal{A} & 0 & -(\mathcal{B}-\mathcal{C}) & 0
\end{bmatrix},
\end{equation}
where $\vec{\xi}=[\hat{\mathcal{M}}_{1x},\hat{\mathcal{M}}_{1y},\hat{\mathcal{M}}_{2x},\hat{\mathcal{M}}_{2y}]^{\textsc{t}}$ and $\omega_0=4\gamma J_1/\mu_s$. System \eqref{eq:LL-lin-rec1} has solution $\vec{\xi}=\vec{\xi}_0e^{-i\omega t}$, which is nontrivial ($\vec{\xi}_0\ne\vec{0}$) if $\omega=i\omega_0\lambda_\nu$ with $\lambda_\nu$ being the eigenvalues of matrix $\mathbb{M}$. The eigenvalues $\lambda_\nu$ are imaginary and compose two complex-conjugated pairs. The pair of the non-negative eigenfrequencies are
\begin{equation}\label{eq:dispersion}
\omega_{\pm}(\vec{k})=\omega_0\left(\sqrt{\mathcal{B}^2_{\vec{k}}-\mathcal{A}^2_{\vec{k}}}\pm\mathcal{C}_{\vec{k}}\right).
\end{equation}
Dispersion \eqref{eq:dispersion} is shown in Fig.~\ref{fig:disp} by dashed lines. It is important to emphasize a very good agreement between spectra obtained using the full-scale DFT calculations and using the simplified model \eqref{eq:H-discr}, see Fig.~\ref{fig:disp}(c).

For the case $ka_0\ll1$ and $\kappa\ll1$, dispersion \eqref{eq:dispersion} can be approximated as 
\begin{subequations}
\begin{align}
\label{eq:disp-approx}    &\omega_{\pm}(\vec{k})\approx\sqrt{\omega^2_{\textsc{afmr}}+c^2k^2}\pm\Lambda k_xk_y,\\
\label{eq:disp-approx-pars}&\omega_{\textsc{afmr}}=\frac{4\gamma\sqrt{J_1K}}{\mu_s},\qquad c=\frac{2\gamma J_1 a_0}{\mu_s}\sqrt{1+2\frac{J_2}{J_1}},\qquad \Lambda=\frac{4\gamma\tilde{J}a_0^2}{\mu_s}.
\end{align}
\end{subequations}

\begin{figure}
    \centering
    \includegraphics[width=0.5\textwidth]{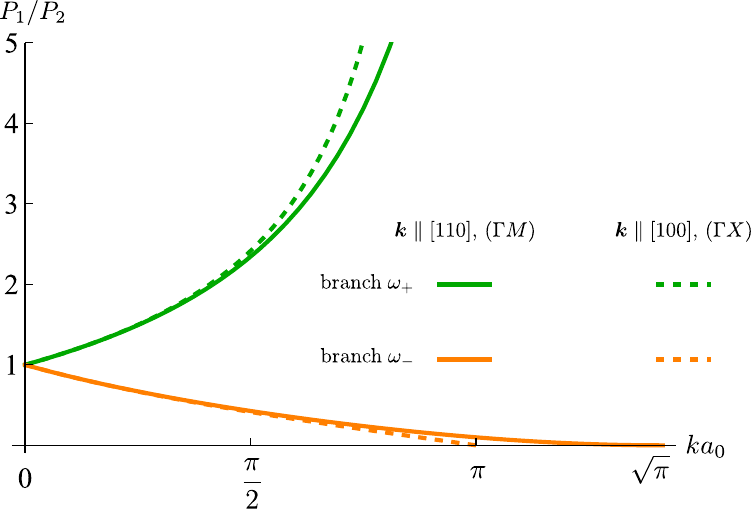}
    \caption{Distribution of the magnon intensity between two sublattices \eqref{eq:P1P2} for various branches and directions of $\vec{k}$-vector. Parameters of Fig.~\ref{fig:disp} were used.}
    \label{fig:P1P2}
\end{figure}

Computation of the corresponding eigenvectors of matrix $\mathbb{M}$ enables us to establish the following relations
\begin{equation}\label{eq:M-rel}
    \hat{\mathcal{M}}_{1y}=\pm i\, \hat{\mathcal{M}}_{1x},\qquad \hat{\mathcal{M}}_{2y}=\pm i\, \hat{\mathcal{M}}_{2x},\qquad \hat{\mathcal{M}}_{2x}=\frac{\hat{\mathcal{M}}_{1x}}{\mathcal{A}_{\vec{k}}}\left(\pm\sqrt{\mathcal{B}^2_{\vec{k}}-\mathcal{A}^2_{\vec{k}}}- \mathcal{B}_{\vec{k}}\right).
\end{equation}
Introducing the magnon intensity for $\nu$-th sublattice $P_\nu=\sqrt{|\hat{\mathcal{M}}_{\nu x}|^2+|\hat{\mathcal{M}}_{\nu y}|^2}$ with $\nu=1,\,2$, we determine the relative distribution of the magnon intensities between the sublattices
\begin{equation}\label{eq:P1P2}
\frac{P_1}{P_2}=\frac{\mathcal{A}_{\vec{k}}}{\left|\pm\sqrt{\mathcal{B}^2_{\vec{k}}-\mathcal{A}^2_{\vec{k}}}- \mathcal{B}_{\vec{k}}\right|}.
\end{equation}
Here the signs ``$\pm$'' correspond to different branches $\omega_\pm$. Interestingly, the distribution of the magnon intensity between two sublattices is not affected by the altermagneism. An example of the evolution of quantity \eqref{eq:P1P2} within the 1st Brillouin zone is shown in Fig.~\ref{fig:P1P2}.

\section{Continuum approximation}
To obtain the continuous approximation of Hamiltonian \eqref{eq:H-discr}, we utilize the Taylor expansion
\begin{equation}
\vec{\mathcal{M}}_\nu(\vec{R}_{\vec{n}}+\delta\vec{R}_\sigma)\approx\vec{\mathcal{M}}_\nu(\vec{R}_{\vec{n}})+(\delta\vec{R}_\sigma)_\alpha\partial_\alpha\vec{\mathcal{M}}_{\nu}(\vec{R}_{\vec{n}})+\frac12(\delta\vec{R}_\sigma)_\alpha(\delta\vec{R}_\sigma)_\beta\partial^2_{\alpha\beta}\vec{\mathcal{M}}_{\nu}(\vec{R}_{\vec{n}})
\end{equation}
with $\alpha,\beta\in\{x,y\}$ and replace the summation by integration: $\sum_{\vec{R}_{\vec{n}}}(\dots)\to\frac{1}{a_0^2}\int(\dots)\dd x\dd y$. This enables us to present hamiltonian \eqref{eq:H-discr} in the following form
\begin{equation}\label{eq:H-cnt}
\begin{split}
\mathcal{H}=\int\Bigl\{&\frac{4J_1}{a_0^2}\vec{\mathcal{M}}_1\cdot\vec{\mathcal{M}}_2-\frac{J_1}{2}\partial_\alpha\vec{\mathcal{M}}_1\cdot\partial_\alpha\vec{\mathcal{M}}_2+\frac{J_2}{2}\left(\partial_\alpha\vec{\mathcal{M}}_1\cdot\partial_\alpha\vec{\mathcal{M}}_1+\partial_\alpha\vec{\mathcal{M}}_2\cdot\partial_\alpha\vec{\mathcal{M}}_2\right)\\
&+2\tilde{J}\left(\partial_x\vec{\mathcal{M}}_1\cdot\partial_y\vec{\mathcal{M}}_1-\partial_x\vec{\mathcal{M}}_2\cdot\partial_y\vec{\mathcal{M}}_2\right)-\frac{K}{a_0^2}\left(\mathcal{M}_{1z}^2+\mathcal{M}_{2z}^2\right)\Bigr\}\dd x\dd y+\text{const}.
\end{split}
\end{equation}
The continuous form of Equations of motion \eqref{eq:LL} is as follows 
\begin{equation}\label{eq:LL-cnt}
\partial_t\vec{\mathcal{M}}_\nu=\frac{\gamma}{\mathcal{M}_s}\left[\vec{\mathcal{M}}_\nu\times\frac{\delta\mathcal{H}}{\delta\vec{\mathcal{M}}_\nu}\right],\qquad \nu=1,\,2,
\end{equation}
where $\mathcal{M}_s=\mu_s/a_0^2$.
Next, we introduce the dimensionless N{\'e}el vector $\vec{n}=(\vec{\mathcal{M}}_1-\vec{\mathcal{M}}_2)/2$ and the magnetization vector $\vec{m}=(\vec{\mathcal{M}}_1+\vec{\mathcal{M}}_2)/2$. Taking into account that $\vec{n}^2+\vec{m}^2=1$ and $\vec{n}\cdot\vec{m}=0$, we present hamiltonian \eqref{eq:H-cnt} in form 
\begin{equation}\label{eq:H}
\mathcal{H}=\int\left[H_{ex}\mathcal{M}_s\vec{m}^2+A_{\textsc{afm}}\,\partial_\alpha\vec{n}\cdot\partial_\alpha\vec{n}+A_{\textsc{alt}}(\partial_x\vec{m}\cdot\partial_y\vec{n}+\partial_y\vec{m}\cdot\partial_x\vec{n})-H_{an}\mathcal{M}_sn_z^2\right]\dd x\dd y,
\end{equation}
where we neglected all terms quadratic in $m$ except the uniform exchange, and introduced the notations $H_{ex}=8J_1/\mu_s$ for the exchange field, $A_{\textsc{afm}}=\frac{J_1}{2}+J_2$ for the antiferromagnetic stiffness, $A_{\textsc{alt}}=4\tilde{J}$ for the altermagnetic stiffness, and $H_{an}=2K/\mu_s$ for the anisotropy field. Introducing the volume saturation magnetization $M_s=2\mu_s/(a_0^2c_0)$ with $c_0$ being the magnetic nit cell size in $z$-direction, one can write a volume generalization of \eqref{eq:H}:
\begin{equation}\label{eq:H-volume}
\mathcal{H}^{\textsc{3d}}=\int\left[\frac12H_{ex}M_s\vec{m}^2+c_0^{-1}A_{\textsc{afm}}\,\partial_i\vec{n}\cdot\partial_i\vec{n}+c_0^{-1}A_{\textsc{alt}}(\partial_x\vec{m}\cdot\partial_y\vec{n}+\partial_y\vec{m}\cdot\partial_x\vec{n})-\frac12H_{an}M_sn_z^2\right]\dd x\dd y\dd z,
\end{equation}
where $i\in\{x,y,z\}$. Form of \eqref{eq:H-volume} enables one to establish the relations $A_{\textsc{afm}}=A_{\text{iso}}M_s^2c_0$, and $A_{\textsc{alt}}=A_{\text{ani}}M_s^2c_0$, where $A_{\text{iso}}$ and $A_{\text{ani}}$ are the exchange stiffnesses used in the main text.

In terms of $\vec{n}$ and $\vec{m}$, equations \eqref{eq:LL-cnt} obtain the following form 
\begin{equation}\label{eq:LL-nm}
\dot{\vec{n}}=\frac{\gamma}{2\mathcal{M}_s}\left[\vec{m}\times\frac{\delta\mathcal{H}}{\delta\vec{n}}+\vec{n}\times\frac{\delta\mathcal{H}}{\delta\vec{m}}\right],\qquad \dot{\vec{m}}=\frac{\gamma}{2\mathcal{M}_s}\left[\vec{n}\times\frac{\delta\mathcal{H}}{\delta\vec{n}}+\vec{m}\times\frac{\delta\mathcal{H}}{\delta\vec{m}}\right],
\end{equation}
where the dot indicates the time derivative.

In the first order in small parameter $|\vec{m}|\ll1$ and taking into account that $\sqrt{H_{an}/H_{ex}}\ll1$, we obtain from \eqref{eq:LL-nm}
\begin{subequations}
\begin{align}\label{eq:dynamics_n}
&\bigl[\ddot{\vec{n}}-c^2\nabla^2\vec{n}-\omega_{\textsc{afmr}}^2n_z\vec{e}_z+\Lambda\bigl(2\vec{n}\times\partial^2_{xy}\dot{\vec{n}}+\dot{\vec{n}}\times\partial^2_{xy}\vec{n}+\partial_x\vec{n}\times\partial_y\dot{\vec{n}}+\partial_y\vec{n}\times\partial_x\dot{\vec{n}}\bigr)\bigr]\times\vec{n}=0,\\
& \label{eq:dynamics_m}   
\vec{m}=\frac{1}{\gamma H_{\text{ex}}}\left[\dot{\vec{n}}+\Lambda\vec{n}\times\partial^2_{xy}\vec{n}\right]\times\vec{n}.
\end{align}
\end{subequations}
Here $\omega_{\textsc{afmr}}=\gamma\sqrt{H_{ex}H_{an}}$, $c=\gamma\sqrt{H_{ex}A_{\textsc{afm}}/\mathcal{M}_s}=\gamma\sqrt{2H_{ex}A_{\text{iso}}M_s}$ and $\Lambda=\gamma A_{\textsc{alt}}/\mathcal{M}_s=2\gamma A_{\text{ani}}M_s$ are the same as in \eqref{eq:disp-approx-pars}. Constants $\omega_{\textsc{afmr}}$ and $c$ determine typical length scale of the system: $\ell=c/\omega_{\textsc{afmr}}=a_0\sqrt{(J_1/2+J_2)/(2K)}$.

Linearization of Eq.~\eqref{eq:dynamics_n} with respect to small components $n_x$ and $n_y$ on the top of the ground state $\vec{n}=\vec{e}_z$ results in the following set of equations
\begin{equation}\label{eq:nxy-lin}
    \begin{split}
        &\ddot{n}_y-c^2\nabla^2n_y+\omega_{\textsc{afmr}}^2n_y+2\Lambda\partial_{xy}\dot{n}_x=0,\\
        &\ddot{n}_x-c^2\nabla^2n_x+\omega_{\textsc{afmr}}^2n_x-2\Lambda\partial_{xy}\dot{n}_y=0.
    \end{split}
\end{equation}
Equations \eqref{eq:nxy-lin} have solutions in form of a circularly polarized wave $n_x=n_x^0\cos(\vec{k}\cdot\vec{r}-\omega t+\phi_0)$, $n_y=n_y^0\sin(\vec{k}\cdot\vec{r}-\omega t+\phi_0)$ which turns Eqs.~\eqref{eq:nxy-lin} to $\hat{\mathbb{L}}\vec{n}^0=0$. Here 
\begin{equation}
  \hat{\mathbb{L}}=\begin{bmatrix}
      \omega^2-\omega_{\textsc{afmr}}^2-c^2k^2 & 2\Lambda k_x k_y \omega \\
      2\Lambda k_x k_y\omega & \omega^2-\omega_{\textsc{afmr}}^2-c^2k^2
  \end{bmatrix}.  
\end{equation}
The condition $\det(\hat{\mathbb{L}})=0$ for the solution nontriviality ($\vec{n}^0\ne\vec{0}$) results in the dispersion relation
\begin{equation}\label{eq:disp-det}
    \omega=\sqrt{\omega_{\textsc{afmr}}^2+c^2k^2+\Lambda^2 k_x^2k_y^2}\pm\Lambda k_xk_y\approx\sqrt{\omega_{\textsc{afmr}}^2+c^2k^2}\pm\Lambda k_xk_y,
\end{equation}
which coincides with the dispersion \eqref{eq:disp-approx} previously obtained from the discrete model. Note that the terms of order $\mathcal{O}(\Lambda^2)$ can not be kept in dispersion \eqref{eq:disp-det} because they were neglected in Eq.~\eqref{eq:dynamics_n}.

One obtains equation \eqref{eq:dynamics_n} as  the Euler-Lagrange equation for the action $\mathcal{S}=\int\dd t\int\dd x \dd y\mathscr{L}$ with Lagrangian
\begin{equation}\label{eq:Ln}
\mathscr{L}=\frac12\dot{\vec{n}}^2-\frac{c^2}{2}\partial_\alpha\vec{n}\cdot\partial_\alpha\vec{n}-\frac{\omega_{\textsc{afmr}}^2}{2}(1-n_z^2)-\Lambda\left[\vec{n}\times\dot{\vec{n}}\right]\cdot\partial_{xy}^2\vec{n}
\end{equation}
if the constraint $|\vec{n}|=1$ is applied. As it follows from \eqref{eq:Ln}, the one-dimensional solutions $\vec{n}(x,t)$ and $\vec{n}(y,t)$ are not affected by the altermgnetic $\Lambda$-term. Let us now consider the reference frame $x'=(x+y)/\sqrt{2}$ and $y'=(-x+y)/\sqrt{2}$ which is rotated by $\pi/4$. In this case, $\partial^2_{xy}=\frac{1}{2}(\partial_{x'x'}^2-\partial_{y'y'}^2)$. For the case of one-dimensional solutions $\vec{n}(\xi,t)$ with $\xi=x'$, the effective Lagrangian can be presented in form
\begin{equation}\label{eq:L-eff}
\mathscr{L}=\frac12\left\{\dot{\vec{n}}^2-c^2\vec{n}'^2-\omega_{\textsc{afmr}}^2(1-n_z^2)-\Lambda \,\left[\vec{n}\times\dot{\vec{n}}\right]\cdot\vec{n}''\right\}
\end{equation}
where prime denotes derivative with respect to $\xi$. 

\paragraph{Travelling wave solutions.} In what follows, we consider a traveling wave solution $\vec{n}=\vec{n}(\xi-vt)$ of Eq.~\eqref{eq:dynamics_n}. It is instructive to introduce a new variable $\xi_v=(\xi-vt)/(\ell\sqrt{1-v^2/c^2})$. In this case, Lagrangian \eqref{eq:L-eff} obtains the following form
\begin{equation}\label{eq:L-eff-TW}
\begin{split}
&\mathscr{L}^{\text{eff}}=\frac{\omega_{\textsc{afmr}}^2}{2}\left[-(\partial_{\xi_v}\vec{n})^2-(1-n_z^2)+\sigma_{v}\,\vec{n}\cdot(\partial_{\xi_v}\vec{n}\times\partial_{\xi_v}^2\vec{n})\right],\\
&\sigma_v=\varepsilon \frac{v/c}{(1-v^2/c^2)^{3/2}},\qquad \varepsilon=\frac{\Lambda\omega_{\textsc{afmr}}}{c^2}=\sqrt{\frac{H_{an}}{H_{ex}}}\frac{A_{\textsc{alt}}}{A_{\textsc{afm}}}=\frac{4\tilde{J}K^{1/2}}{J_1^{3/2}\left(1+2\frac{J_2}{J_1}\right)}
\end{split}
\end{equation}
Note that $\sigma_v$ is the only parameter that controls the shape of the traveling-wave solutions.

\section{Domain wall dynamics and the Walker breakdown}
Here we consider domain wall solutions (DW) of Lagrangian \eqref{eq:L-eff}. 

\subsection{Dynamics below the Walker breakdown (translational motion)}
\begin{figure*}
    \includegraphics[width=\textwidth]{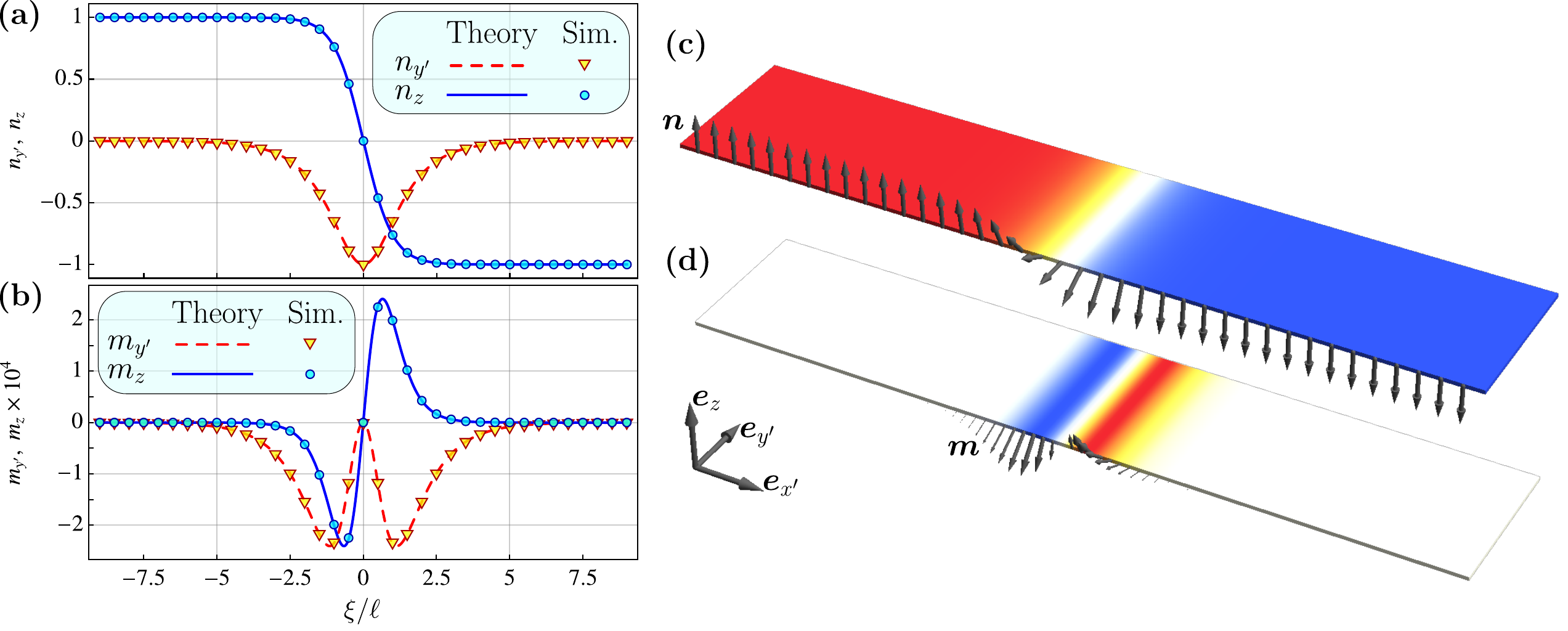}
    \caption{Profiles of a static Bloch DW ($\phi_{\textsc{dw}}=-\pi/2$) for $\varepsilon=0.05$ and $\sqrt{H_{an}/H_{ex}} = 0.025$. Panels (a) and (b) show the distributions of $y'$ and $z$ components of N{\'e}el and magnetization vectors, respectively. Lines are plotted for predictions~\eqref{eq:DW-simple} and~\eqref{eq:m-DW-static}. Symbols correspond to the results of simulations. (c) and (d) correspond to the distribution of $z$ component of N{\'e}el and magnetization vectors, respectively.}\label{fig:static_DW}
\end{figure*}

Let's start with the solutions in the form of the traveling waves $\vec{n}=\vec{n}(\xi-vt)$. In the moving frame of reference, these solutions are described by Lagrangian \eqref{eq:L-eff-TW}. For $\sigma_v=0$, it results in a known DW solution
\begin{equation}\label{eq:DW-simple}
    \theta_{\textsc{dw}}=2\arctan e^{p\xi_v},\qquad \phi_{\textsc{dw}}=\text{const},
\end{equation}
where $\xi_v=(\xi-vt)/(\ell\sqrt{1-v^2/c^2})$, and $p=\pm1$ is the DW topological charge. Thus, $\ell=x_{\textsc{dw}}$ is thickness of the static ($v=0$) DW. Remarkably, due to the altermagnetic effects, the static DW has a nonvanishing magnetization
\begin{equation}\label{eq:m-DW-static}
    \vec{m}(\xi)=\sqrt{\frac{H_{an}}{H_{ex}}}\frac{\varepsilon}{2}\left[\frac{\sinh^2(\xi/\ell)}{\cosh^3(\xi/\ell)}\left(\vec{e}_{x'}\cos\phi_{\textsc{dw}}+\vec{e}_{y'}\sin\phi_{\textsc{dw}}\right)+p\frac{\sinh(\xi/\ell)}{\cosh^3(\xi/\ell)}\vec{e}_{z}\right].
\end{equation}
For the case $v=0$ and $\varepsilon\ne0$, the structure of DW \eqref{eq:DW-simple}-\eqref{eq:m-DW-static} is shown in Fig.~\ref{fig:static_DW}.

In the following, we consider the altermagnetic $\sigma_v$-term in \eqref{eq:L-eff-TW} as a perturbation, and look for the solutions in the following form
\begin{equation}\label{eq:sol-pert}
\theta=\theta_{\textsc{dw}}+\vartheta,\qquad \phi=\phi_{\textsc{dw}}+\frac{\varphi}{\sin\theta_{\textsc{dw}}},
\end{equation}
where the small deviations $\vartheta$ and $\varphi$ are of the same order as $\sigma_v$. We substitute \eqref{eq:sol-pert} in the Euler-Lagrange equations generated by Lagrangian \eqref{eq:L-eff-TW}. In the first order of the perturbation theory we obtain
\begin{align}
\label{eq:theta1}    &\partial_{\xi_v}^2\vartheta-\cos(2\theta_{\textsc{dw}})\vartheta=0,\\
\label{eqLphi1}    &\partial_{\xi_v}^2\varphi-\cos(2\theta_{\textsc{dw}})\varphi=-\frac{\sigma_v}{2}\left[(\partial_{\xi_v}\theta_{\textsc{dw}})^3+2\partial_{\xi_v}^3\theta_{\textsc{dw}}\right].
\end{align}
Eq.~\eqref{eq:theta1} results in $\vartheta\equiv0$, and with the help of \eqref{eq:DW-simple} we find the general bounded solution for $\varphi$:
\begin{equation}\label{eq:phi1-gen}
    \varphi=\frac{\sigma_vp}{2}\frac{\ln\cosh \xi_v+\mathcal{C}}{\cosh \xi_v}=\frac{p\varepsilon}{2}\frac{v/c}{(1-v^2/c^2)^{3/2}}\frac{\ln\cosh[(\xi-vt)/(\ell\sqrt{1-v^2/c^2})]+\mathcal{C}}{\cosh[(\xi-vt)/(\ell\sqrt{1-v^2/c^2})]},
\end{equation}
with $\mathcal{C}$ being an arbitrary constant. This arbitrariness is related to undefined value of $\phi_{\textsc{dw}}$ in the DW solution.

\begin{figure*}
    \includegraphics[width=0.85\textwidth]{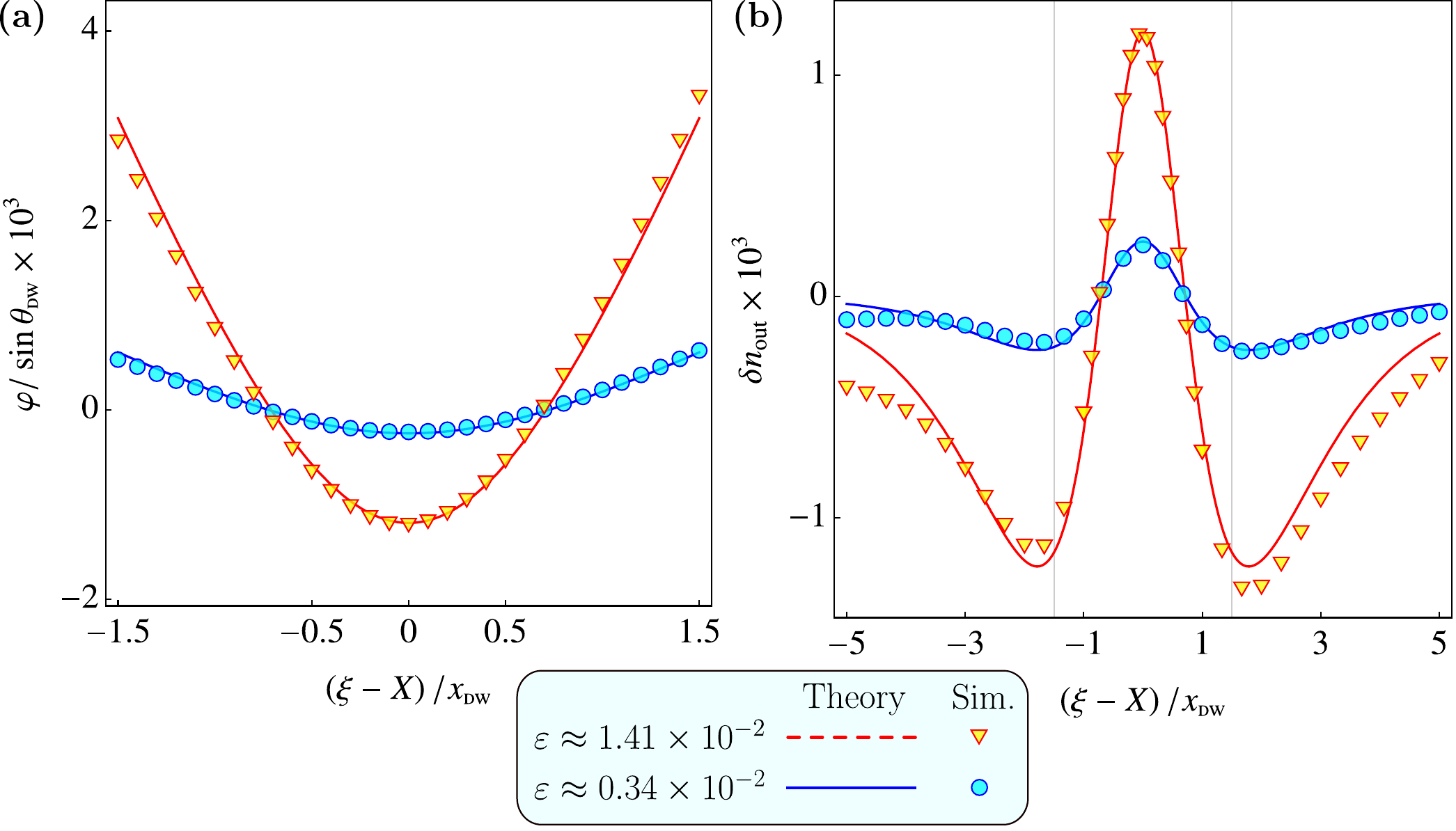}
    \caption{Deformation of domain wall profile as a function of coordinate for domain wall velocity $v=0.4c$: \textbf{(a)} -- deformation of a domain wall phase $\varphi$; \textbf{(b)} -- out of plane component $\delta n_{\text{out}}$ of N\'{e}el vector for domain wall.  In \textbf{(a)} and \textbf{(b)} symbols correspond to the data obtained by means of numerical simulations for different altemagnetic exchange constants: dots for $A_\textsc{alt}/A_\textsc{afm} = 0.1\ (\varepsilon\approx 0.34\times 10^{-2})$ and truangles for $A_\textsc{alt}/A_\textsc{afm} = 0.43$ ($\varepsilon\approx 1.41\times 10^{-2}$ -- case of RuO$_2$). Solid and dashed lines are plotted with analitycal prediction~\eqref{eq:phi1-gen}. In all simulations we considered stripes with $\sqrt{H_\textsc{x}/H_\textsc{a}} \approx 31$. For details of numerical simulations see Supplementary Materials~\ref{sec:sim}.}\label{fig:deformation_DW}
\end{figure*}

The component of $\vec{n}$, which is perpendicular to the plane of the rotation of $\vec{n}$ in the domain wall, is $\delta n_{\text{out}}=\vec{n}\cdot\vec{\nu}\approx\varphi_1$. Here $\vec{\nu}=-\vec{e}_{x'}\sin\phi_{\textsc{dw}}+\vec{e}_{y'}\cos\phi_{\textsc{dw}}$ is the unit vector, which is perpendicular to the rotation plane. The profiles for $\varphi$ and $\delta n_{\text{out}}$ are shown in Fig.~\ref{fig:deformation_DW}. The criterion $\delta n_{\text{out}}\approx1$ can be used for estimation of the critical velocity of the DW structural instability. This leads to $\sigma_v\approx1$ and we obtain the estimation
\begin{equation}\label{eq:vlim-est}
    v_{\text{lim}}\approx c\,\sqrt{1-\left(\frac{\varepsilon}{2}\right)^{2/3}}\approx c\left(1-\frac{\varepsilon^{2/3}}{2^{5/3}}\right).
\end{equation}
Estimation \eqref{eq:vlim-est} is presented in the main text in Eq.~(3).

\subsection{Dynamics above the Walker breakdown (oscillatory dynamics)}

To consider the domain wall dynamics in the general case we use the approach of collective variables \cite{Rajaraman82,Tveten13}. Following this approach we assume that the solution can be approximated as $\vec{n}(\vec{r},t)=\vec{n}_0(\vec{r},q^1(t),\,q^2(t),\dots)$, where $\vec{n}_0$ is a known function and the collective variables $q^i(t)$ determine the shape of the profile $\vec{n}_0$. By substituting the ansatz $\vec{n}_0$ into Lagrangian \eqref{eq:Ln} and integrating over the spatial coordinates we get the effective Lagrange function $\mathcal{L}=\int\mathscr{L}\dd\vec{r}$ in terms of the collective variables: 
\begin{equation}\label{eq:L-cv}
\mathcal{L}=\frac12g_{ij}\dot{q}^i\dot{q}^j+A_i\dot{q}^i-U(q^i)
\end{equation}
Here $g_{ij}=g_{ji}$ is the Riemann metric of the configuration space of the collective variables, $A_i$ is the gauge potential and $U$ is the potential energy, which  are determined as follows
\begin{equation}\label{eq:components}
\begin{split}
&g_{ij}=\int\left(\frac{\partial\vec{n}_0}{\partial q^i}\cdot\frac{\partial\vec{n}_0}{\partial q^j}\right)\dd x\dd y,\qquad A_i=\Lambda\int\left[\frac{\partial\vec{n}_0}{\partial q^i}\times\vec{n}_0\right]\cdot\partial_{xy}^2\vec{n}_0\,\dd x\dd y,\\
&U=\int\left[\frac{c^2}{2}\partial_\alpha\vec{n}_0\cdot\partial_\alpha\vec{n}_0+\frac{\omega_{\textsc{afmr}}^2}{2}\left(1-n_{0z}^2\right)\right]\dd x\dd y.
    \end{split}
\end{equation}
Note, that in the general case, $g_{ij}$ as well as $A_i$ can be a function of the collective variables. 


The Euler-Lagrange equation 
\begin{equation}\label{eq:EL-general}
    -\frac{\dd}{\dd t}\frac{\partial\mathcal{L}}{\partial\dot{q}^i}+\frac{\partial\mathcal{L}}{\partial q^i}=0
\end{equation}
that corresponds to \eqref{eq:L-cv} is as follows
\begin{equation}\label{eq:EL}
g_{ij}\ddot{q}^j+\Gamma_{ijk}\dot{q}^j\dot{q}^k+f_{ij}\dot{q}^j+\frac{\partial U}{\partial q^i}=0
\end{equation}
where $\Gamma_{ijk}=\frac12(\partial_jg_{ik}+\partial_kg_{ij}-\partial_ig_{jk})$ is the Levi-Civita connection on the configuration space and $f_{ij}=\partial_iA_j-\partial_jA_i$ is the gauge strength. Here we use the notation $\partial_i=\partial/\partial q^i$ for the sake of simplicity.

\begin{figure}
    \centering
    \includegraphics[width=\textwidth]{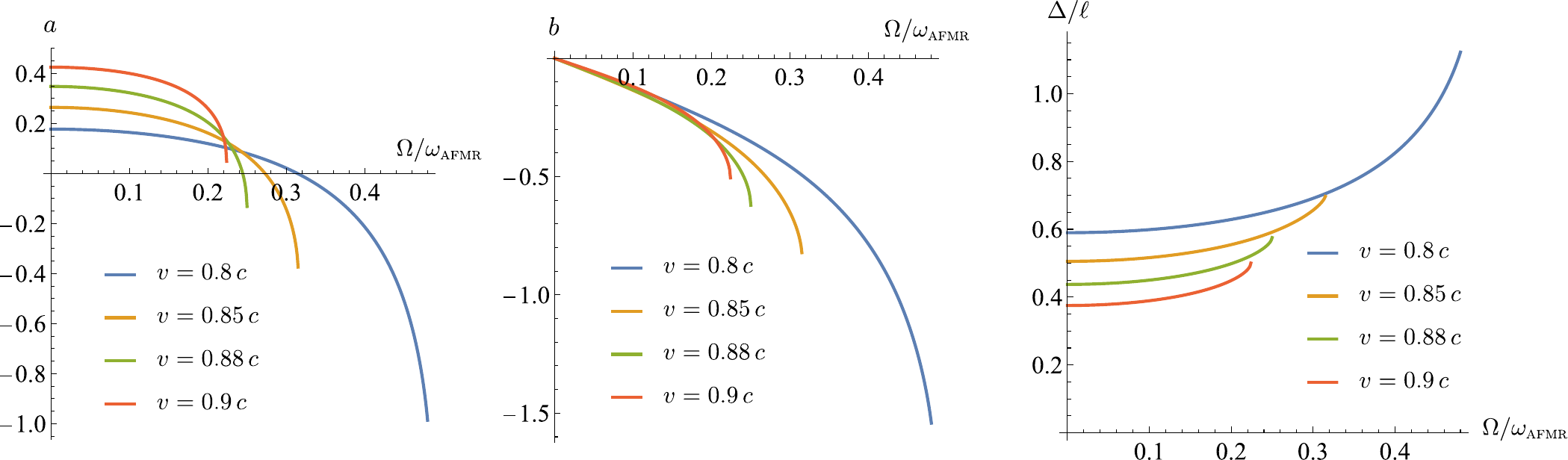}
    \caption{Equilibrium values of parameters $a$, $b$, $\Delta$ obtained by means of numerical solutions of Eqs.~\eqref{eq:abD} are shown as functions of $\Omega$ for different values of $v$. For all plots $\Lambda=0.1c^2/\omega_{\textsc{afmr}}$ (corresponds to $\varepsilon=0.1$). Here $\ell=c/\omega_{\textsc{afmr}}=x_{\textsc{dw}}$ is the thickness of a static domain wall.}
    \label{fig:abD}
\end{figure}

Let us now consider the following ansatz for $\vec{n}_0=\sin\theta_0(\vec{e}_{x'}\cos\phi_0+\vec{e}_{y'}\sin\phi_0)+\vec{e}_{z}\cos\theta_0$:
\begin{equation}\label{eq:ansatz-dw}
    \cos\theta_0(x',t)=-p\tanh\frac{x'-X(t)}{\Delta(t)},\qquad\phi_0(x',t)=\Phi(t)+a(t)\ln\cosh\frac{x'-X(t)}{\Delta(t)}+b(t)\frac{x'-X(t)}{\Delta(t)},
\end{equation}
which represents the moving domain wall. Here $X,\,\Phi,\,a,\,b$ and $\Delta$ are the collective variables previously denoted as $q_i$. $X$ denotes the DW position, $\Phi$ is the DW phase, $a$ is the amplitude of DW deformation induced by altermagnetism, and $b$ is the amplitude of the coordinate-dependent part of the phase of a precessing DW \cite{Kosevich90}. Now, taking into account that $\partial_{xy}^2=\frac12(\partial_{x'x'}^2-\partial_{y'y'}^2)$, we obtain from \eqref{eq:components} the following metric:
\begin{equation}\label{eq:gabD}
\begin{matrix}
    g_{XX}=\frac{2}{\Delta}\left(1+\frac{a^2}{3}+b^2\right), & g_{X\Phi}=-2b, & g_{Xa}=-2b\mathfrak{c}_1, & g_{Xb}=-a, & g_{X\Delta}=\frac{2ab}{\Delta};\\
    & g_{\Phi\Phi}=2\Delta, & g_{\Phi a}=2\Delta\mathfrak{c}_1, & g_{\Phi b}=0, & g_{\Phi\Delta}=-a;\\
    & & g_{aa}=2\Delta\mathfrak{c}_2, & g_{ab}=0, & g_{a\Delta}=-a\mathfrak{c}_4;\\
    & & & g_{bb}=2\Delta\mathfrak{c}_3, & g_{b\Delta}=2b\mathfrak{c}_3;\\
    & & & & g_{\Delta\Delta}=\frac{2}{\Delta}\left[(1+\mathfrak{c}_3)\frac{a^2}{3}+\mathfrak{c}_3(1+b^2)\right].
\end{matrix}
\end{equation}
Constants $\mathfrak{c}_i$ take the following values: $\mathfrak{c}_1=1-\ln2\approx0.307$, $\mathfrak{c}_2=2-\frac{\pi^2}{12}-\ln2(2-\ln2)\approx0.272$, $\mathfrak{c}_3=\frac{\pi^2}{12}\approx0.822$,  $\mathfrak{c}_4=\frac12(3-2\ln2)\approx0.807$. Components of the gauge potential are as follows
\begin{equation}\label{eq:AiD}
\begin{split}
    &A_X=\frac{p\Lambda}{\Delta^2}a\left(\frac13-\frac{a^2}{5}-b^2\right),\qquad A_\Phi=\frac{2p\Lambda}{3\Delta}ab,\qquad A_a=\frac{p\Lambda\mathfrak{c}_5}{\Delta}ab,\\ &A_b=\frac{p\Lambda}{\Delta}\left(-\frac12+\frac{a^2}{3}+\frac{b^2}{2}\right),\qquad A_\Delta=-\frac{p\Lambda}{\Delta^2}b\left(\frac12+a^2+\frac{b^2}{2}\right).
    \end{split}
\end{equation}
Here $\mathfrak{c}_5=2(4-3\ln2)/9\approx0.427$.
Finally, the scalar potential $U$ in \eqref{eq:components} has the following explicit form
\begin{equation}\label{eq:UD}
    U=\Delta\left[\omega_{\textsc{afmr}}^2+\frac{c^2}{\Delta^2}\left(1+\frac{a^2}{3}+b^2\right)\right].
\end{equation}
With the metric \eqref{eq:gabD}, gauge field \eqref{eq:AiD}, and potential \eqref{eq:UD} we construct the effective Lagrange function \eqref{eq:L-cv} and formulate equations of motion for the collective variables. The obtained set of five nonlinear and cumbersome equations has a solution $X=vt$, $\Phi=\Omega t$, $a=\text{const}$, $b=\text{const}$, $\Delta=\text{const}$ which corresponds to a stationary dynamics. In this case, two equations of the system, namely $\delta\mathcal{S}/\delta X=0$ and $\delta\mathcal{S}/\delta\Phi=0$, are turned to identities, while the rest three equations determine the equilibrium values of parameters $a,\,b,\,\Delta$ for the given values of $v$ and $\Omega$:
\begin{subequations}\label{eq:abD}
    \begin{align}
        &a(c^2-v^2)=p\Lambda\left[\frac{v}{2\Delta}\left(1-\frac95a^2-b^2\right)+\Omega b\right],\\
        &b(c^2-v^2)+V\Omega\Delta=p\Lambda a\left(\frac{\Omega}{3}-\frac{bv}{\Delta}\right),\\
        &(c^2-v^2)\left(1+\frac{a^2}{3}+b^2\right)-(\omega_{\textsc{afmr}}^2-\Omega^2)\Delta^2=\frac{2p\Lambda}{3}\left[\frac{va}{\Delta}\left(1-\frac35a^2-3b^2\right)+ab\Omega\right].
    \end{align}
\end{subequations}
In the limit case of an antiferromagnet ($\Lambda=0$), equations \eqref{eq:abD} have a solution
\begin{equation}
    a=0,\qquad b=-\frac{v\Omega}{\omega_{\textsc{afmr}}c\sqrt{1-\frac{v^2}{c^2}-\frac{\Omega^2}{\omega_{\textsc{afmr}}^2}}},\qquad \Delta=\frac{c}{\omega_{\textsc{afmr}}}\frac{1-\frac{v^2}{c^2}}{\sqrt{1-\frac{v^2}{c^2}-\frac{\Omega^2}{\omega_{\textsc{afmr}}^2}}},
\end{equation}
which coincides with the previously obtained results for the moving antiferromagnetic domain wall \cite{Kosevich90}. For $\Lambda\ne0$, we solve equations \eqref{eq:abD} numerically, see Fig.~\ref{fig:abD}.

Next, we consider the energy of the system
\begin{equation}\label{eq:EvsOmega}
    E=\frac12g_{ij}\dot{q}^i\dot{q}^j+U=\frac{c^2+v^2}{\Delta}\left(1+\frac{a^2}{3}+b^2\right)+(\omega_{\textsc{afmr}}^2+\Omega^2)\Delta-2v\Omega b.
\end{equation}

\begin{figure}
    \centering
    \includegraphics[width=\textwidth]{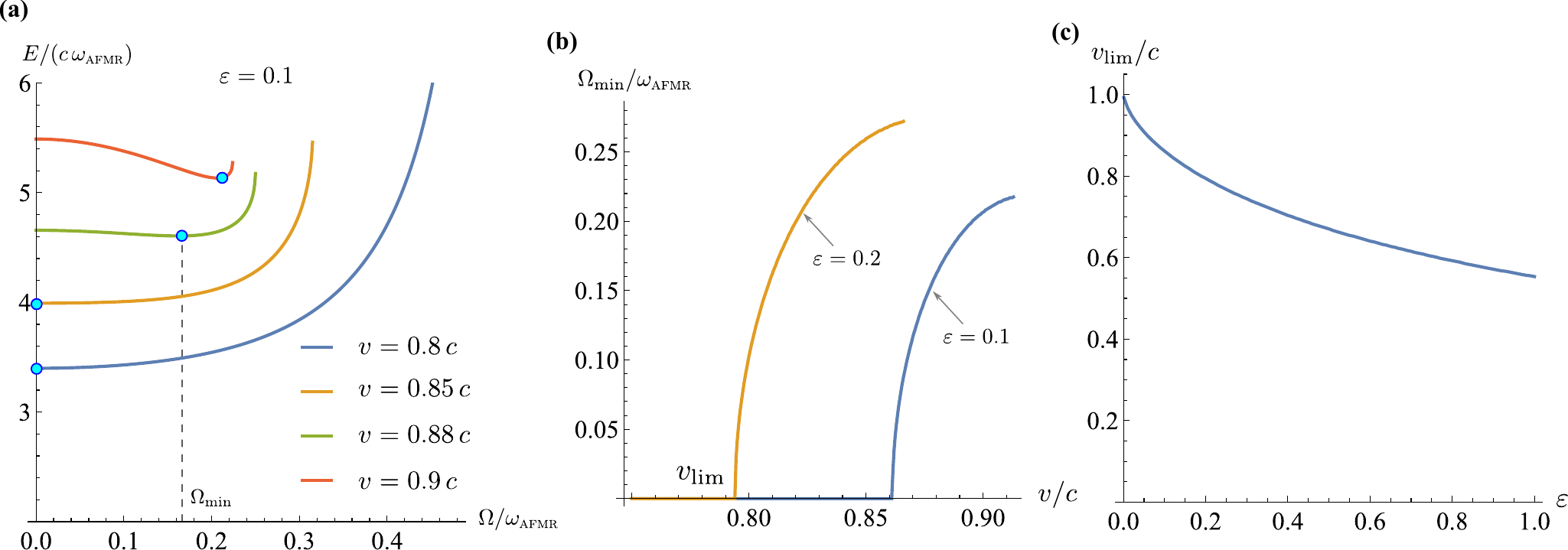}
    \caption{(a) -- dependencies $E(\Omega)$ for different domain wall velocities obtained with the help of \eqref{eq:EvsOmega} for the results shown in Fig.~\ref{fig:abD}. (b) -- frequency of the domain wall precession, which corresponds to the energy minimum for the given domain wall velocity. (c) -- dependence of the critical (Walker) velocity on the altermagnetic parameter $\varepsilon=\Lambda\omega_{\textsc{afmr}}/c^2$.}
    \label{fig:EOmega}
\end{figure}

Having dependencies $a(v,\Omega)$, $b(v,\Omega)$, and $\Delta(v,\Omega)$ shown in Fig.~\ref{fig:abD}, with the help of \eqref{eq:EvsOmega} we now compute the dependence $E(v,\Omega)$, see Fig.~\ref{fig:EOmega}(a).
Figure~\ref{fig:EOmega}(a) demonstrates the existence of a critical velocity $v_{\text{lim}}<c$ such that for $v<v_{\text{lim}}$ and $v>v_{\text{lim}}$ the energy minimum takes place for $\Omega=0$ and $\Omega>0$, respectively. The dependencies of the frequency $\Omega_{\text{min}}$ that  correspond to the energy minimum have a root-like form, see Fig.~\ref{fig:EOmega}(b). On this basis, we expect the precession of the domain wall, whose velocity exceeds the critical value $v>v_{\text{lim}}$ . The dependence of the critical velocity $v_{\text{lim}}$ on the altermagnetic parameter $\varepsilon$ is shown in Fig.~\ref{fig:EOmega}(c). In the antiferromagnetic limit ($\varepsilon\to0$), one has $v_{\text{lim}}=c$, i.e. the critical velocity coincides with the magnon velocity.

\section{Details of numerical simulations}\label{sec:sim}

The dynamics of magnetic moments are described by the Landau--Lifshits equations
\begin{equation}\label{eq:LLG}
    \left(1+\alpha_\textsc{g}^2\right)\frac{\mathrm{d}\vec{\mathcal{M}}_\nu}{\mathrm{d}t} = \frac{\gamma}{\mu_s}\left[\vec{\mathcal{M}}_\nu\times\frac{\partial\mathcal{H}}{\partial\vec{\mathcal{M}}_\nu}\right]+\alpha_\textsc{g}\frac{\gamma}{\mu_s}\vec{\mathcal{M}}_\nu\times\left[\vec{\mathcal{M}}_\nu\times\frac{\partial\mathcal{H}}{\partial\vec{\mathcal{M}}_\nu}\right],
\end{equation}
where $\alpha_\textsc{g}$ is the Gilbert damping parameter and $\mathcal{H}$ is defined in~\eqref{eq:H-discr}. The dynamical problem is considered as a set of $3N_1 N_2$ ordinary differential equations~\eqref{eq:LLG} with respect to $3N_1 N_2$ unknown functions $\mathcal{M}_\nu^\textsc{x}(t),\ \mathcal{M}_\nu^\textsc{y}(t),\ \mathcal{M}_\nu^\textsc{z}(t)$. Parameters $N_1$ and $N_2$ define the size of the system. For the given initial conditions, the set of time evolution equations~\eqref{eq:LLG} is integrated numerically using the Runge–-Kutta method in Python. During the integration process, the condition $\left|\vec{\mathcal{M}}_\nu(t)\right| = 1$ is controlled.

\subsection{Simulation of spin waves}\label{sec:sim-SW}
To simulate the spin waves we considered a system with a size of $N_1\times N_2 = 500\times500$ magnetic moments. The simulations are carried out in two steps. In the first step, we simulate the dynamics of the system in the external magnetic field $\vec{H}_\nu = H_0 \text{sinc}\left(2\pi\ \vec{k}\cdot\vec{R}_\nu\right)\text{sinc}\left[2\pi \left(t-t_0\right)\right]$, where $t_0 = t_\text{sim} / 2$ is a center of temporal part of field profile and $H_0$ is an amplitude of the applied field.

In the second step we performe a space-time Fourier transformation for the complex-valued parameter $M_\nu^\textsc{x}(t)+ i\ M_\nu^\textsc{y}(t)$. The resulting eigenfrequencies are plotted in Figs.~\ref{fig:disp} and 1 of the main text. The simulation parameters are presented in Tab.~\ref{tab:disp-iso}.
\begin{table}[h]
    \centering
    \begin{tabular}{|l|l||l|l|}
         \multicolumn{2}{|c||}{Spin-lattice parameters} & \multicolumn{2}{c|}{Continuum parameters} \\
         \hline
         \hline
         \cellcolor{yellow!25}Lattice constant &\cellcolor{yellow!25} $a_0 = 0.448$ nm& Exchange stiffness & $A_\textsc{afm} = 7.4$ meV\\
         Magnetic moment & $\mu=\mu_\textsc{b}$ & Altermagnetic exchange  & $A_\textsc{alt} = 3.2$ meV\\
         Inter-layer AFM exchange & $J_1 = 11.1$ meV &\cellcolor{yellow!25}$\Lambda$-term &\cellcolor{yellow!25}$\Lambda = 1.96$ rad nm$^2$/ps \\
         Intra-layer FM exchange & $J_2 = 1.88$ meV  & Exchange field&$H_\textsc{x} = 1534$ T\\
         Anisotropic exchange & $\tilde{J} = 0.8$ meV &\cellcolor{yellow!25}Magnon velocity &\cellcolor{yellow!25}$c = 35$ nm/ps\\
         Uniaxial anisotropy (SW) & $K^\textsc{sw} = 0$ meV & Anisotropy field (SW)& $H_\textsc{a}^\textsc{sw} = 0$ T\\
         Uniaxial anisotropy (DW) & $K^\textsc{dw} = 0.047$ meV & Anisotropy field (DW)& $H_\textsc{a}^\textsc{dw} = 1.63$ T\\
         \hline
    \end{tabular}
    \caption{Parameters used in spin-lattice simulations and corresponding parameters for the continuum model. Cells colored in light yellow color correspond to parameters obtained by fitting the data from DFT calculation. Index SW corresponds to anisotropy parameters used in simulations of spinwaves, while DW corresponds to parameters used in simulations of domain walls.}
    \label{tab:disp-iso}
\end{table}

\subsection{Statics and dynamics of the domain wall}\label{sec:sim-DW}

To simulate the domain wall properties we considered a system with a size of $N_1\times N_2 = 10000\times30$ magnetic moments with periodic boundary conditions along $N_2$ direction.

\paragraph{Static domain wall.}
First, we consider the static domain wall. The numerical experiment is performed in two steps. Initially, we relax the domain wall structure in an over-damped regime ($\alpha_\textsc{g} = 0.75$). A domain wall is placed in the center of the sample. After relaxation, we extract the corresponding order parameters of the system, i.e. N{\'e}el $\vec{n}$ and magnetization $\vec{m}$ vectors. The resulting components of order parameters are presented in Fig.~\ref{fig:static_DW} and Fig.~2 of main text by symbols. The corresponding parameters of simulations are presented in~Tab.~\ref{tab:disp-iso}. 

\paragraph{Free domain wall motion~($\alpha_\textsc{g} = 0$).} 
Here we consider the case of a free domain wall motion with the vanishing damping. The initial distribution of the magnetic moments $\vec{\mathcal{M}}_\nu$ is calculated using Eqs.~\eqref{eq:sol-pert} and~\eqref{eq:phi1-gen}. We consider Bloch domain walls, therefore an initial domain wall phase is chosen as $\phi_\textsc{dw} = -\pi/2$.  The domain wall position $q$, phase $\Phi$, and profile of deformation are extracted from simulation data by fitting of $n_z = -\tanh\left[\left(\xi-q\right)/x_\textsc{dw}\right]$, solution of $\tan\Phi = n_y(q) / n_x(q)$, and fitting of $\Phi-\Phi(q) = \varphi_1$, respectively. The resulting domain wall deformation and its amplitude are presented in Figs.~\ref{fig:deformation_DW} and~4 of the main text.

The domain wall phase precession as a function of the domain wall velocity are obtained by linear fitting of domain wall phase $\Phi_\text{fit} = \Omega t$. The frequencies as functions of velocity are plotted in Fig.~4 of the main text.

\section{Interaction between the domain wall and the magnetic tip }\label{sec:magnetic_tip}
The interaction energy between the tip and the domain wall is calculated as the Zeeman energy
\begin{equation}
    U(X,Z)=-\int \mathbf{m}(\mathbf{r};X,Z)\cdot\mathbf{B}(\mathbf{r})d\mathbf{r},
\end{equation}
where the coordinates $X$ and $Z$ define the position of the domain wall center with respect to the magnetic tip. The domain wall is aligned along $Y$ direction. For the calculations we assume that $Y$ is parallel to [110], where the altermagnetic effect is maximal. We further assume the domain wall plane is parallel to $Z$ axes and passes through the full sample thickness, which is 20~nm. The magnetic field of the tip is modelled according to \cite{Suter2004}:
\begin{equation}
    \mathbf{B}=\frac{\mu_0M_\mathrm{sat}R_s^3}{\left(X^2+Y^2+Z^2\right)^{3/2}}\left[ZX\hat{X}+ZY\hat{Y}+(6Z^2-5)\hat{Z}\right],
\end{equation}
where $\mu_0M_\mathrm{sat}=2.26$~T is a saturation magnetization of the magnetic tip, $R_s=100$~nm is the radius of the tip, which is assumed to be a spherical particle.  

\begin{figure*}[h]
\includegraphics[width=0.5\textwidth]{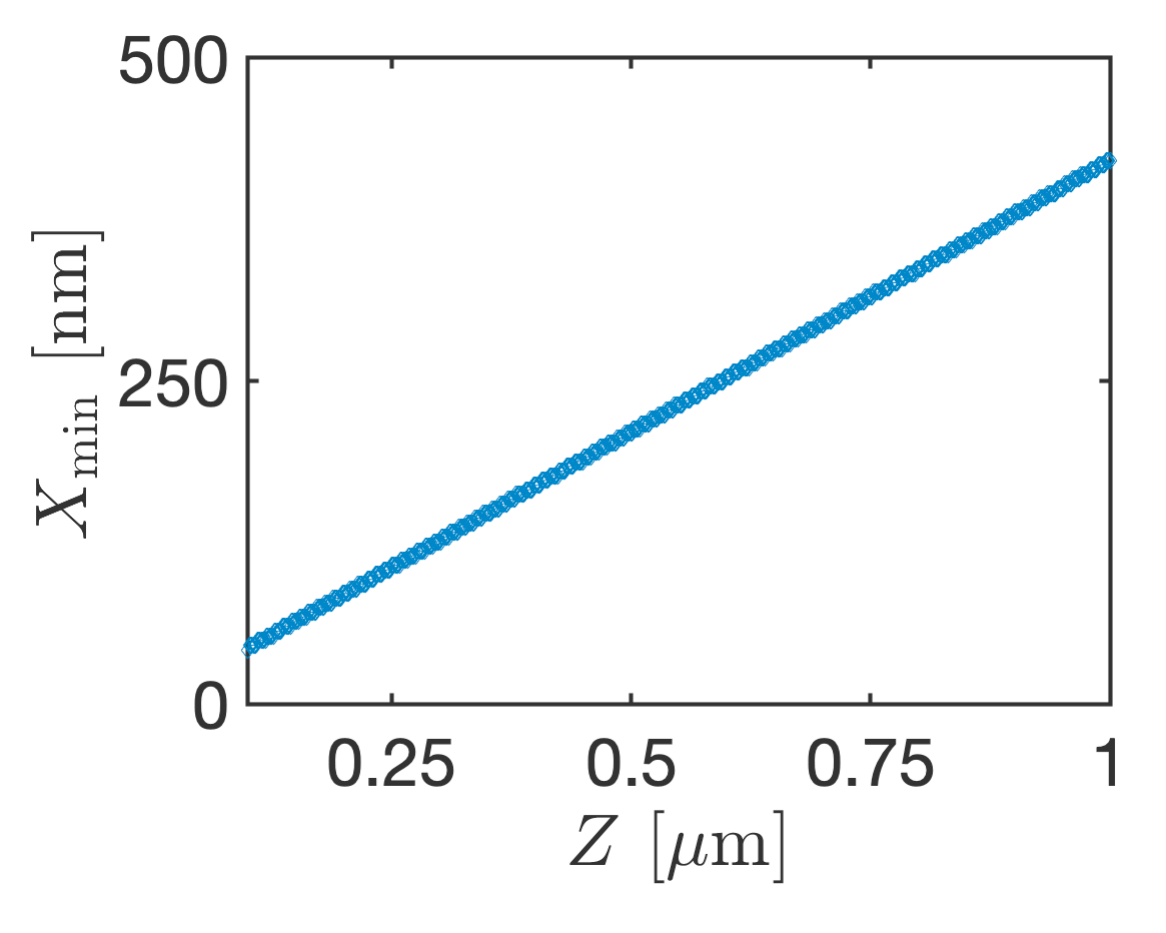}
\caption{Equilibrium position of the domain wall center ($X_\mathrm{min}$) as a function of the vertical distance $Z$ between the tip and the sample surface.}
\label{fig_position_vs_Z}
\end{figure*}

The equilibrium position of the domain wall (Fig.~\ref{fig_position_vs_Z}) is calculated from the minimum condition for $U(X,Z)$ at a given $Z$. The vertical force $F_Z$ is calculated in assumption that the position of the domain wall is fixed (pinned domain wall).
\begin{figure*}[h]
\includegraphics[width=0.5\textwidth]{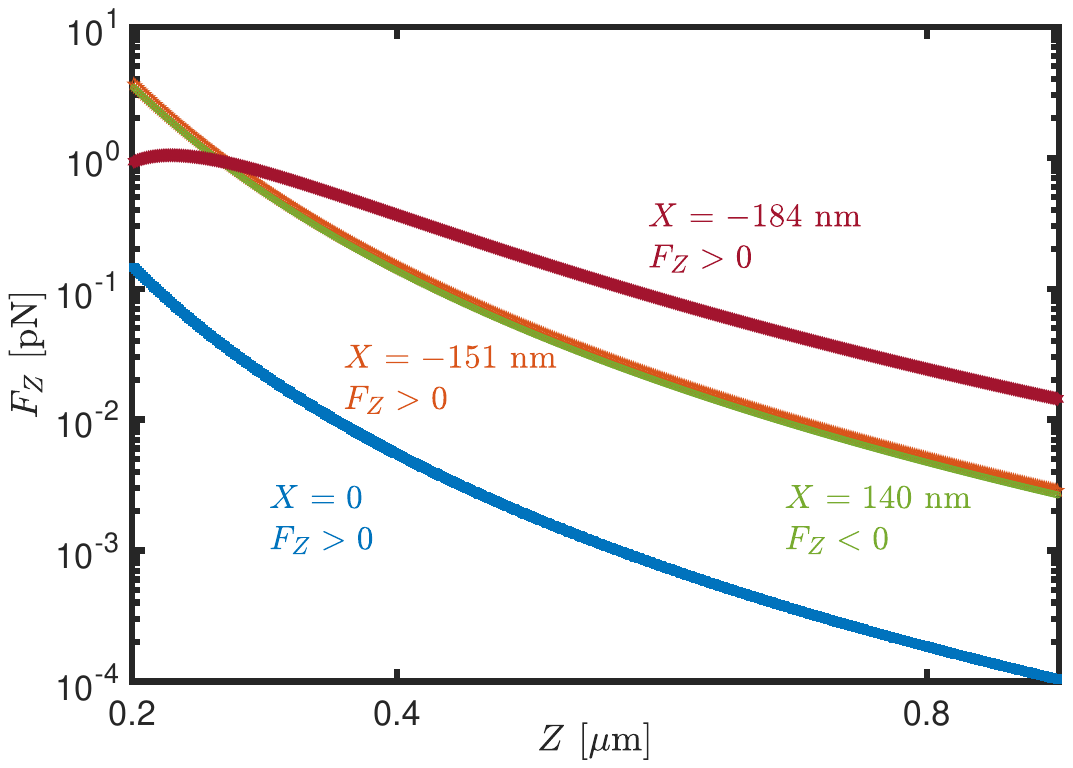}
\caption{Absolute value of the vertical force $F_Z$ as a function of $Z$ for different positions $X$ of the domain wall. The force at the opposite sites of the tip has opposite sign. The curve corresponding to $X=-184$~nm is shown in Fig.3 of the main text.}
\label{vertical_force_vs_Z}
\end{figure*}
%


\begin{thebibliography}{48}%
\makeatletter
\providecommand \@ifxundefined [1]{%
 \@ifx{#1\undefined}
}%
\providecommand \@ifnum [1]{%
 \ifnum #1\expandafter \@firstoftwo
 \else \expandafter \@secondoftwo
 \fi
}%
\providecommand \@ifx [1]{%
 \ifx #1\expandafter \@firstoftwo
 \else \expandafter \@secondoftwo
 \fi
}%
\providecommand \natexlab [1]{#1}%
\providecommand \enquote  [1]{``#1''}%
\providecommand \bibnamefont  [1]{#1}%
\providecommand \bibfnamefont [1]{#1}%
\providecommand \citenamefont [1]{#1}%
\providecommand \href@noop [0]{\@secondoftwo}%
\providecommand \href [0]{\begingroup \@sanitize@url \@href}%
\providecommand \@href[1]{\@@startlink{#1}\@@href}%
\providecommand \@@href[1]{\endgroup#1\@@endlink}%
\providecommand \@sanitize@url [0]{\catcode `\\12\catcode `\$12\catcode
  `\&12\catcode `\#12\catcode `\^12\catcode `\_12\catcode `\%12\relax}%
\providecommand \@@startlink[1]{}%
\providecommand \@@endlink[0]{}%
\providecommand \url  [0]{\begingroup\@sanitize@url \@url }%
\providecommand \@url [1]{\endgroup\@href {#1}{\urlprefix }}%
\providecommand \urlprefix  [0]{URL }%
\providecommand \Eprint [0]{\href }%
\providecommand \doibase [0]{https://doi.org/}%
\providecommand \selectlanguage [0]{\@gobble}%
\providecommand \bibinfo  [0]{\@secondoftwo}%
\providecommand \bibfield  [0]{\@secondoftwo}%
\providecommand \translation [1]{[#1]}%
\providecommand \BibitemOpen [0]{}%
\providecommand \bibitemStop [0]{}%
\providecommand \bibitemNoStop [0]{.\EOS\space}%
\providecommand \EOS [0]{\spacefactor3000\relax}%
\providecommand \BibitemShut  [1]{\csname bibitem#1\endcsname}%
\let\auto@bib@innerbib\@empty
\bibitem [{\citenamefont {{\v{S}}mejkal}\ \emph {et~al.}(2020)\citenamefont
  {{\v{S}}mejkal}, \citenamefont {Gonz{\'{a}}lez-Hern{\'{a}}ndez},
  \citenamefont {Jungwirth},\ and\ \citenamefont {Sinova}}]{Smejkal2020}%
  \BibitemOpen
  \bibfield  {author} {\bibinfo {author} {\bibfnamefont {L.}~\bibnamefont
  {{\v{S}}mejkal}}, \bibinfo {author} {\bibfnamefont {R.}~\bibnamefont
  {Gonz{\'{a}}lez-Hern{\'{a}}ndez}}, \bibinfo {author} {\bibfnamefont
  {T.}~\bibnamefont {Jungwirth}},\ and\ \bibinfo {author} {\bibfnamefont
  {J.}~\bibnamefont {Sinova}},\ }\bibfield  {title} {\bibinfo {title} {{Crystal
  time-reversal symmetry breaking and spontaneous Hall effect in collinear
  antiferromagnets}},\ }\href {https://doi.org/10.1126/sciadv.aaz8809}
  {\bibfield  {journal} {\bibinfo  {journal} {Science Advances}\ }\textbf
  {\bibinfo {volume} {6}},\ \bibinfo {pages} {eaaz8809} (\bibinfo {year}
  {2020})},\ \Eprint {https://arxiv.org/abs/1901.00445} {arXiv:1901.00445}
  \BibitemShut {NoStop}%
\bibitem [{\citenamefont {{\v{S}}mejkal}\ \emph
  {et~al.}(2022{\natexlab{a}})\citenamefont {{\v{S}}mejkal}, \citenamefont
  {Sinova},\ and\ \citenamefont {Jungwirth}}]{Smejkal2021a}%
  \BibitemOpen
  \bibfield  {author} {\bibinfo {author} {\bibfnamefont {L.}~\bibnamefont
  {{\v{S}}mejkal}}, \bibinfo {author} {\bibfnamefont {J.}~\bibnamefont
  {Sinova}},\ and\ \bibinfo {author} {\bibfnamefont {T.}~\bibnamefont
  {Jungwirth}},\ }\bibfield  {title} {\bibinfo {title} {{Beyond Conventional
  Ferromagnetism and Antiferromagnetism: A Phase with Nonrelativistic Spin and
  Crystal Rotation Symmetry}},\ }\href
  {https://doi.org/10.1103/PhysRevX.12.031042} {\bibfield  {journal} {\bibinfo
  {journal} {Physical Review X}\ }\textbf {\bibinfo {volume} {12}},\ \bibinfo
  {pages} {031042} (\bibinfo {year} {2022}{\natexlab{a}})}\BibitemShut
  {NoStop}%
\bibitem [{\citenamefont {{\v{S}}mejkal}\ \emph
  {et~al.}(2022{\natexlab{b}})\citenamefont {{\v{S}}mejkal}, \citenamefont
  {Sinova},\ and\ \citenamefont {Jungwirth}}]{Smejkal2022a}%
  \BibitemOpen
  \bibfield  {author} {\bibinfo {author} {\bibfnamefont {L.}~\bibnamefont
  {{\v{S}}mejkal}}, \bibinfo {author} {\bibfnamefont {J.}~\bibnamefont
  {Sinova}},\ and\ \bibinfo {author} {\bibfnamefont {T.}~\bibnamefont
  {Jungwirth}},\ }\bibfield  {title} {\bibinfo {title} {{Emerging Research
  Landscape of Altermagnetism}},\ }\href
  {https://doi.org/10.1103/PhysRevX.12.040501} {\bibfield  {journal} {\bibinfo
  {journal} {Physical Review X}\ }\textbf {\bibinfo {volume} {12}},\ \bibinfo
  {pages} {040501} (\bibinfo {year} {2022}{\natexlab{b}})},\ \Eprint
  {https://arxiv.org/abs/2204.10844} {arXiv:2204.10844} \BibitemShut {NoStop}%
\bibitem [{\citenamefont {Guo}\ \emph {et~al.}(2023)\citenamefont {Guo},
  \citenamefont {Liu}, \citenamefont {Janson}, \citenamefont {Fulga},
  \citenamefont {van~den Brink},\ and\ \citenamefont {Facio}}]{Guo2023b}%
  \BibitemOpen
  \bibfield  {author} {\bibinfo {author} {\bibfnamefont {Y.}~\bibnamefont
  {Guo}}, \bibinfo {author} {\bibfnamefont {H.}~\bibnamefont {Liu}}, \bibinfo
  {author} {\bibfnamefont {O.}~\bibnamefont {Janson}}, \bibinfo {author}
  {\bibfnamefont {I.~C.}\ \bibnamefont {Fulga}}, \bibinfo {author}
  {\bibfnamefont {J.}~\bibnamefont {van~den Brink}},\ and\ \bibinfo {author}
  {\bibfnamefont {J.~I.}\ \bibnamefont {Facio}},\ }\bibfield  {title} {\bibinfo
  {title} {{Spin-split collinear antiferromagnets: A large-scale ab-initio
  study}},\ }\href {https://doi.org/10.1016/j.mtphys.2023.100991} {\bibfield
  {journal} {\bibinfo  {journal} {Materials Today Physics}\ }\textbf {\bibinfo
  {volume} {32}},\ \bibinfo {pages} {100991} (\bibinfo {year} {2023})},\
  \Eprint {https://arxiv.org/abs/2207.07592} {arXiv:2207.07592} \BibitemShut
  {NoStop}%
\bibitem [{\citenamefont {Feng}\ \emph {et~al.}(2022)\citenamefont {Feng},
  \citenamefont {Zhou}, \citenamefont {{\v{S}}mejkal}, \citenamefont {Wu},
  \citenamefont {Zhu}, \citenamefont {Guo}, \citenamefont
  {Gonz{\'{a}}lez-Hern{\'{a}}ndez}, \citenamefont {Wang}, \citenamefont {Yan},
  \citenamefont {Qin}, \citenamefont {Zhang}, \citenamefont {Wu}, \citenamefont
  {Chen}, \citenamefont {Meng}, \citenamefont {Liu}, \citenamefont {Xia},
  \citenamefont {Sinova}, \citenamefont {Jungwirth},\ and\ \citenamefont
  {Liu}}]{Feng2022}%
  \BibitemOpen
  \bibfield  {author} {\bibinfo {author} {\bibfnamefont {Z.}~\bibnamefont
  {Feng}}, \bibinfo {author} {\bibfnamefont {X.}~\bibnamefont {Zhou}}, \bibinfo
  {author} {\bibfnamefont {L.}~\bibnamefont {{\v{S}}mejkal}}, \bibinfo {author}
  {\bibfnamefont {L.}~\bibnamefont {Wu}}, \bibinfo {author} {\bibfnamefont
  {Z.}~\bibnamefont {Zhu}}, \bibinfo {author} {\bibfnamefont {H.}~\bibnamefont
  {Guo}}, \bibinfo {author} {\bibfnamefont {R.}~\bibnamefont
  {Gonz{\'{a}}lez-Hern{\'{a}}ndez}}, \bibinfo {author} {\bibfnamefont
  {X.}~\bibnamefont {Wang}}, \bibinfo {author} {\bibfnamefont {H.}~\bibnamefont
  {Yan}}, \bibinfo {author} {\bibfnamefont {P.}~\bibnamefont {Qin}}, \bibinfo
  {author} {\bibfnamefont {X.}~\bibnamefont {Zhang}}, \bibinfo {author}
  {\bibfnamefont {H.}~\bibnamefont {Wu}}, \bibinfo {author} {\bibfnamefont
  {H.}~\bibnamefont {Chen}}, \bibinfo {author} {\bibfnamefont {Z.}~\bibnamefont
  {Meng}}, \bibinfo {author} {\bibfnamefont {L.}~\bibnamefont {Liu}}, \bibinfo
  {author} {\bibfnamefont {Z.}~\bibnamefont {Xia}}, \bibinfo {author}
  {\bibfnamefont {J.}~\bibnamefont {Sinova}}, \bibinfo {author} {\bibfnamefont
  {T.}~\bibnamefont {Jungwirth}},\ and\ \bibinfo {author} {\bibfnamefont
  {Z.}~\bibnamefont {Liu}},\ }\bibfield  {title} {\bibinfo {title} {{An
  anomalous Hall effect in altermagnetic ruthenium dioxide}},\ }\href
  {https://doi.org/10.1038/s41928-022-00866-z} {\bibfield  {journal} {\bibinfo
  {journal} {Nature Electronics}\ }\textbf {\bibinfo {volume} {5}},\ \bibinfo
  {pages} {735} (\bibinfo {year} {2022})},\ \Eprint
  {https://arxiv.org/abs/2002.08712} {arXiv:2002.08712} \BibitemShut {NoStop}%
\bibitem [{\citenamefont {Sato}\ \emph {et~al.}(2023)\citenamefont {Sato},
  \citenamefont {Haddad}, \citenamefont {Fulga}, \citenamefont {Assaad},\ and\
  \citenamefont {van~den Brink}}]{Sato2023a}%
  \BibitemOpen
  \bibfield  {author} {\bibinfo {author} {\bibfnamefont {T.}~\bibnamefont
  {Sato}}, \bibinfo {author} {\bibfnamefont {S.}~\bibnamefont {Haddad}},
  \bibinfo {author} {\bibfnamefont {I.~C.}\ \bibnamefont {Fulga}}, \bibinfo
  {author} {\bibfnamefont {F.~F.}\ \bibnamefont {Assaad}},\ and\ \bibinfo
  {author} {\bibfnamefont {J.}~\bibnamefont {van~den Brink}},\ }\bibfield
  {title} {\bibinfo {title} {{Altermagnetic anomalous Hall effect emerging from
  electronic correlations}},\ }\href {http://arxiv.org/abs/2312.16290}
  {\bibfield  {journal} {\bibinfo  {journal} {Arxiv Preprint}\ } (\bibinfo
  {year} {2023})},\ \Eprint {https://arxiv.org/abs/2312.16290}
  {arXiv:2312.16290} \BibitemShut {NoStop}%
\bibitem [{\citenamefont {Naka}\ \emph {et~al.}(2019)\citenamefont {Naka},
  \citenamefont {Hayami}, \citenamefont {Kusunose}, \citenamefont {Yanagi},
  \citenamefont {Motome},\ and\ \citenamefont {Seo}}]{Naka2019}%
  \BibitemOpen
  \bibfield  {author} {\bibinfo {author} {\bibfnamefont {M.}~\bibnamefont
  {Naka}}, \bibinfo {author} {\bibfnamefont {S.}~\bibnamefont {Hayami}},
  \bibinfo {author} {\bibfnamefont {H.}~\bibnamefont {Kusunose}}, \bibinfo
  {author} {\bibfnamefont {Y.}~\bibnamefont {Yanagi}}, \bibinfo {author}
  {\bibfnamefont {Y.}~\bibnamefont {Motome}},\ and\ \bibinfo {author}
  {\bibfnamefont {H.}~\bibnamefont {Seo}},\ }\bibfield  {title} {\bibinfo
  {title} {{Spin current generation in organic antiferromagnets}},\ }\href
  {https://doi.org/10.1038/s41467-019-12229-y} {\bibfield  {journal} {\bibinfo
  {journal} {Nature Communications}\ }\textbf {\bibinfo {volume} {10}},\
  \bibinfo {pages} {4305} (\bibinfo {year} {2019})},\ \Eprint
  {https://arxiv.org/abs/1902.02506} {arXiv:1902.02506} \BibitemShut {NoStop}%
\bibitem [{\citenamefont {Gonz{\'{a}}lez-Hern{\'{a}}ndez}\ \emph
  {et~al.}(2021)\citenamefont {Gonz{\'{a}}lez-Hern{\'{a}}ndez}, \citenamefont
  {{\v{S}}mejkal}, \citenamefont {V{\'{y}}born{\'{y}}}, \citenamefont {Yahagi},
  \citenamefont {Sinova}, \citenamefont {Jungwirth},\ and\ \citenamefont
  {{\v{Z}}elezn{\'{y}}}}]{Gonzalez-Hernandez2021}%
  \BibitemOpen
  \bibfield  {author} {\bibinfo {author} {\bibfnamefont {R.}~\bibnamefont
  {Gonz{\'{a}}lez-Hern{\'{a}}ndez}}, \bibinfo {author} {\bibfnamefont
  {L.}~\bibnamefont {{\v{S}}mejkal}}, \bibinfo {author} {\bibfnamefont
  {K.}~\bibnamefont {V{\'{y}}born{\'{y}}}}, \bibinfo {author} {\bibfnamefont
  {Y.}~\bibnamefont {Yahagi}}, \bibinfo {author} {\bibfnamefont
  {J.}~\bibnamefont {Sinova}}, \bibinfo {author} {\bibfnamefont
  {T.}~\bibnamefont {Jungwirth}},\ and\ \bibinfo {author} {\bibfnamefont
  {J.}~\bibnamefont {{\v{Z}}elezn{\'{y}}}},\ }\bibfield  {title} {\bibinfo
  {title} {{Efficient Electrical Spin Splitter Based on Nonrelativistic
  Collinear Antiferromagnetism}},\ }\href
  {https://doi.org/10.1103/PhysRevLett.126.127701} {\bibfield  {journal}
  {\bibinfo  {journal} {Physical Review Letters}\ }\textbf {\bibinfo {volume}
  {126}},\ \bibinfo {pages} {127701} (\bibinfo {year} {2021})},\ \Eprint
  {https://arxiv.org/abs/2002.07073} {arXiv:2002.07073} \BibitemShut {NoStop}%
\bibitem [{\citenamefont {Krempask{\'{y}}}\ \emph {et~al.}(2024)\citenamefont
  {Krempask{\'{y}}}, \citenamefont {{\v{S}}mejkal}, \citenamefont {D'Souza},
  \citenamefont {Hajlaoui}, \citenamefont {Springholz}, \citenamefont
  {Uhl{\'{i}}řov{\'{a}}}, \citenamefont {Alarab}, \citenamefont
  {Constantinou}, \citenamefont {Strocov}, \citenamefont {Usanov},
  \citenamefont {Pudelko}, \citenamefont {Gonz{\'{a}}lez-Hern{\'{a}}ndez},
  \citenamefont {{Birk Hellenes}}, \citenamefont {Jansa}, \citenamefont
  {Reichlov{\'{a}}}, \citenamefont {{\v{S}}ob{\'{a}}ň}, \citenamefont
  {{Gonzalez Betancourt}}, \citenamefont {Wadley}, \citenamefont {Sinova},
  \citenamefont {Kriegner}, \citenamefont {Min{\'{a}}r}, \citenamefont {Dil},\
  and\ \citenamefont {Jungwirth}}]{Krempasky2024}%
  \BibitemOpen
  \bibfield  {author} {\bibinfo {author} {\bibfnamefont {J.}~\bibnamefont
  {Krempask{\'{y}}}}, \bibinfo {author} {\bibfnamefont {L.}~\bibnamefont
  {{\v{S}}mejkal}}, \bibinfo {author} {\bibfnamefont {S.~W.}\ \bibnamefont
  {D'Souza}}, \bibinfo {author} {\bibfnamefont {M.}~\bibnamefont {Hajlaoui}},
  \bibinfo {author} {\bibfnamefont {G.}~\bibnamefont {Springholz}}, \bibinfo
  {author} {\bibfnamefont {K.}~\bibnamefont {Uhl{\'{i}}řov{\'{a}}}}, \bibinfo
  {author} {\bibfnamefont {F.}~\bibnamefont {Alarab}}, \bibinfo {author}
  {\bibfnamefont {P.~C.}\ \bibnamefont {Constantinou}}, \bibinfo {author}
  {\bibfnamefont {V.}~\bibnamefont {Strocov}}, \bibinfo {author} {\bibfnamefont
  {D.}~\bibnamefont {Usanov}}, \bibinfo {author} {\bibfnamefont {W.~R.}\
  \bibnamefont {Pudelko}}, \bibinfo {author} {\bibfnamefont {R.}~\bibnamefont
  {Gonz{\'{a}}lez-Hern{\'{a}}ndez}}, \bibinfo {author} {\bibfnamefont
  {A.}~\bibnamefont {{Birk Hellenes}}}, \bibinfo {author} {\bibfnamefont
  {Z.}~\bibnamefont {Jansa}}, \bibinfo {author} {\bibfnamefont
  {H.}~\bibnamefont {Reichlov{\'{a}}}}, \bibinfo {author} {\bibfnamefont
  {Z.}~\bibnamefont {{\v{S}}ob{\'{a}}ň}}, \bibinfo {author} {\bibfnamefont
  {R.~D.}\ \bibnamefont {{Gonzalez Betancourt}}}, \bibinfo {author}
  {\bibfnamefont {P.}~\bibnamefont {Wadley}}, \bibinfo {author} {\bibfnamefont
  {J.}~\bibnamefont {Sinova}}, \bibinfo {author} {\bibfnamefont
  {D.}~\bibnamefont {Kriegner}}, \bibinfo {author} {\bibfnamefont
  {J.}~\bibnamefont {Min{\'{a}}r}}, \bibinfo {author} {\bibfnamefont {J.~H.}\
  \bibnamefont {Dil}},\ and\ \bibinfo {author} {\bibfnamefont {T.}~\bibnamefont
  {Jungwirth}},\ }\bibfield  {title} {\bibinfo {title} {{Altermagnetic lifting
  of Kramers spin degeneracy}},\ }\href
  {https://doi.org/10.1038/s41586-023-06907-7} {\bibfield  {journal} {\bibinfo
  {journal} {Nature}\ }\textbf {\bibinfo {volume} {626}},\ \bibinfo {pages}
  {517} (\bibinfo {year} {2024})},\ \Eprint {https://arxiv.org/abs/2308.10681}
  {arXiv:2308.10681} \BibitemShut {NoStop}%
\bibitem [{\citenamefont {Fedchenko}\ \emph {et~al.}(2024)\citenamefont
  {Fedchenko}, \citenamefont {Minar}, \citenamefont {Akashdeep}, \citenamefont
  {D'Souza}, \citenamefont {Vasilyev}, \citenamefont {Tkach}, \citenamefont
  {Odenbreit}, \citenamefont {Nguyen}, \citenamefont {Kutnyakhov},
  \citenamefont {Wind}, \citenamefont {Wenthaus}, \citenamefont {Scholz},
  \citenamefont {Rossnagel}, \citenamefont {Hoesch}, \citenamefont
  {Aeschlimann}, \citenamefont {Stadtmueller}, \citenamefont {Klaeui},
  \citenamefont {Schoenhense}, \citenamefont {Jakob}, \citenamefont
  {Jungwirth}, \citenamefont {Smejkal}, \citenamefont {Sinova},\ and\
  \citenamefont {Elmers}}]{Fedchenko2024}%
  \BibitemOpen
  \bibfield  {author} {\bibinfo {author} {\bibfnamefont {O.}~\bibnamefont
  {Fedchenko}}, \bibinfo {author} {\bibfnamefont {J.}~\bibnamefont {Minar}},
  \bibinfo {author} {\bibfnamefont {A.}~\bibnamefont {Akashdeep}}, \bibinfo
  {author} {\bibfnamefont {S.~W.}\ \bibnamefont {D'Souza}}, \bibinfo {author}
  {\bibfnamefont {D.}~\bibnamefont {Vasilyev}}, \bibinfo {author}
  {\bibfnamefont {O.}~\bibnamefont {Tkach}}, \bibinfo {author} {\bibfnamefont
  {L.}~\bibnamefont {Odenbreit}}, \bibinfo {author} {\bibfnamefont {Q.~L.}\
  \bibnamefont {Nguyen}}, \bibinfo {author} {\bibfnamefont {D.}~\bibnamefont
  {Kutnyakhov}}, \bibinfo {author} {\bibfnamefont {N.}~\bibnamefont {Wind}},
  \bibinfo {author} {\bibfnamefont {L.}~\bibnamefont {Wenthaus}}, \bibinfo
  {author} {\bibfnamefont {M.}~\bibnamefont {Scholz}}, \bibinfo {author}
  {\bibfnamefont {K.}~\bibnamefont {Rossnagel}}, \bibinfo {author}
  {\bibfnamefont {M.}~\bibnamefont {Hoesch}}, \bibinfo {author} {\bibfnamefont
  {M.}~\bibnamefont {Aeschlimann}}, \bibinfo {author} {\bibfnamefont
  {B.}~\bibnamefont {Stadtmueller}}, \bibinfo {author} {\bibfnamefont
  {M.}~\bibnamefont {Klaeui}}, \bibinfo {author} {\bibfnamefont
  {G.}~\bibnamefont {Schoenhense}}, \bibinfo {author} {\bibfnamefont
  {G.}~\bibnamefont {Jakob}}, \bibinfo {author} {\bibfnamefont
  {T.}~\bibnamefont {Jungwirth}}, \bibinfo {author} {\bibfnamefont
  {L.}~\bibnamefont {Smejkal}}, \bibinfo {author} {\bibfnamefont
  {J.}~\bibnamefont {Sinova}},\ and\ \bibinfo {author} {\bibfnamefont {H.~J.}\
  \bibnamefont {Elmers}},\ }\bibfield  {title} {\bibinfo {title} {{Observation
  of time-reversal symmetry breaking in the band structure of altermagnetic
  RuO{\$}{\_}2{\$}}},\ }\href {https://doi.org/10.1126/sciadv.adj4883}
  {\bibfield  {journal} {\bibinfo  {journal} {Science Advances}\ }\textbf
  {\bibinfo {volume} {10}},\ \bibinfo {pages} {31} (\bibinfo {year} {2024})},\
  \Eprint {https://arxiv.org/abs/2306.02170} {arXiv:2306.02170} \BibitemShut
  {NoStop}%
\bibitem [{\citenamefont {Lee}\ \emph {et~al.}(2024)\citenamefont {Lee},
  \citenamefont {Lee}, \citenamefont {Jung}, \citenamefont {Jung},
  \citenamefont {Kim}, \citenamefont {Lee}, \citenamefont {Seok}, \citenamefont
  {Kim}, \citenamefont {Park}, \citenamefont {{\v{S}}mejkal}, \citenamefont
  {Kang},\ and\ \citenamefont {Kim}}]{Lee2024}%
  \BibitemOpen
  \bibfield  {author} {\bibinfo {author} {\bibfnamefont {S.}~\bibnamefont
  {Lee}}, \bibinfo {author} {\bibfnamefont {S.}~\bibnamefont {Lee}}, \bibinfo
  {author} {\bibfnamefont {S.}~\bibnamefont {Jung}}, \bibinfo {author}
  {\bibfnamefont {J.}~\bibnamefont {Jung}}, \bibinfo {author} {\bibfnamefont
  {D.}~\bibnamefont {Kim}}, \bibinfo {author} {\bibfnamefont {Y.}~\bibnamefont
  {Lee}}, \bibinfo {author} {\bibfnamefont {B.}~\bibnamefont {Seok}}, \bibinfo
  {author} {\bibfnamefont {J.}~\bibnamefont {Kim}}, \bibinfo {author}
  {\bibfnamefont {B.~G.}\ \bibnamefont {Park}}, \bibinfo {author}
  {\bibfnamefont {L.}~\bibnamefont {{\v{S}}mejkal}}, \bibinfo {author}
  {\bibfnamefont {C.-J.}\ \bibnamefont {Kang}},\ and\ \bibinfo {author}
  {\bibfnamefont {C.}~\bibnamefont {Kim}},\ }\bibfield  {title} {\bibinfo
  {title} {{Broken Kramers Degeneracy in Altermagnetic MnTe}},\ }\href
  {https://doi.org/10.1103/PhysRevLett.132.036702} {\bibfield  {journal}
  {\bibinfo  {journal} {Physical Review Letters}\ }\textbf {\bibinfo {volume}
  {132}},\ \bibinfo {pages} {036702} (\bibinfo {year} {2024})},\ \Eprint
  {https://arxiv.org/abs/2308.11180} {arXiv:2308.11180} \BibitemShut {NoStop}%
\bibitem [{\citenamefont {Osumi}\ \emph {et~al.}(2023)\citenamefont {Osumi},
  \citenamefont {Souma}, \citenamefont {Aoyama}, \citenamefont {Yamauchi},
  \citenamefont {Honma}, \citenamefont {Nakayama}, \citenamefont {Takahashi},
  \citenamefont {Ohgushi},\ and\ \citenamefont {Sato}}]{Osumi2023}%
  \BibitemOpen
  \bibfield  {author} {\bibinfo {author} {\bibfnamefont {T.}~\bibnamefont
  {Osumi}}, \bibinfo {author} {\bibfnamefont {S.}~\bibnamefont {Souma}},
  \bibinfo {author} {\bibfnamefont {T.}~\bibnamefont {Aoyama}}, \bibinfo
  {author} {\bibfnamefont {K.}~\bibnamefont {Yamauchi}}, \bibinfo {author}
  {\bibfnamefont {A.}~\bibnamefont {Honma}}, \bibinfo {author} {\bibfnamefont
  {K.}~\bibnamefont {Nakayama}}, \bibinfo {author} {\bibfnamefont
  {T.}~\bibnamefont {Takahashi}}, \bibinfo {author} {\bibfnamefont
  {K.}~\bibnamefont {Ohgushi}},\ and\ \bibinfo {author} {\bibfnamefont
  {T.}~\bibnamefont {Sato}},\ }\bibfield  {title} {\bibinfo {title}
  {{Observation of Giant Band Splitting in Altermagnetic MnTe}},\ }\href
  {https://arxiv.org/abs/2308.10117v1 http://arxiv.org/abs/2308.10117}
  {\bibfield  {journal} {\bibinfo  {journal} {Arxiv Preprint}\ } (\bibinfo
  {year} {2023})},\ \Eprint {https://arxiv.org/abs/2308.10117}
  {arXiv:2308.10117} \BibitemShut {NoStop}%
\bibitem [{\citenamefont {{\v{S}}mejkal}\ \emph {et~al.}(2023)\citenamefont
  {{\v{S}}mejkal}, \citenamefont {Marmodoro}, \citenamefont {Ahn},
  \citenamefont {Gonz{\'{a}}lez-Hern{\'{a}}ndez}, \citenamefont {Turek},
  \citenamefont {Mankovsky}, \citenamefont {Ebert}, \citenamefont {D'Souza},
  \citenamefont {{\v{S}}ipr}, \citenamefont {Sinova},\ and\ \citenamefont
  {Jungwirth}}]{Smejkal2023}%
  \BibitemOpen
  \bibfield  {author} {\bibinfo {author} {\bibfnamefont {L.}~\bibnamefont
  {{\v{S}}mejkal}}, \bibinfo {author} {\bibfnamefont {A.}~\bibnamefont
  {Marmodoro}}, \bibinfo {author} {\bibfnamefont {K.-H.}\ \bibnamefont {Ahn}},
  \bibinfo {author} {\bibfnamefont {R.}~\bibnamefont
  {Gonz{\'{a}}lez-Hern{\'{a}}ndez}}, \bibinfo {author} {\bibfnamefont
  {I.}~\bibnamefont {Turek}}, \bibinfo {author} {\bibfnamefont
  {S.}~\bibnamefont {Mankovsky}}, \bibinfo {author} {\bibfnamefont
  {H.}~\bibnamefont {Ebert}}, \bibinfo {author} {\bibfnamefont {S.~W.}\
  \bibnamefont {D'Souza}}, \bibinfo {author} {\bibfnamefont {O.}~\bibnamefont
  {{\v{S}}ipr}}, \bibinfo {author} {\bibfnamefont {J.}~\bibnamefont {Sinova}},\
  and\ \bibinfo {author} {\bibfnamefont {T.}~\bibnamefont {Jungwirth}},\
  }\bibfield  {title} {\bibinfo {title} {{Chiral magnons in altermagnetic
  RuO2}},\ }\href {https://doi.org/10.1103/PhysRevLett.131.256703} {\bibfield
  {journal} {\bibinfo  {journal} {Physical Review Letters}\ }\textbf {\bibinfo
  {volume} {131}},\ \bibinfo {pages} {256703} (\bibinfo {year}
  {2023})}\BibitemShut {NoStop}%
\bibitem [{\citenamefont {Gohlke}\ \emph {et~al.}(2023)\citenamefont {Gohlke},
  \citenamefont {Corticelli}, \citenamefont {Moessner}, \citenamefont
  {McClarty},\ and\ \citenamefont {Mook}}]{Gohlke2023}%
  \BibitemOpen
  \bibfield  {author} {\bibinfo {author} {\bibfnamefont {M.}~\bibnamefont
  {Gohlke}}, \bibinfo {author} {\bibfnamefont {A.}~\bibnamefont {Corticelli}},
  \bibinfo {author} {\bibfnamefont {R.}~\bibnamefont {Moessner}}, \bibinfo
  {author} {\bibfnamefont {P.~A.}\ \bibnamefont {McClarty}},\ and\ \bibinfo
  {author} {\bibfnamefont {A.}~\bibnamefont {Mook}},\ }\bibfield  {title}
  {\bibinfo {title} {{Spurious Symmetry Enhancement in Linear Spin Wave Theory
  and Interaction-Induced Topology in Magnons}},\ }\href
  {https://doi.org/10.1103/PhysRevLett.131.186702} {\bibfield  {journal}
  {\bibinfo  {journal} {Physical Review Letters}\ }\textbf {\bibinfo {volume}
  {131}},\ \bibinfo {pages} {186702} (\bibinfo {year} {2023})}\BibitemShut
  {NoStop}%
\bibitem [{\citenamefont {Gomonay}\ \emph {et~al.}(2018)\citenamefont
  {Gomonay}, \citenamefont {Yamamoto},\ and\ \citenamefont
  {Sinova}}]{Gomonay2018a}%
  \BibitemOpen
  \bibfield  {author} {\bibinfo {author} {\bibfnamefont {O.}~\bibnamefont
  {Gomonay}}, \bibinfo {author} {\bibfnamefont {K.}~\bibnamefont {Yamamoto}},\
  and\ \bibinfo {author} {\bibfnamefont {J.}~\bibnamefont {Sinova}},\
  }\bibfield  {title} {\bibinfo {title} {{Spin caloric effects in
  antiferromagnets assisted by an external spin current}},\ }\href
  {https://doi.org/10.1088/1361-6463/aac56b} {\bibfield  {journal} {\bibinfo
  {journal} {Journal of Physics D: Applied Physics}\ }\textbf {\bibinfo
  {volume} {51}},\ \bibinfo {pages} {264004} (\bibinfo {year} {2018})},\
  \Eprint {https://arxiv.org/abs/1803.07949} {arXiv:1803.07949} \BibitemShut
  {NoStop}%
\bibitem [{\citenamefont {Rezende}\ \emph {et~al.}(2019)\citenamefont
  {Rezende}, \citenamefont {Azevedo},\ and\ \citenamefont
  {Rodr{\'{i}}guez-Su{\'{a}}rez}}]{Rezende2019}%
  \BibitemOpen
  \bibfield  {author} {\bibinfo {author} {\bibfnamefont {S.~M.}\ \bibnamefont
  {Rezende}}, \bibinfo {author} {\bibfnamefont {A.}~\bibnamefont {Azevedo}},\
  and\ \bibinfo {author} {\bibfnamefont {R.~L.}\ \bibnamefont
  {Rodr{\'{i}}guez-Su{\'{a}}rez}},\ }\bibfield  {title} {\bibinfo {title}
  {{Introduction to antiferromagnetic magnons}},\ }\href
  {https://doi.org/10.1063/1.5109132} {\bibfield  {journal} {\bibinfo
  {journal} {Journal of Applied Physics}\ }\textbf {\bibinfo {volume} {126}},\
  \bibinfo {pages} {151101} (\bibinfo {year} {2019})}\BibitemShut {NoStop}%
\bibitem [{Note1()}]{Note1}%
  \BibitemOpen
  \bibinfo {note} {~Strictly speaking, in the case of altermagnets, we should
  call it altermagnetic resonance. However, since the difference between alter-
  and antiferromagnetic dynamics disappears at the $\Gamma $ point, we retain
  the traditional term antiferromagnetic resonance.}\BibitemShut {Stop}%
\bibitem [{\citenamefont {Hodt}\ and\ \citenamefont {Linder}(2023)}]{Hodt2023}%
  \BibitemOpen
  \bibfield  {author} {\bibinfo {author} {\bibfnamefont {E.~W.}\ \bibnamefont
  {Hodt}}\ and\ \bibinfo {author} {\bibfnamefont {J.}~\bibnamefont {Linder}},\
  }\bibfield  {title} {\bibinfo {title} {{Spin pumping in an altermagnet/normal
  metal bilayer}},\ }\href {http://arxiv.org/abs/2310.15220} {\bibfield
  {journal} {\bibinfo  {journal} {Arxiv Preprint}\ } (\bibinfo {year}
  {2023})},\ \Eprint {https://arxiv.org/abs/2310.15220} {arXiv:2310.15220}
  \BibitemShut {NoStop}%
\bibitem [{\citenamefont {Brekke}\ \emph {et~al.}(2023)\citenamefont {Brekke},
  \citenamefont {Brataas},\ and\ \citenamefont {Sudb{\o}}}]{Brekke2023}%
  \BibitemOpen
  \bibfield  {author} {\bibinfo {author} {\bibfnamefont {B.}~\bibnamefont
  {Brekke}}, \bibinfo {author} {\bibfnamefont {A.}~\bibnamefont {Brataas}},\
  and\ \bibinfo {author} {\bibfnamefont {A.}~\bibnamefont {Sudb{\o}}},\
  }\bibfield  {title} {\bibinfo {title} {{Two-dimensional altermagnets:
  Superconductivity in a minimal microscopic model}},\ }\href
  {https://doi.org/10.1103/PhysRevB.108.224421} {\bibfield  {journal} {\bibinfo
   {journal} {Physical Review B}\ }\textbf {\bibinfo {volume} {108}},\ \bibinfo
  {pages} {224421} (\bibinfo {year} {2023})},\ \Eprint
  {https://arxiv.org/abs/2308.08606} {arXiv:2308.08606} \BibitemShut {NoStop}%
\bibitem [{\citenamefont {Bhowal}\ and\ \citenamefont
  {Spaldin}(2022)}]{Bhowal2022}%
  \BibitemOpen
  \bibfield  {author} {\bibinfo {author} {\bibfnamefont {S.}~\bibnamefont
  {Bhowal}}\ and\ \bibinfo {author} {\bibfnamefont {N.~A.}\ \bibnamefont
  {Spaldin}},\ }\bibfield  {title} {\bibinfo {title} {{Magnetic octupoles as
  the order parameter for unconventional antiferromagnetism}},\ }\href
  {http://arxiv.org/abs/2212.03756} {\bibfield  {journal} {\bibinfo  {journal}
  {Arxiv Preprint}\ } (\bibinfo {year} {2022})},\ \Eprint
  {https://arxiv.org/abs/2212.03756} {arXiv:2212.03756} \BibitemShut {NoStop}%
\bibitem [{\citenamefont {Nazaretski}\ \emph {et~al.}(2008)\citenamefont
  {Nazaretski}, \citenamefont {Akhadov}, \citenamefont {Martin}, \citenamefont
  {Pelekhov}, \citenamefont {Hammel},\ and\ \citenamefont
  {Movshovich}}]{Nazaretski2008}%
  \BibitemOpen
  \bibfield  {author} {\bibinfo {author} {\bibfnamefont {E.}~\bibnamefont
  {Nazaretski}}, \bibinfo {author} {\bibfnamefont {E.~A.}\ \bibnamefont
  {Akhadov}}, \bibinfo {author} {\bibfnamefont {I.}~\bibnamefont {Martin}},
  \bibinfo {author} {\bibfnamefont {D.~V.}\ \bibnamefont {Pelekhov}}, \bibinfo
  {author} {\bibfnamefont {P.~C.}\ \bibnamefont {Hammel}},\ and\ \bibinfo
  {author} {\bibfnamefont {R.}~\bibnamefont {Movshovich}},\ }\bibfield  {title}
  {\bibinfo {title} {{Spatial characterization of the magnetic field profile of
  a probe tip used in magnetic resonance force microscopy}},\ }\href
  {https://doi.org/10.1063/1.2937401} {\bibfield  {journal} {\bibinfo
  {journal} {Applied Physics Letters}\ }\textbf {\bibinfo {volume} {92}},\
  \bibinfo {pages} {214104} (\bibinfo {year} {2008})}\BibitemShut {NoStop}%
\bibitem [{\citenamefont {Bhallamudi}\ \emph {et~al.}(2013)\citenamefont
  {Bhallamudi}, \citenamefont {Wolfe}, \citenamefont {Amin}, \citenamefont
  {Labanowski}, \citenamefont {Berger}, \citenamefont {Stroud}, \citenamefont
  {Sinova},\ and\ \citenamefont {Hammel}}]{Bhallamudi2013}%
  \BibitemOpen
  \bibfield  {author} {\bibinfo {author} {\bibfnamefont {V.~P.}\ \bibnamefont
  {Bhallamudi}}, \bibinfo {author} {\bibfnamefont {C.~S.}\ \bibnamefont
  {Wolfe}}, \bibinfo {author} {\bibfnamefont {V.~P.}\ \bibnamefont {Amin}},
  \bibinfo {author} {\bibfnamefont {D.~E.}\ \bibnamefont {Labanowski}},
  \bibinfo {author} {\bibfnamefont {A.~J.}\ \bibnamefont {Berger}}, \bibinfo
  {author} {\bibfnamefont {D.}~\bibnamefont {Stroud}}, \bibinfo {author}
  {\bibfnamefont {J.}~\bibnamefont {Sinova}},\ and\ \bibinfo {author}
  {\bibfnamefont {P.~C.}\ \bibnamefont {Hammel}},\ }\bibfield  {title}
  {\bibinfo {title} {{Experimental Demonstration of Scanned Spin-Precession
  Microscopy}},\ }\href {https://doi.org/10.1103/PhysRevLett.111.117201}
  {\bibfield  {journal} {\bibinfo  {journal} {Phys. Rev. Lett.}\ }\textbf
  {\bibinfo {volume} {111}},\ \bibinfo {pages} {117201} (\bibinfo {year}
  {2013})}\BibitemShut {NoStop}%
\bibitem [{\citenamefont {Rugar}\ \emph {et~al.}(1992)\citenamefont {Rugar},
  \citenamefont {Yannoni},\ and\ \citenamefont {Sidles}}]{Rugar1992}%
  \BibitemOpen
  \bibfield  {author} {\bibinfo {author} {\bibfnamefont {D.}~\bibnamefont
  {Rugar}}, \bibinfo {author} {\bibfnamefont {C.~S.}\ \bibnamefont {Yannoni}},\
  and\ \bibinfo {author} {\bibfnamefont {J.~A.}\ \bibnamefont {Sidles}},\
  }\bibfield  {title} {\bibinfo {title} {{Mechanical detection of magnetic
  resonance}},\ }\href {https://doi.org/10.1038/360563a0} {\bibfield  {journal}
  {\bibinfo  {journal} {Nature}\ }\textbf {\bibinfo {volume} {360}},\ \bibinfo
  {pages} {563} (\bibinfo {year} {1992})}\BibitemShut {NoStop}%
\bibitem [{\citenamefont {Yuan}\ \emph {et~al.}(2018)\citenamefont {Yuan},
  \citenamefont {Wang}, \citenamefont {Yung},\ and\ \citenamefont
  {Wang}}]{Yuan2018}%
  \BibitemOpen
  \bibfield  {author} {\bibinfo {author} {\bibfnamefont {H.~Y.}\ \bibnamefont
  {Yuan}}, \bibinfo {author} {\bibfnamefont {X.~S.}\ \bibnamefont {Wang}},
  \bibinfo {author} {\bibfnamefont {M.-H.}\ \bibnamefont {Yung}},\ and\
  \bibinfo {author} {\bibfnamefont {X.~R.}\ \bibnamefont {Wang}},\ }\bibfield
  {title} {\bibinfo {title} {{Rock-and-roll skyrmion propagation under
  parametric pumping}},\ }\href {https://doi.org/arXiv:1804.07202v1} {\ ,\
  \bibinfo {pages} {1} (\bibinfo {year} {2018})},\ \Eprint
  {https://arxiv.org/abs/1804.07202} {arXiv:1804.07202} \BibitemShut {NoStop}%
\bibitem [{\citenamefont {Tveten}\ \emph {et~al.}(2016)\citenamefont {Tveten},
  \citenamefont {M{\"{u}}ller}, \citenamefont {Linder},\ and\ \citenamefont
  {Brataas}}]{Tveten2016}%
  \BibitemOpen
  \bibfield  {author} {\bibinfo {author} {\bibfnamefont {E.~G.}\ \bibnamefont
  {Tveten}}, \bibinfo {author} {\bibfnamefont {T.}~\bibnamefont
  {M{\"{u}}ller}}, \bibinfo {author} {\bibfnamefont {J.}~\bibnamefont
  {Linder}},\ and\ \bibinfo {author} {\bibfnamefont {A.}~\bibnamefont
  {Brataas}},\ }\bibfield  {title} {\bibinfo {title} {{Intrinsic magnetization
  of antiferromagnetic textures}},\ }\href
  {https://doi.org/10.1103/PhysRevB.93.104408} {\bibfield  {journal} {\bibinfo
  {journal} {Physical Review B}\ }\textbf {\bibinfo {volume} {93}},\ \bibinfo
  {pages} {104408} (\bibinfo {year} {2016})},\ \Eprint
  {https://arxiv.org/abs/1506.06561} {arXiv:1506.06561} \BibitemShut {NoStop}%
\bibitem [{\citenamefont {Haldane}(1983)}]{Haldane1983a}%
  \BibitemOpen
  \bibfield  {author} {\bibinfo {author} {\bibfnamefont {F.~D.~M.}\
  \bibnamefont {Haldane}},\ }\bibfield  {title} {\bibinfo {title} {{Nonlinear
  field theory of large-spin Heisenberg antiferromagnets: Semiclassically
  quantized solitons of the one-dimensional easy-axis N{\'{e}}el state}},\
  }\href {https://doi.org/10.1103/PhysRevLett.50.1153} {\bibfield  {journal}
  {\bibinfo  {journal} {Physical Review Letters}\ }\textbf {\bibinfo {volume}
  {50}},\ \bibinfo {pages} {1153} (\bibinfo {year} {1983})}\BibitemShut
  {NoStop}%
\bibitem [{\citenamefont {Fradkin}\ and\ \citenamefont
  {Stone}(1988)}]{Fradkin1988}%
  \BibitemOpen
  \bibfield  {author} {\bibinfo {author} {\bibfnamefont {E.}~\bibnamefont
  {Fradkin}}\ and\ \bibinfo {author} {\bibfnamefont {M.}~\bibnamefont
  {Stone}},\ }\bibfield  {title} {\bibinfo {title} {{Topological terms in one-
  and two-dimensional quantum Heisenberg antiferromagnets}},\ }\href
  {https://doi.org/10.1103/PhysRevB.38.7215} {\bibfield  {journal} {\bibinfo
  {journal} {Physical Review B}\ }\textbf {\bibinfo {volume} {38}},\ \bibinfo
  {pages} {7215} (\bibinfo {year} {1988})}\BibitemShut {NoStop}%
\bibitem [{\citenamefont {Papanicolaou}(1995)}]{Papanicolaou1995}%
  \BibitemOpen
  \bibfield  {author} {\bibinfo {author} {\bibfnamefont {N.}~\bibnamefont
  {Papanicolaou}},\ }\bibfield  {title} {\bibinfo {title} {{Antiferromagnetic
  domain walls}},\ }\href {https://doi.org/10.1103/PhysRevB.51.15062}
  {\bibfield  {journal} {\bibinfo  {journal} {Physical Review B}\ }\textbf
  {\bibinfo {volume} {51}},\ \bibinfo {pages} {15062} (\bibinfo {year}
  {1995})}\BibitemShut {NoStop}%
\bibitem [{\citenamefont {Ivanov}\ and\ \citenamefont
  {Kolezhuk}(1995)}]{Ivanov1995a}%
  \BibitemOpen
  \bibfield  {author} {\bibinfo {author} {\bibfnamefont {B.~A.}\ \bibnamefont
  {Ivanov}}\ and\ \bibinfo {author} {\bibfnamefont {A.~K.}\ \bibnamefont
  {Kolezhuk}},\ }\bibfield  {title} {\bibinfo {title} {{Solitons with internal
  degrees of freedom in 1D Heisenberg antiferromagnets}},\ }\href
  {https://doi.org/10.1103/PhysRevLett.74.1859} {\bibfield  {journal} {\bibinfo
   {journal} {Physical Review Letters}\ }\textbf {\bibinfo {volume} {74}},\
  \bibinfo {pages} {1859} (\bibinfo {year} {1995})}\BibitemShut {NoStop}%
\bibitem [{\citenamefont {Urban}\ \emph {et~al.}(2006)\citenamefont {Urban},
  \citenamefont {Putilin}, \citenamefont {Wigen}, \citenamefont {Liou},
  \citenamefont {Cross}, \citenamefont {Hammel},\ and\ \citenamefont
  {Roukes}}]{Urban2006}%
  \BibitemOpen
  \bibfield  {author} {\bibinfo {author} {\bibfnamefont {R.}~\bibnamefont
  {Urban}}, \bibinfo {author} {\bibfnamefont {A.}~\bibnamefont {Putilin}},
  \bibinfo {author} {\bibfnamefont {P.~E.}\ \bibnamefont {Wigen}}, \bibinfo
  {author} {\bibfnamefont {S.-H.}\ \bibnamefont {Liou}}, \bibinfo {author}
  {\bibfnamefont {M.~C.}\ \bibnamefont {Cross}}, \bibinfo {author}
  {\bibfnamefont {P.~C.}\ \bibnamefont {Hammel}},\ and\ \bibinfo {author}
  {\bibfnamefont {M.~L.}\ \bibnamefont {Roukes}},\ }\bibfield  {title}
  {\bibinfo {title} {{Perturbation of magnetostatic modes observed by
  ferromagnetic resonance force microscopy}},\ }\href
  {https://doi.org/10.1103/PhysRevB.73.212410} {\bibfield  {journal} {\bibinfo
  {journal} {Physical Review B}\ }\textbf {\bibinfo {volume} {73}},\ \bibinfo
  {pages} {212410} (\bibinfo {year} {2006})}\BibitemShut {NoStop}%
\bibitem [{\citenamefont {Gomonay}\ \emph {et~al.}(2016)\citenamefont
  {Gomonay}, \citenamefont {Jungwirth},\ and\ \citenamefont
  {Sinova}}]{Gomonay2016}%
  \BibitemOpen
  \bibfield  {author} {\bibinfo {author} {\bibfnamefont {O.}~\bibnamefont
  {Gomonay}}, \bibinfo {author} {\bibfnamefont {T.}~\bibnamefont {Jungwirth}},\
  and\ \bibinfo {author} {\bibfnamefont {J.}~\bibnamefont {Sinova}},\
  }\bibfield  {title} {\bibinfo {title} {{High Antiferromagnetic Domain Wall
  Velocity Induced by N{\'{e}}el Spin-Orbit Torques}},\ }\href
  {https://doi.org/10.1103/PhysRevLett.117.017202} {\bibfield  {journal}
  {\bibinfo  {journal} {Physical Review Letters}\ }\textbf {\bibinfo {volume}
  {117}},\ \bibinfo {pages} {017202} (\bibinfo {year} {2016})},\ \Eprint
  {https://arxiv.org/abs/1602.06766} {arXiv:1602.06766} \BibitemShut {NoStop}%
\bibitem [{\citenamefont {Gomonai}\ \emph {et~al.}(1990)\citenamefont
  {Gomonai}, \citenamefont {Ivanov},\ and\ \citenamefont
  {L'vov}}]{Gomonai1990}%
  \BibitemOpen
  \bibfield  {author} {\bibinfo {author} {\bibfnamefont {E.~V.}\ \bibnamefont
  {Gomonai}}, \bibinfo {author} {\bibfnamefont {B.~A.}\ \bibnamefont
  {Ivanov}},\ and\ \bibinfo {author} {\bibfnamefont {V.~A.}\ \bibnamefont
  {L'vov}},\ }\bibfield  {title} {\bibinfo {title} {{Symmetry and dynamics of
  domain walls}},\ }\href@noop {} {\bibfield  {journal} {\bibinfo  {journal}
  {Sov. Phys. JETP}\ }\textbf {\bibinfo {volume} {70}},\ \bibinfo {pages} {174}
  (\bibinfo {year} {1990})}\BibitemShut {NoStop}%
\bibitem [{\citenamefont {Schryer}\ and\ \citenamefont
  {Walker}(1974)}]{Schryer1974a}%
  \BibitemOpen
  \bibfield  {author} {\bibinfo {author} {\bibfnamefont {N.~L.}\ \bibnamefont
  {Schryer}}\ and\ \bibinfo {author} {\bibfnamefont {L.~R.}\ \bibnamefont
  {Walker}},\ }\bibfield  {title} {\bibinfo {title} {{The motion of 180°
  domain walls in uniform dc magnetic fields}},\ }\href
  {https://doi.org/10.1063/1.1663252} {\bibfield  {journal} {\bibinfo
  {journal} {Journal of Applied Physics}\ }\textbf {\bibinfo {volume} {45}},\
  \bibinfo {pages} {5406} (\bibinfo {year} {1974})}\BibitemShut {NoStop}%
\bibitem [{\citenamefont {Lin}\ \emph {et~al.}()\citenamefont {Lin},
  \citenamefont {Chen}, \citenamefont {Lu}, \citenamefont {Liang},
  \citenamefont {Feng}, \citenamefont {Yamagami}, \citenamefont {Osiecki},
  \citenamefont {Leandersson}, \citenamefont {Thiagarajan}, \citenamefont
  {Liu}, \citenamefont {Felser},\ and\ \citenamefont {Ma}}]{Lin2024}%
  \BibitemOpen
  \bibfield  {author} {\bibinfo {author} {\bibfnamefont {Z.}~\bibnamefont
  {Lin}}, \bibinfo {author} {\bibfnamefont {D.}~\bibnamefont {Chen}}, \bibinfo
  {author} {\bibfnamefont {W.}~\bibnamefont {Lu}}, \bibinfo {author}
  {\bibfnamefont {X.}~\bibnamefont {Liang}}, \bibinfo {author} {\bibfnamefont
  {S.}~\bibnamefont {Feng}}, \bibinfo {author} {\bibfnamefont {K.}~\bibnamefont
  {Yamagami}}, \bibinfo {author} {\bibfnamefont {J.}~\bibnamefont {Osiecki}},
  \bibinfo {author} {\bibfnamefont {M.}~\bibnamefont {Leandersson}}, \bibinfo
  {author} {\bibfnamefont {B.}~\bibnamefont {Thiagarajan}}, \bibinfo {author}
  {\bibfnamefont {J.}~\bibnamefont {Liu}}, \bibinfo {author} {\bibfnamefont
  {C.}~\bibnamefont {Felser}},\ and\ \bibinfo {author} {\bibfnamefont
  {J.}~\bibnamefont {Ma}},\ }\href {https://doi.org/10.48550} {\emph {\bibinfo
  {title} {{Observation of Giant Spin Splitting and d-wave Spin Texture in Room
  Temperature Altermagnet RuO2}}}},\ \bibinfo {type} {Tech. Rep.}\BibitemShut
  {Stop}%
\bibitem [{\citenamefont {Bose}\ \emph {et~al.}(2022)\citenamefont {Bose},
  \citenamefont {Schreiber}, \citenamefont {Jain}, \citenamefont {Shao},
  \citenamefont {Nair}, \citenamefont {Sun}, \citenamefont {Zhang},
  \citenamefont {Muller}, \citenamefont {Tsymbal}, \citenamefont {Schlom},\
  and\ \citenamefont {Ralph}}]{Bose2022}%
  \BibitemOpen
  \bibfield  {author} {\bibinfo {author} {\bibfnamefont {A.}~\bibnamefont
  {Bose}}, \bibinfo {author} {\bibfnamefont {N.~J.}\ \bibnamefont {Schreiber}},
  \bibinfo {author} {\bibfnamefont {R.}~\bibnamefont {Jain}}, \bibinfo {author}
  {\bibfnamefont {D.-F.}\ \bibnamefont {Shao}}, \bibinfo {author}
  {\bibfnamefont {H.~P.}\ \bibnamefont {Nair}}, \bibinfo {author}
  {\bibfnamefont {J.}~\bibnamefont {Sun}}, \bibinfo {author} {\bibfnamefont
  {X.~S.}\ \bibnamefont {Zhang}}, \bibinfo {author} {\bibfnamefont {D.~A.}\
  \bibnamefont {Muller}}, \bibinfo {author} {\bibfnamefont {E.~Y.}\
  \bibnamefont {Tsymbal}}, \bibinfo {author} {\bibfnamefont {D.~G.}\
  \bibnamefont {Schlom}},\ and\ \bibinfo {author} {\bibfnamefont {D.~C.}\
  \bibnamefont {Ralph}},\ }\bibfield  {title} {\bibinfo {title} {{Tilted spin
  current generated by the collinear antiferromagnet ruthenium dioxide}},\
  }\href {https://doi.org/10.1038/s41928-022-00744-8} {\bibfield  {journal}
  {\bibinfo  {journal} {Nature Electronics}\ }\textbf {\bibinfo {volume} {5}},\
  \bibinfo {pages} {267} (\bibinfo {year} {2022})},\ \Eprint
  {https://arxiv.org/abs/2108.09150} {arXiv:2108.09150} \BibitemShut {NoStop}%
\bibitem [{\citenamefont {Karube}\ \emph {et~al.}(2022)\citenamefont {Karube},
  \citenamefont {Tanaka}, \citenamefont {Sugawara}, \citenamefont {Kadoguchi},
  \citenamefont {Kohda},\ and\ \citenamefont {Nitta}}]{Karube2022}%
  \BibitemOpen
  \bibfield  {author} {\bibinfo {author} {\bibfnamefont {S.}~\bibnamefont
  {Karube}}, \bibinfo {author} {\bibfnamefont {T.}~\bibnamefont {Tanaka}},
  \bibinfo {author} {\bibfnamefont {D.}~\bibnamefont {Sugawara}}, \bibinfo
  {author} {\bibfnamefont {N.}~\bibnamefont {Kadoguchi}}, \bibinfo {author}
  {\bibfnamefont {M.}~\bibnamefont {Kohda}},\ and\ \bibinfo {author}
  {\bibfnamefont {J.}~\bibnamefont {Nitta}},\ }\bibfield  {title} {\bibinfo
  {title} {{Observation of Spin-Splitter Torque in Collinear Antiferromagnetic
  RuO2}},\ }\href {https://doi.org/10.1103/PhysRevLett.129.137201} {\bibfield
  {journal} {\bibinfo  {journal} {Physical Review Letters}\ }\textbf {\bibinfo
  {volume} {129}},\ \bibinfo {pages} {137201} (\bibinfo {year} {2022})},\
  \Eprint {https://arxiv.org/abs/2111.07487} {arXiv:2111.07487} \BibitemShut
  {NoStop}%
\bibitem [{\citenamefont {Bai}\ \emph {et~al.}(2022)\citenamefont {Bai},
  \citenamefont {Han}, \citenamefont {Feng}, \citenamefont {Zhou},
  \citenamefont {Su}, \citenamefont {Wang}, \citenamefont {Liao}, \citenamefont
  {Zhu}, \citenamefont {Chen}, \citenamefont {Pan}, \citenamefont {Fan},\ and\
  \citenamefont {Song}}]{Bai2022}%
  \BibitemOpen
  \bibfield  {author} {\bibinfo {author} {\bibfnamefont {H.}~\bibnamefont
  {Bai}}, \bibinfo {author} {\bibfnamefont {L.}~\bibnamefont {Han}}, \bibinfo
  {author} {\bibfnamefont {X.~Y.}\ \bibnamefont {Feng}}, \bibinfo {author}
  {\bibfnamefont {Y.~J.}\ \bibnamefont {Zhou}}, \bibinfo {author}
  {\bibfnamefont {R.~X.}\ \bibnamefont {Su}}, \bibinfo {author} {\bibfnamefont
  {Q.}~\bibnamefont {Wang}}, \bibinfo {author} {\bibfnamefont {L.~Y.}\
  \bibnamefont {Liao}}, \bibinfo {author} {\bibfnamefont {W.~X.}\ \bibnamefont
  {Zhu}}, \bibinfo {author} {\bibfnamefont {X.~Z.}\ \bibnamefont {Chen}},
  \bibinfo {author} {\bibfnamefont {F.}~\bibnamefont {Pan}}, \bibinfo {author}
  {\bibfnamefont {X.~L.}\ \bibnamefont {Fan}},\ and\ \bibinfo {author}
  {\bibfnamefont {C.}~\bibnamefont {Song}},\ }\bibfield  {title} {\bibinfo
  {title} {{Observation of Spin Splitting Torque in a Collinear Antiferromagnet
  RuO2}},\ }\href {https://doi.org/10.1103/PhysRevLett.128.197202} {\bibfield
  {journal} {\bibinfo  {journal} {Physical Review Letters}\ }\textbf {\bibinfo
  {volume} {128}},\ \bibinfo {pages} {197202} (\bibinfo {year} {2022})},\
  \Eprint {https://arxiv.org/abs/2109.05933} {arXiv:2109.05933} \BibitemShut
  {NoStop}%
\bibitem [{\citenamefont {Zhou}\ \emph {et~al.}(2024)\citenamefont {Zhou},
  \citenamefont {Feng}, \citenamefont {Zhang}, \citenamefont {{\v{S}}mejkal},
  \citenamefont {Sinova}, \citenamefont {Mokrousov},\ and\ \citenamefont
  {Yao}}]{Zhou2023}%
  \BibitemOpen
  \bibfield  {author} {\bibinfo {author} {\bibfnamefont {X.}~\bibnamefont
  {Zhou}}, \bibinfo {author} {\bibfnamefont {W.}~\bibnamefont {Feng}}, \bibinfo
  {author} {\bibfnamefont {R.-W.}\ \bibnamefont {Zhang}}, \bibinfo {author}
  {\bibfnamefont {L.}~\bibnamefont {{\v{S}}mejkal}}, \bibinfo {author}
  {\bibfnamefont {J.}~\bibnamefont {Sinova}}, \bibinfo {author} {\bibfnamefont
  {Y.}~\bibnamefont {Mokrousov}},\ and\ \bibinfo {author} {\bibfnamefont
  {Y.}~\bibnamefont {Yao}},\ }\bibfield  {title} {\bibinfo {title} {{Crystal
  Thermal Transport in Altermagnetic RuO2}},\ }\href
  {https://doi.org/10.1103/PhysRevLett.132.056701} {\bibfield  {journal}
  {\bibinfo  {journal} {Physical Review Letters}\ }\textbf {\bibinfo {volume}
  {132}},\ \bibinfo {pages} {056701} (\bibinfo {year} {2024})},\ \Eprint
  {https://arxiv.org/abs/2305.01410} {arXiv:2305.01410} \BibitemShut {NoStop}%
\bibitem [{\citenamefont {Jeong}\ \emph {et~al.}(2024)\citenamefont {Jeong},
  \citenamefont {Choi}, \citenamefont {Nair}, \citenamefont {Buiarelli},
  \citenamefont {Pourbahari}, \citenamefont {Oh}, \citenamefont {Bassim},
  \citenamefont {Seo}, \citenamefont {Choi}, \citenamefont {Fernandes},
  \citenamefont {Birol}, \citenamefont {Zhao}, \citenamefont {Lee},\ and\
  \citenamefont {Jalan}}]{Jeong2024}%
  \BibitemOpen
  \bibfield  {author} {\bibinfo {author} {\bibfnamefont {S.~G.}\ \bibnamefont
  {Jeong}}, \bibinfo {author} {\bibfnamefont {I.~H.}\ \bibnamefont {Choi}},
  \bibinfo {author} {\bibfnamefont {S.}~\bibnamefont {Nair}}, \bibinfo {author}
  {\bibfnamefont {L.}~\bibnamefont {Buiarelli}}, \bibinfo {author}
  {\bibfnamefont {B.}~\bibnamefont {Pourbahari}}, \bibinfo {author}
  {\bibfnamefont {J.~Y.}\ \bibnamefont {Oh}}, \bibinfo {author} {\bibfnamefont
  {N.}~\bibnamefont {Bassim}}, \bibinfo {author} {\bibfnamefont
  {A.}~\bibnamefont {Seo}}, \bibinfo {author} {\bibfnamefont {W.~S.}\
  \bibnamefont {Choi}}, \bibinfo {author} {\bibfnamefont {R.~M.}\ \bibnamefont
  {Fernandes}}, \bibinfo {author} {\bibfnamefont {T.}~\bibnamefont {Birol}},
  \bibinfo {author} {\bibfnamefont {L.}~\bibnamefont {Zhao}}, \bibinfo {author}
  {\bibfnamefont {J.~S.}\ \bibnamefont {Lee}},\ and\ \bibinfo {author}
  {\bibfnamefont {B.}~\bibnamefont {Jalan}},\ }\bibfield  {title} {\bibinfo
  {title} {{Altermagnetic Polar Metallic phase in Ultra-Thin
  Epitaxially-Strained RuO2 Films}},\ }\href {http://arxiv.org/abs/2405.05838}
  {\ ,\ \bibinfo {pages} {1} (\bibinfo {year} {2024})},\ \Eprint
  {https://arxiv.org/abs/2405.05838} {arXiv:2405.05838} \BibitemShut {NoStop}%
\bibitem [{\citenamefont {Ryden}\ and\ \citenamefont
  {Lawson}(1970)}]{Ryden1970a}%
  \BibitemOpen
  \bibfield  {author} {\bibinfo {author} {\bibfnamefont {W.~D.}\ \bibnamefont
  {Ryden}}\ and\ \bibinfo {author} {\bibfnamefont {A.~W.}\ \bibnamefont
  {Lawson}},\ }\bibfield  {title} {\bibinfo {title} {{Magnetic Susceptibility
  of IrO2 and RuO2}},\ }\href {https://doi.org/10.1063/1.1672908} {\bibfield
  {journal} {\bibinfo  {journal} {The Journal of Chemical Physics}\ }\textbf
  {\bibinfo {volume} {52}},\ \bibinfo {pages} {6058} (\bibinfo {year}
  {1970})}\BibitemShut {NoStop}%
\bibitem [{\citenamefont {Riga}\ \emph {et~al.}(1977)\citenamefont {Riga},
  \citenamefont {Tenret-No{\"{e}}l}, \citenamefont {Pireaux}, \citenamefont
  {Caudano}, \citenamefont {Verbist},\ and\ \citenamefont
  {Gobillon}}]{Riga1977}%
  \BibitemOpen
  \bibfield  {author} {\bibinfo {author} {\bibfnamefont {J.}~\bibnamefont
  {Riga}}, \bibinfo {author} {\bibfnamefont {C.}~\bibnamefont
  {Tenret-No{\"{e}}l}}, \bibinfo {author} {\bibfnamefont {J.~J.}\ \bibnamefont
  {Pireaux}}, \bibinfo {author} {\bibfnamefont {R.}~\bibnamefont {Caudano}},
  \bibinfo {author} {\bibfnamefont {J.~J.}\ \bibnamefont {Verbist}},\ and\
  \bibinfo {author} {\bibfnamefont {Y.}~\bibnamefont {Gobillon}},\ }\bibfield
  {title} {\bibinfo {title} {{Electronic structure of rutile
  oxides TiO$_2$, RuO$_2$ and IrO$_2$ studied by x-ray photoelectron spectroscopy}},\
  }\href {https://doi.org/10.1088/0031-8949/16/5-6/027} {\bibfield  {journal}
  {\bibinfo  {journal} {Physica Scripta}\ }\textbf {\bibinfo {volume} {16}},\
  \bibinfo {pages} {351} (\bibinfo {year} {1977})}\BibitemShut {NoStop}%
\bibitem [{\citenamefont {Hiraishi}\ \emph {et~al.}(2024)\citenamefont
  {Hiraishi}, \citenamefont {Okabe}, \citenamefont {Koda}, \citenamefont
  {Kadono}, \citenamefont {Muroi}, \citenamefont {Hirai},\ and\ \citenamefont
  {Hiroi}}]{Hiraishi2024}%
  \BibitemOpen
  \bibfield  {author} {\bibinfo {author} {\bibfnamefont {M.}~\bibnamefont
  {Hiraishi}}, \bibinfo {author} {\bibfnamefont {H.}~\bibnamefont {Okabe}},
  \bibinfo {author} {\bibfnamefont {A.}~\bibnamefont {Koda}}, \bibinfo {author}
  {\bibfnamefont {R.}~\bibnamefont {Kadono}}, \bibinfo {author} {\bibfnamefont
  {T.}~\bibnamefont {Muroi}}, \bibinfo {author} {\bibfnamefont
  {D.}~\bibnamefont {Hirai}},\ and\ \bibinfo {author} {\bibfnamefont
  {Z.}~\bibnamefont {Hiroi}},\ }\bibfield  {title} {\bibinfo {title}
  {{Nonmagnetic Ground State in RuO$_2$  Revealed by
  Muon Spin Rotation}},\ }\href
  {https://doi.org/10.1103/PhysRevLett.132.166702} {\bibfield  {journal}
  {\bibinfo  {journal} {Physical Review Letters}\ }\textbf {\bibinfo {volume}
  {132}},\ \bibinfo {pages} {166702} (\bibinfo {year} {2024})},\ \Eprint
  {https://arxiv.org/abs/2403.10028} {arXiv:2403.10028} \BibitemShut {NoStop}%
\bibitem [{\citenamefont {Smolyanyuk}\ \emph {et~al.}(2024)\citenamefont
  {Smolyanyuk}, \citenamefont {Mazin}, \citenamefont {Garcia-Gassull},\ and\
  \citenamefont {Valent{\'{i}}}}]{Smolyanyuk2024}%
  \BibitemOpen
  \bibfield  {author} {\bibinfo {author} {\bibfnamefont {A.}~\bibnamefont
  {Smolyanyuk}}, \bibinfo {author} {\bibfnamefont {I.~I.}\ \bibnamefont
  {Mazin}}, \bibinfo {author} {\bibfnamefont {L.}~\bibnamefont
  {Garcia-Gassull}},\ and\ \bibinfo {author} {\bibfnamefont {R.}~\bibnamefont
  {Valent{\'{i}}}},\ }\bibfield  {title} {\bibinfo {title} {{Fragility of the
  magnetic order in the prototypical altermagnet RuO$_2$}},\ }\href
  {https://doi.org/10.1103/PhysRevB.109.134424} {\bibfield  {journal} {\bibinfo
   {journal} {Physical Review B}\ }\textbf {\bibinfo {volume} {109}},\ \bibinfo
  {pages} {134424} (\bibinfo {year} {2024})},\ \Eprint
  {https://arxiv.org/abs/2310.06909} {arXiv:2310.06909} \BibitemShut {NoStop}%
\bibitem [{\citenamefont {Dzialoshinskii}(1958)}]{Dzialoshinskii1958}%
  \BibitemOpen
  \bibfield  {author} {\bibinfo {author} {\bibfnamefont {I.~E.}\ \bibnamefont
  {Dzialoshinskii}},\ }\href
  {http://www.jetp.ac.ru/cgi-bin/dn/e{\_}006{\_}06{\_}1120.pdf} {\emph
  {\bibinfo {title} {JETP}}},\ \bibinfo {type} {Tech. Rep.}\ (\bibinfo {year}
  {1958})\BibitemShut {NoStop}%
\bibitem [{\citenamefont {Dzyaloshinskii}(1958)}]{Dzyaloshinskii1957}%
  \BibitemOpen
  \bibfield  {author} {\bibinfo {author} {\bibfnamefont {I.}~\bibnamefont
  {Dzyaloshinskii}},\ }\bibfield  {title} {\bibinfo {title} {{A thermodynamic
  theory of "weak" ferromagnetism of antiferromagnetics}},\ }\href@noop {}
  {\bibfield  {journal} {\bibinfo  {journal} {Journal of Physics and Chemistry
  of Solids}\ }\textbf {\bibinfo {volume} {4}},\ \bibinfo {pages} {241}
  (\bibinfo {year} {1958})}\BibitemShut {NoStop}%
\bibitem [{\citenamefont {McClarty}\ and\ \citenamefont
  {Rau}(2023)}]{McClarty2023}%
  \BibitemOpen
  \bibfield  {author} {\bibinfo {author} {\bibfnamefont {P.~A.}\ \bibnamefont
  {McClarty}}\ and\ \bibinfo {author} {\bibfnamefont {J.~G.}\ \bibnamefont
  {Rau}},\ }\bibfield  {title} {\bibinfo {title} {{Landau Theory of
  Altermagnetism}},\ }\href {http://arxiv.org/abs/2308.04484} {\  (\bibinfo
  {year} {2023})},\ \Eprint {https://arxiv.org/abs/2308.04484}
  {arXiv:2308.04484} \BibitemShut {NoStop}%
\bibitem [{\citenamefont {Hals}\ \emph {et~al.}(2011)\citenamefont {Hals},
  \citenamefont {Tserkovnyak},\ and\ \citenamefont {Brataas}}]{Hals2011}%
  \BibitemOpen
  \bibfield  {author} {\bibinfo {author} {\bibfnamefont {K.~M.~D.}\
  \bibnamefont {Hals}}, \bibinfo {author} {\bibfnamefont {Y.}~\bibnamefont
  {Tserkovnyak}},\ and\ \bibinfo {author} {\bibfnamefont {A.}~\bibnamefont
  {Brataas}},\ }\bibfield  {title} {\bibinfo {title} {{Phenomenology of
  Current-Induced Dynamics in Antiferromagnets}},\ }\href
  {https://doi.org/10.1103/PhysRevLett.106.107206} {\bibfield  {journal}
  {\bibinfo  {journal} {Physical Review Letters}\ }\textbf {\bibinfo {volume}
  {106}},\ \bibinfo {pages} {107206} (\bibinfo {year} {2011})},\ \Eprint
  {https://arxiv.org/abs/1012.5655} {arXiv:1012.5655} \BibitemShut {NoStop}%
\bibitem [{\citenamefont {Suter}(2004)}]{Suter2004}%
  \BibitemOpen
  \bibfield  {author} {\bibinfo {author} {\bibfnamefont {A.}~\bibnamefont
  {Suter}},\ }\bibfield  {title} {\bibinfo {title} {{The magnetic resonance
  force microscope}},\ }\href {https://doi.org/10.1016/j.pnmrs.2004.06.001}
  {\bibfield  {journal} {\bibinfo  {journal} {Progress in Nuclear Magnetic
  Resonance Spectroscopy}\ }\textbf {\bibinfo {volume} {45}},\ \bibinfo {pages}
  {239} (\bibinfo {year} {2004})}\BibitemShut {NoStop}%
\end{thebibliography}

\begin{thebibliography}{4}%
\makeatletter
\providecommand \@ifxundefined [1]{%
 \@ifx{#1\undefined}
}%
\providecommand \@ifnum [1]{%
 \ifnum #1\expandafter \@firstoftwo
 \else \expandafter \@secondoftwo
 \fi
}%
\providecommand \@ifx [1]{%
 \ifx #1\expandafter \@firstoftwo
 \else \expandafter \@secondoftwo
 \fi
}%
\providecommand \natexlab [1]{#1}%
\providecommand \enquote  [1]{``#1''}%
\providecommand \bibnamefont  [1]{#1}%
\providecommand \bibfnamefont [1]{#1}%
\providecommand \citenamefont [1]{#1}%
\providecommand \href@noop [0]{\@secondoftwo}%
\providecommand \href [0]{\begingroup \@sanitize@url \@href}%
\providecommand \@href[1]{\@@startlink{#1}\@@href}%
\providecommand \@@href[1]{\endgroup#1\@@endlink}%
\providecommand \@sanitize@url [0]{\catcode `\\12\catcode `\$12\catcode
  `\&12\catcode `\#12\catcode `\^12\catcode `\_12\catcode `\%12\relax}%
\providecommand \@@startlink[1]{}%
\providecommand \@@endlink[0]{}%
\providecommand \url  [0]{\begingroup\@sanitize@url \@url }%
\providecommand \@url [1]{\endgroup\@href {#1}{\urlprefix }}%
\providecommand \urlprefix  [0]{URL }%
\providecommand \Eprint [0]{\href }%
\providecommand \doibase [0]{https://doi.org/}%
\providecommand \selectlanguage [0]{\@gobble}%
\providecommand \bibinfo  [0]{\@secondoftwo}%
\providecommand \bibfield  [0]{\@secondoftwo}%
\providecommand \translation [1]{[#1]}%
\providecommand \BibitemOpen [0]{}%
\providecommand \bibitemStop [0]{}%
\providecommand \bibitemNoStop [0]{.\EOS\space}%
\providecommand \EOS [0]{\spacefactor3000\relax}%
\providecommand \BibitemShut  [1]{\csname bibitem#1\endcsname}%
\let\auto@bib@innerbib\@empty
\bibitem [{\citenamefont {Rajaraman}(1982)}]{Rajaraman82}%
  \BibitemOpen
  \bibfield  {author} {\bibinfo {author} {\bibfnamefont {R.}~\bibnamefont
  {Rajaraman}},\ }\href@noop {} {\emph {\bibinfo {title} {Solitons and
  {I}nstanton}}}\ (\bibinfo  {publisher} {North--Holland},\ \bibinfo {address}
  {Amsterdam},\ \bibinfo {year} {1982})\BibitemShut {NoStop}%
\bibitem [{\citenamefont {Tveten}\ \emph {et~al.}(2013)\citenamefont {Tveten},
  \citenamefont {Qaiumzadeh}, \citenamefont {Tretiakov},\ and\ \citenamefont
  {Brataas}}]{Tveten13}%
  \BibitemOpen
  \bibfield  {author} {\bibinfo {author} {\bibfnamefont {E.~G.}\ \bibnamefont
  {Tveten}}, \bibinfo {author} {\bibfnamefont {A.}~\bibnamefont {Qaiumzadeh}},
  \bibinfo {author} {\bibfnamefont {O.~A.}\ \bibnamefont {Tretiakov}},\ and\
  \bibinfo {author} {\bibfnamefont {A.}~\bibnamefont {Brataas}},\ }\bibfield
  {title} {\bibinfo {title} {Staggered dynamics in antiferromagnets by
  collective coordinates},\ }\href
  {https://doi.org/10.1103/PhysRevLett.110.127208} {\bibfield  {journal}
  {\bibinfo  {journal} {Physical Review Letters}\ }\textbf {\bibinfo {volume}
  {110}},\ \bibinfo {pages} {134446} (\bibinfo {year} {2013})}\BibitemShut
  {NoStop}%
\bibitem [{\citenamefont {Kosevich}\ \emph {et~al.}(1990)\citenamefont
  {Kosevich}, \citenamefont {Ivanov},\ and\ \citenamefont
  {Kovalev}}]{Kosevich90}%
  \BibitemOpen
  \bibfield  {author} {\bibinfo {author} {\bibfnamefont {A.~M.}\ \bibnamefont
  {Kosevich}}, \bibinfo {author} {\bibfnamefont {B.~A.}\ \bibnamefont
  {Ivanov}},\ and\ \bibinfo {author} {\bibfnamefont {A.~S.}\ \bibnamefont
  {Kovalev}},\ }\bibfield  {title} {\bibinfo {title} {Magnetic solitons},\
  }\href {https://doi.org/10.1016/0370-1573(90)90130-T} {\bibfield  {journal}
  {\bibinfo  {journal} {Physics Reports}\ }\textbf {\bibinfo {volume} {194}},\
  \bibinfo {pages} {117} (\bibinfo {year} {1990})}\BibitemShut {NoStop}%
\bibitem [{\citenamefont {Suter}()}]{Suter2004}%
  \BibitemOpen
  \bibfield  {author} {\bibinfo {author} {\bibfnamefont {A.}~\bibnamefont
  {Suter}},\ }\bibfield  {title} {\bibinfo {title} {The magnetic resonance
  force microscope},\ }\href
  {https://doi.org/https://doi.org/10.1016/j.pnmrs.2004.06.001} {\bibfield
  {journal} {\bibinfo  {journal} {Prog. Nucl. Magn. Reson. Spectrosc.}\
  }\textbf {\bibinfo {volume} {45}},\ \bibinfo {pages} {239}}\BibitemShut
  {NoStop}%
\end{thebibliography}

\end{document}